\documentclass[fleqn,usenatbib]{mnras}

\usepackage{newtxtext,newtxmath}

\usepackage[T1]{fontenc}
\usepackage{ae,aecompl}

\usepackage{amsfonts}
\usepackage{amsmath}
\usepackage[ngerman, british]{babel}
\usepackage{booktabs}
\usepackage{color}
\usepackage{comment}
\usepackage{enumitem}
\usepackage{graphicx}
\usepackage[utf8]{inputenc}
\usepackage[version=4]{mhchem}
\usepackage{natbib}
\usepackage[]{tabularx}

\newcommand{\mum}{\,\mu\hbox{m}}

\title[Modelling HD~131488  ]{The debris disc of HD~131488 -- Bringing together thermal emission and scattered light\thanks{Based on observations collected at the European Southern Observatory under ESO programme 0101.C-0753(B)}}

\author[N. Pawellek et al.]{Nicole Pawellek$^{1,2}$ \thanks{E-mail: nicole.pawellek@univie.ac.at}, 
Attila Mo\'or$^{2,3}$,
Florian Kirchschlager$^{4,5}$,
Julien Milli$^{6}$,
\newauthor
\'Agnes K\'osp\'al$^{2,3,7,13}$,
P\'eter \'Abrah\'am$^{2,3,7}$,
Sebastian Marino$^{8}$,
Mark Wyatt$^{9}$,
\newauthor
Isabel Rebollido$^{10}$,
A. Meredith Hughes$^{11}$,
Faustine Cantalloube$^{12}$,
Thomas Henning$^{13}$
\\
$^{1}$ Institut f\"ur Astrophysik, Universit\"at Wien, T\"urkenschanzstra\ss{}e 17, 1180 Vienna, Austria\\
$^{2}$ Konkoly Observatory, Research Centre for Astronomy and Earth Sciences, E\"otv\"os Lor\'and Research Network (ELKH),\\ 
Konkoly-Thege Mikl\'os \'ut 15-17, 1121 Budapest, Hungary \\
$^{3}$ CSFK, MTA Centre of Excellence, Budapest, Konkoly-Thege Mikl\'os \'ut 15-17., 1121, Hungary \\
$^{4}$ Department of Physics and Astronomy, University College London, Gower Street, London WC1E 6BT, UK\\
$^{5}$ Sterrenkundig Observatorium, Ghent University, Krijgslaan 281-S9, B9000 Gent, Belgium\\
$^{6}$ Univ. Grenoble Alpes, CNRS, IPAG, 38000 Grenoble, France \\
$^{7}$ ELTE E\"otv\"os Lor\'and University, Institute of Physics, P\'azm\'any P\'eter s\'et\'any 1/A, 1117 Budapest, Hungary\\
$^{8}$ Department for Physics and Astronomy, University of Exeter, Stocker Road, EX4 4QL Exeter, UK\\
$^{9}$ Institute of Astronomy, University of Cambridge, Madingley Road, CB3 0HA Cambridge, UK\\
$^{10}$ Centro de Astrobiolog\'ia (CAB, CSIC-INTA), Camino Bajo del Castillo s/n, Villanueva de la Cañada, 28692 Madrid, Spain\\
$^{11}$ Astronomy Department and Van Vleck Observatory, Wesleyan University, 96 Foss Hill Drive, Middletown, CT 06459, USA\\
$^{12}$ Aix Marseille Univ, CNRS, CNES, LAM, Marseille, France\\
$^{13}$ Max-Planck-Institut f\"ur Astronomie, K\"onigstuhl 17, 69117 Heidelberg, Germany\\
}

\date{Accepted XXX. Received YYY; in original form ZZZ}

\pubyear{2019}

\begin{document}
\label{firstpage}
\pagerange{\pageref{firstpage}--\pageref{lastpage}}
\maketitle

\begin{abstract}
We show the first SPHERE/IRDIS and IFS data of the \ce{CO}-rich debris disc around HD~131488.
We use N-body simulations to model both the scattered light images and the SED of the disc in a self-consistent way.
We apply the Henyey-Greenstein approximation, Mie theory, and the Discrete Dipole Approximation to model the emission of individual dust grains. 
Our study shows that only when gas drag is taken into account can we find a model that is consistent with scattered light as well as thermal emission data of the disc. The models suggest a gas surface density of $2\times10^{-5}~M_\oplus/$au$^2$ which is in agreement with estimates from ALMA observations. Thus, our modelling procedure allows us to roughly constrain the expected amount of gas in a debris disc without actual gas measurements. 
We also show that the shallow size distribution of the dust leads to a significant contribution of large particles to the overall amount of scattered light.
The scattering phase function indicates a dust porosity of $\sim0.2\ldots 0.6$ which is in agreement with a pebble pile scenario for planetesimal growth. 

\end{abstract}




\section{Introduction}

Circumstellar debris discs are optically thin collections of solids ranging from planetesimal size bodies down to dust grains. All of the components are thought to be part of a collisional cascade in which larger objects are gradually ground to smaller particles through mutual destructive collisions \citep{wyatt-2008}. 
Observations are only sensitive to the lowest mass end of the population: thermal emission of dust is detectable at infrared (IR) and millimetre wavelengths, while the stellar light scattered by the disc is mostly observable in the optical/near-IR regime. Besides the gravitational force exerted by the star 
and possible planets, the observed second generation grains are also subject to additional non-gravitational forces related to stellar radiation and wind \citep{krivov-2010}. 
Depending inversely on their size the stellar radiation pressure can push dust grains on more and more eccentric orbits forming an extended halo of barely bound particles outside the planetesimal belt. Below a certain size dust is blown out from the system by this force. By causing an inward migration of grains the 
Poynting-Robertson effect and stellar wind drag can affect the spatial distribution of dust as well.

The presence of gas can also influence the dynamics and the spatial 
distribution of dust particles.
Recently, detections of far-IR \ion{O}{i}, \ion{C}{ii} and particularly 
millimeter \ce{CO} lines revealed gas 
in some 20 debris discs 
\citep[e.g.,][]{dent-et-al-2014, marino-et-al-2016, lieman-sifry-et-al-2016, moor-et-al-2017, matra-et-al-2019, schneiderman-et-al-2021}.
In most of these 
systems the observed gas is likely secondary and released through collisions 
of large volatile-rich bodies \citep{kral-et-al-2017, kral-et-al-2019, marino-et-al-2020}. Remarkably, as observations 
of less abundant \ce{CO} isotopologues implied, in a subset of this sample the mass 
of \ce{CO} gas is on a par with that of less massive protoplanetary discs 
\citep{kospal-et-al-2013, pericaud-et-al-2017, moor-et-al-2019,rebollido-et-al-2022}. All of these \ce{CO}-rich debris discs 
surround young, 5--50\,Myr old, A-type stars; their observed gas material is at 
least partly co-located with the cold dust in these systems. Though we can 
measure only a few constituents of the complete gas mixture it is probable that the 
total gas mass is at least comparable to that of dust measured at millimeter 
wavelengths \citep[e.g.][]{moor-et-al-2017}.   

High spatial resolution scattered light images 
of several  \ce{CO}-rich debris disks revealed complex structures in the distribution 
of those small dust grains that could be most affected by gas. 
Optical and near-IR observations of HD~141569A  
have revealed complex morphology with two rings at $\sim$245 and 400\,au as 
well as spiral features in the disc \citep[][and references therein]{biller-et-al-2015}. 
Millimeter interferometric \ce{CO} line observations of the system showed that the inner ring is located just
at the outer edge of the gas disc \citep{flaherty-et-al-2016,difolco-et-al-2020}.  
Recent imaging with VLT/SPHERE showed additional concentric ringlets
between 47 and 93\,au cospatial with the gas disc \citep{perrot-et-al-2016}. 
By observing the disc around HD\,131835 with SPHERE, \citet{feldt-et-al-2017} 
also discovered concentric dust rings that are co-located with the  
circumstellar gas material.

Though some of these structures could be the result of perturbations
by planetary or stellar companions \citep[e.g.][]{augereau-papaloizou-2004, feldt-et-al-2017}, the presence of 
gas in these systems provides alternative explanations. 
In an optically thin gaseous debris disc, the combined effect of stellar radiation 
and gas drag induces radial drift of dust. 
Assuming gas pressure decreases with radius, small dust particles migrate outward and can form a 
narrow ring at the outer edge of the gas disc \citep{takeuchi-artymowicz-2001}, as in the case of HD\,141569A 
\citep{flaherty-et-al-2016}. 
Considering heating of gas by photo-electrons from nearby dust grains 
\citet{klahr-lin-2005} and \citet{besla-wu-2007} found that this effect 
can lead to strong local dust enhancements via a clumping instability.
Depending on the gas and dust surface density  
such photoelectric instability can result in sharp concentric rings 
providing a feasible explanation for such features in HD\,131835 and 
HD\,141569A systems \citep{richert-et-al-2018}. 

The usage of scattered light data is not limited to structural analysis, 
multiwavelength measurements allow to investigate the grain properties as well.
Based on VLT/SPHERE imaging of the gaseous debris disc around HD\,32297, \citet{bhowmik-et-al-2019} 
reported the presence of copious amount of grains smaller than the blowout size in this system 
and proposed that their pile-up is related to gas drag and/or avalanche mechanisms. 
According to this scenario by slowing down the motion of small unbound grains -- that 
otherwise would leave the system on the orbital timescale
\citep[][]{meyer-et-al-2007} -- gas drag 
can result in an overabundance of such particles with respect to a 
gas free case. Interestingly, the colour of the gas-bearing debris discs around HD~36546 \citep{lawson-et-al-2021} and HD~141569 \citep{singh-et-al-2021} also suggests the presence of copious submicron-sized or highly porous grains.

To further explore gas-dust interactions in an optically thin environment, in this paper we present the first spatially resolved scattered light images of the gaseous debris disc around HD\,131488 obtained with the SPHERE instrument.
HD\,131488 is an A1-type star at a distance of 154.0$\pm$2.5\,pc \citep{gaia-collaboration-et-al-2016, lindegren-et-al-2018, bailer-jones-et-al-2018}, 
that likely belongs to the $\sim$16\,Myr old Upper Centaurus Lupus subgroup of the 
Scorpius-Centaurus association \citep{melis-et-al-2013, pecaut-mamajek-2016}. The 
infrared excess emission of the system was first identified by \citet{melis-et-al-2013}. 
Based on its SED, the disc has a high fractional luminosity, the dust material is likely 
distributed in two belts \citep{melis-et-al-2013}.  
Using the ALMA interferometer at 1.3\,mm, recently the disc was successfully resolved in continuum and in $J=2-1$ rotational transitions of \ce{^{12}CO}, \ce{^{13}CO} 
and \ce{C^{18}O} lines \citep{moor-et-al-2017}. 
Also, the star shows a gaseous \ce{CaII} absorption associated with its circumstellar environment \citep{rebollido-et-al-2018}.

Analysis of the continuum observation 
implied that large cold dust grains are confined in a ring with a radius of 
$\sim$0\farcs57 ($\sim$88\,au). HD\,131488 has the highest C$^{18}$O line luminosity 
of any gas-bearing debris disc found to date, in fact its measured $L_{\rm C^{18}O}$ 
is even $\sim$1.5$\times$ higher than that of the well known protoplanetary disc 
around the Herbig\,Ae star, HD\,100453 \citep{vanderplas-et-al-2019}
and $\sim$2$\times$ higher than that in TW\,Hya \citep{favre-et-al-2013}. 
The outstandingly high CO mass makes HD\,131488 an ideal choice for a detailed study of the gas-dust interaction.

In \S~\ref{sec:observations} we will discuss the observations of the disc around HD~131488, and the data reduction of the scattered light data. \S~\ref{sec:modelling} will give an overview of the theoretical background used to generate our disc models starting with orbital parameters, scattered light models, and grain composition up to generating the final model images. We will discuss the grain size distribution (\S~\ref{sec:size_distribution}) including the influence of gas present within the disc.
In \S~\ref{sec:results} we present the results of our modelling effort which is then followed by a discussion in \S~\ref{sec:discussion}.

\section{Observations and data reduction}
\label{sec:observations}

\begin{figure*}
	\includegraphics[width=0.8\textwidth]{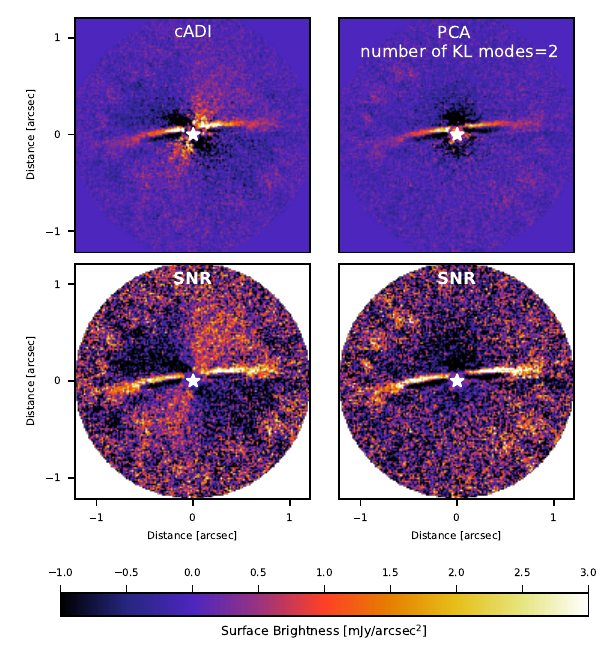}
	\caption{Top: classical ADI-reduced scattered light image (left) and PCA-reduced scattered light image (right) of HD~131488, obtained with IRDIS at $1.6\mu$m (average of the two IRDIS spectral channels). 
 The surface brightness is given in mJy/arcsec$^2$.
	North is up and East is to the left.
 Bottom: respective SNR maps for classical ADI and PCA images.
	 }
	\label{fig:ADI_image}
\end{figure*}

We observed the disc around HD~131488 in the programme 0101.C-0753(B) (PI: A. Mo\'or) on the night of 7$^\text{th}$ April 2018 for one hour with the SPHERE instrument of the VLT \citep{beuzit-et-al-2019,dohlen-et-al-2008}, which is fed with an extreme adaptive optics system to reach a high contrast close to the star. We used the IRDIFS observing mode combining the near-infrared dual-band camera IRDIS \citep{dohlen-et-al-2008}
with the IFS \citep{claudi-et-al-2008}.
The IRDIS observations were carried out using the dual band H23 filter with central wavelengths of 1.593$\mum$ for H2 and 1.667$\mum$ for H3 and a width of 139~nm. The IFS observations dispersed the Y-J band into 39 spectral channels from 958~nm to 1.329$\mum$. 
Both observations used the coronagraph N\_ALC\_YJH\_S \citep{martinez-et-al-2009,carbillet-et-al-2011} with a diameter of 185~mas and were performed in pupil tracking mode to allow for angular differential imaging \citep[ADI,][]{marois-et-al-2006}. The observing conditions were slightly worse than average for the VLT site, with an average DIMM seeing of 0.88$''$ and an average coherence time as measured by the Paranal MASS-DIMM of 3.5~ms. For a star of magnitude G=8, this resulted in an average Strehl in the H band of about 70\%, as estimated by the adaptive optics system, while the direct measurement performed on the average  non-coronagraphic images obtained before and after the coronagraphic sequences indicate a value of 66\%. Despite this performance being lower than average for an instrument like SPHERE, the conditions were very stable, leading to a good dataset with homogeneous quality.

The raw IRDIS and IFS data were pre-processed by the High Contrast Data Centre (HC-DC)\footnote{The HC DC, previously known as the SPHERE DC, performs data reduction on request and also processes all SPHERE public data to make them available publicly. More information is available at \href{https://sphere.osug.fr/spip.php?rubrique16}{https://sphere.osug.fr/spip.php?rubrique16}} \citep{delorme-et-al-2017}.
This pre-processing consists of flat fielding, bad-pixel correction, background subtraction, frame registration, and the IFS wavelength calibration. It uses native recipes from the ESO Data Reduction and Handling software \citep{pavlov-et-al-2008}
complemented by additional recipes developed by the SPHERE Data Center. This pre-processing results in spectro-temporal master cubes of images. For IRDIS, this represents a sequence of 76 images in 2 spectral channels, spanning $28.9^{\circ}$ of field rotation for 34 min effective integration time. For the IFS, this represents a sequence of 60 images in 39 spectral channels, spanning $26.6^{\circ}$ of field rotation for 32 min effective integration time.

We then processed the data with a classical Angular Differential Imaging (ADI; \citealt{marois-et-al-2006})
reduction technique, which consisted of building a model of the coronagraphic image from the median of all pupil-stabilised images, which was then subtracted from each frame before de-rotating and stacking the images. To improve upon this reduction, we also performed a slightly more aggressive data reduction, where the model of the coronagraphic image is constructed using a Principal Component Analysis (PCA; \citealt{soummer-et-al-2012, amara-et-al-2012})
retaining two principal components, a value found to maximise the signal-to-noise (SNR) of the disk. The reduction was performed over the whole frame in a single area extending from 36 mas to 1.23\arcsec{} radially. In Fig.~\ref{fig:ADI_image}, we show the result of both reductions for IRDIS.

The image was normalised to mJy/arcsec$^2$ in the following way. On the non-coronagraphic image, we measured the flux density encircled within a circle of radius 0.1~arcsec, encompassing the PSF core, wings and diffraction spikes from the spiders. Then this flux density is corrected by the transmission of the neutral density filter used to obtain the non-coronagraphic image, and by the ratio between the detector integration time (DIT) of the coronagraphic and non-coronagraphic images, to obtain a reference conversion value. To convert the coronagraphic image from ADU to mJy/arcsec$^2$, the coronagraphic image is divided by the reference conversion value, multiplied by the stellar flux density of HD~131488 found to be 6.9~Jy at the central wavelength of the H band and divided by the pixel surface area in arcsec$^2$. The pixel scale of IRDIS is 0.01225 arcsec/pixel \citep{maire-et-al-2016}. The image in Fig.~\ref{fig:ADI_image} was not corrected by the throughput of the algorithm, which requires a disc model. 

\begin{figure*}
    \includegraphics[width=\textwidth]{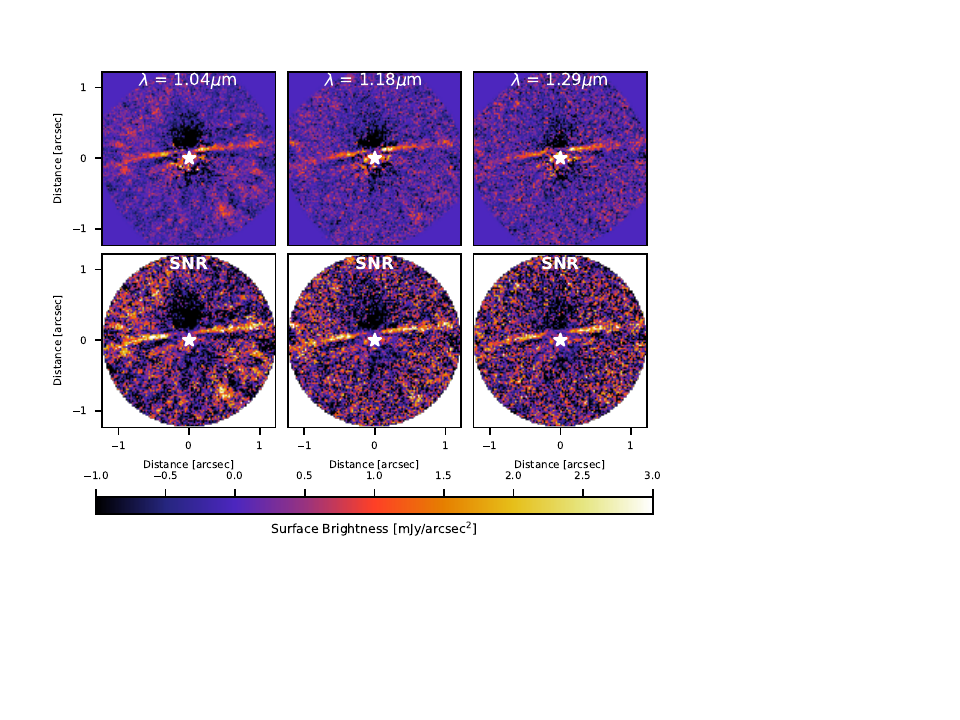}
    \caption{SPHERE/IFS data of HD~131488 reduced with PCA and binned in the following three spectral channels (from left to right): $\lambda = 1.04, 1.18, 1.29\,\mu$m. Top: Surface brightness maps. The surface brightness is given in mJy/arcsec$^2$. North is up and East is to the left. Bottom: SNR maps of the respective surface brightness maps.}
    \label{fig:IFS_data}
\end{figure*}

For the IFS data (Fig.~\ref{fig:IFS_data}), the master cube consists of temporal 60 frames and 39 spectral channels. We binned the spectral channels in three broader channels centred around 1.04~$\mu$m, 1.18~$\mu$m and 1.29~$\mu$m, with a width of  0.16~$\mu$m, 0.12~$\mu$m, and 0.09~$\mu$m respectively. We reduced each spectral channel independently with a PCA algorithm. The disc is clearly detected in each of those three spectral channels.

\subsection{Radial profiles}
\label{sec:radial_profile}
The ADI and PCA reduced H23-band images (Fig.~\ref{fig:ADI_image}) clearly show the detected debris disc of HD~131488 between a radial distance of $0.14\arcsec$ (22~au) and $0.58\arcsec$ (90~au) in both eastern and western directions. Both reduction methods lead to similar results.
To derive the PA and inclination of the disc we used forward modelling applying the Henyey-Greenstein approach (see \S~\ref{sec:HG_results} for details).
We found the PA of the disc in H23-band to be $(97\pm 2)^\circ$ and the inclination to be $(84^{+1.5}_{-2.0})^\circ$. 

\begin{figure*}
\includegraphics[width=\textwidth]{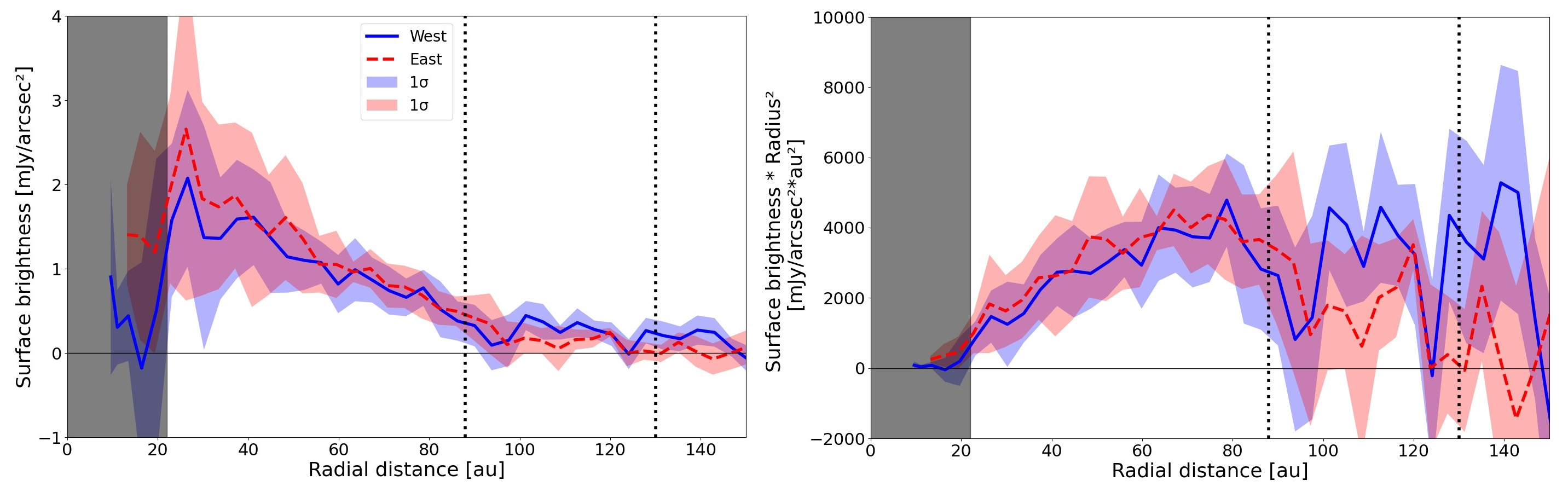}
\caption{Left panel: Surface brightness as a function of radial distance to the star for the SPHERE observations in PCA reduction. The blue solid line shows the western part of the disc, the red dashed line the eastern part. Blue and red shaded areas show the 1$\sigma$ noise level. The grey filled area shows the region where signal and noise are similar. Vertical black dashed lines give the location of the planetesimal belt at 88~au. On the western side a tentative detection of scattered light can be found up to $\sim$130~au. 
Right panel: same as left panel, but multiplied by radial distance squared.
}
\label{fig:Radial_Profile}
\end{figure*}

By fitting the measured ALMA visibilities using a Gaussian ring model \cite{moor-et-al-2017} obtained comparable parameter values in thermal emission: $\text{PA} = (96\pm1)^\circ$ and $i = (82\pm3)^\circ$. That study finds the maximum of the surface brightness at a distance of $(88\pm3)$~au, and a total disc width of ($46\pm12$)~au. 
The location of the peak of surface brightness is similar to the result of our scattered light observations from SPHERE ($\sim90$~au).

We extracted the radial profile of the surface brightness using the same method as described in \cite{choquet-et-al-2017} where H-band data of 49~Cet are analysed. In this method we produce slices along the semi-major axis with a length of 3~pixels above and below that axis 
and a width of 2~pixels. The length was found to provide the best SNR while covering the complete vertical disc extent.
Then we calculate the mean value of the flux density for each slice. 
We estimate the noise level of the images by generating similar slices as for the radial profile itself, but along a line perpendicular to the disc's semi-major axis.
Thus, the slices are located outside of the disc.
Then we calculate the standard deviation of each slice. The result is shown in Fig.~\ref{fig:Radial_Profile}.

Our observations reach an average disc signal-to-noise (SNR) level of 4 using an ADI reduction between 22 and 80~au. This is a stronger detection than using a PCA reduction with a SNR of 3 for the same region. This is caused by a more aggressive reduction process of PCA leading to more over-subtraction of the disc and a lower SNR.
Within a radius of 22~au the noise level is of the same order of magnitude as the disc signal. Thus, we will exclude the inner region from further analyses. 
The right panel of Fig.~(\ref{fig:Radial_Profile}) suggests a possible detection beyond 90~au, especially in the western direction ($\sim130$~au). However, in this region the SNR is low so that the actual extent beyond 90~au remains uncertain. 

\subsection{Presence of planets}
\label{sec:planets}

\begin{figure}
\centering
\includegraphics[width=0.8\columnwidth]{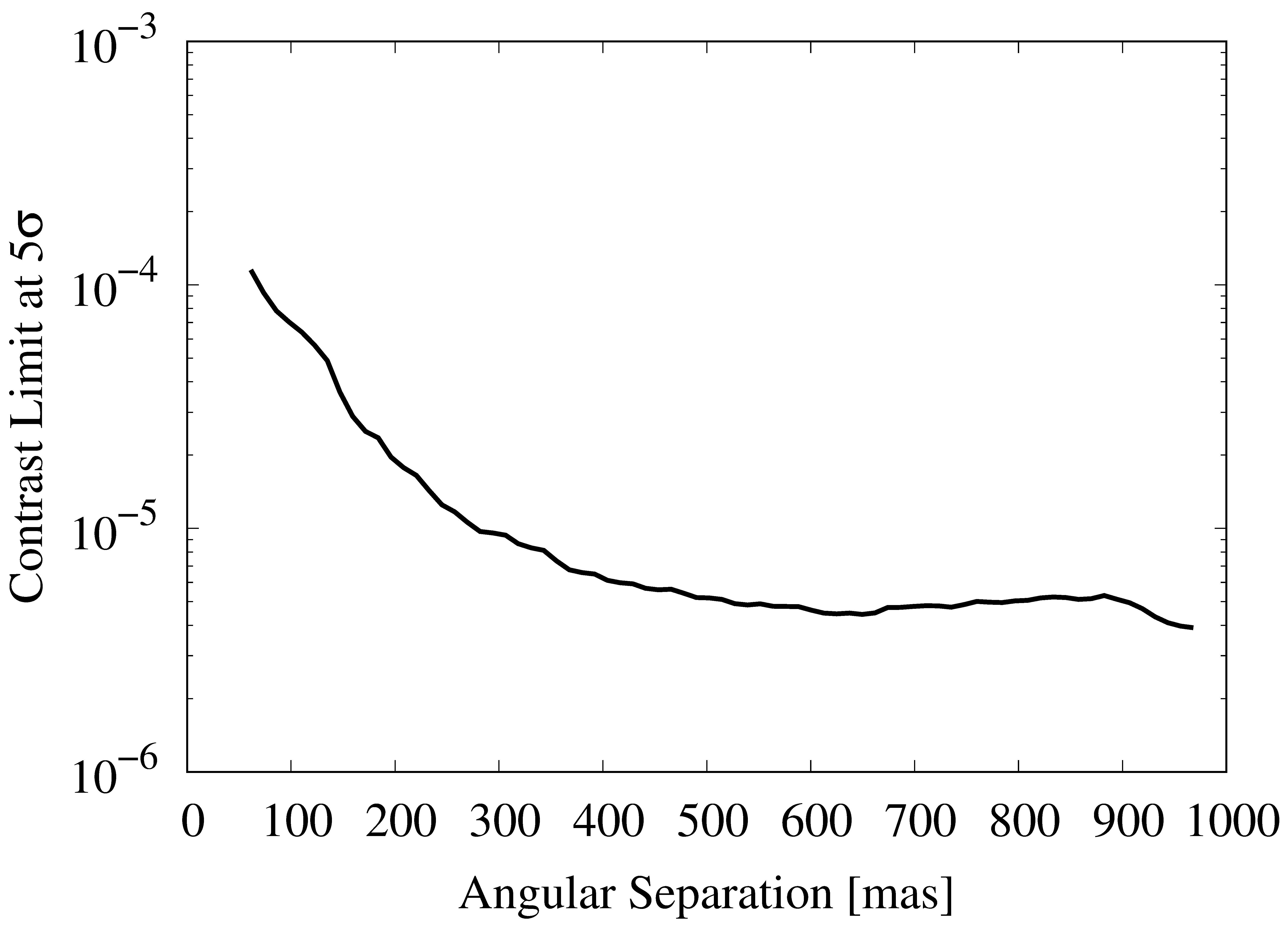}
\caption{Contrast curve for observations of HD131488 inferred from the ANDROMEDA code.}
\label{fig:ANDROMEDA_Results}
\end{figure}

In order to specifically look for point-like sources, such as exoplanets, we made use of the ANgular Differential OptiMal Exoplanet Detection Algorithm \citep[ANDROMEDA,][]{cantalloube-et-al-2015}, as implemented in the High-Contrast Data Centre \citep{delorme-et-al-2017}, which utilises angular differential imaging (ADI) and an inverse problem approach based on a maximum-likelihood estimator. 
It performs a pair-wise subtraction of frames with different rotation angles, models the expected signature that a planetary signal would leave in the residual image (using the off-axis PSF taken before and/or after the observing sequence) and tracks this signal within the pairs of residual images. We used a minimum rotation angle $\delta_\text{min}$ of $1\lambda/D$ between frames within a pair to limit self-subtraction, as recommended in \citet{cantalloube-et-al-2015}. Some bright disc signal remain in the final ANDROMEDA SNR map (maximum SNR of about $9$ for the front part of the ring) because the ring is narrow and its signal may appear point-like. Besides those, there is no point source above the $5\,\sigma$ contrast threshold shown in Fig.~\ref{fig:ANDROMEDA_Results}.

\section{Theoretical Background}
\label{sec:modelling}

In order to analyse the scattered light and thermal emission data of HD~131488, we make use of the MODERATO code \citep[][]{wyatt-et-al-1999, lee-chiang-2016, olofsson-et-al-2019, pawellek-et-al-2019b} calculating the orbits of dust particles influenced by stellar gravity, radiation pressure, and collisional evolution. 
From their position within the disc the code infers the grains' flux density and generates disc images which can be compared to the actual observational images. 

In \S~\ref{sec:orbitparameters} we explain the theoretical approach to calculate the particle orbits. In \S~\ref{sec:scattering_models}-\ref{sec:comparison} we  describe the optical models and parameters, including dust compositions, used to infer the flux density of the dust.  
\S~\ref{sec:size_distribution} focuses on the size distribution of the dust that is influenced by collisional forces and transport processes such as radiation pressure or gas drag. Finally, in \S~\ref{sec:images} we show the resulting images.

\subsection{Orbit parameters}
\label{sec:orbitparameters}

The orbits of dust particles are altered by a number of processes, such as collisions or Poynting-Robertson drag. One of the strongest mechanisms for early-type stars is stellar radiation pressure shaping the overall dust grain distribution of a debris disc hosted by such a star. It is characterised by the radiation pressure parameter, $\beta$, which is defined as the ratio between the radiation force and stellar gravity \citep[][]{burns-et-al-1979}. 
For arbitrary particles (including monomers and agglomerates), $\beta$ is given by
\begin{equation}
    \beta \equiv \frac{\left|\vec{F}_\text{rad}\right|}{\left|\vec{F}_\text{g}\right|}= \frac{1}{4\pi\,G\,c}\times\frac{L_\text{star}}{M_\text{star}}\times\frac{\sigma_\text{grain}\,\overline{Q}_\text{pr}}{m_\text{grain}},
    \label{eq:beta_general}
\end{equation}
where $L_\text{star}$ and $M_\text{star}$ are the stellar luminosity and mass, $G$ the gravitational constant, $c$ the speed of light, $\overline{Q}_\text{pr}$ the radiation pressure efficiency averaged over the stellar spectrum, and $\sigma_\text{grain}$ and $m_\text{grain}$ the particle cross-section and mass. 
In this study, we will focus on spherical particles (including porous grains) so that $\beta$ can be calculated by 
\begin{equation}
 \beta  = \frac{3}{16\pi G c}\times\frac{L_\text{star}}{M_\text{star}}\times \frac{\overline{Q}_\text{pr}}{\varrho\, s\, (1-P)},
 \label{eq:beta}
\end{equation}
 where $\varrho$ is the bulk density of the dust material, $s$ is the grain radius (referred to as \textit{size}), and $P$ is the porosity of the dust material \citep{kirchschlager-wolf-2013}.
The parameter $\overline{Q}_\text{pr}$ is given as
\begin{equation}
\overline{Q}_\text{pr} = \frac{\int Q_\text{pr}(s, \lambda)\, F_\lambda\, \text{d}\lambda}{\int F_\lambda\,\text{d}\lambda},
\end{equation}
where $Q_\text{pr}$ describes the radiation pressure efficiency depending on $s$, wavelength, $\lambda$, and the stellar flux density, $F_\lambda$.
For each grain and wavelength, $Q_\text{pr}$ depends on the absorption and scattering efficiencies, $Q_\text{abs}$ and $Q_\text{sca}$, respectively as well as the asymmetry parameter, $\langle \cos(\vartheta)\rangle$ (also called $g$), and is calculated by
\begin{equation}
 Q_\text{pr}(s,\lambda) = Q_\text{abs}(s,\lambda) + Q_\text{sca}(s,\lambda)\,\left[1 - \langle \cos(\vartheta)\rangle(s, \lambda)\right].
 \label{eq:radiationpressureefficiency}
\end{equation}
The asymmetry parameter depends on the scattering angle, $\vartheta$ and is calculated following \cite{bohren-huffman-1983}:
\begin{equation}
    \langle \cos(\vartheta)\rangle = g = \int\limits_{4\pi} p\, \cos(\vartheta)\,d\Omega,
    \label{eq:asymmetryparameter}
\end{equation}
with $p$ being the phase function and $\Omega$ the solid angle. In total intensity, the phase function is given by $S_{11}$ of the M\"uller matrix \citep[][]{bohren-huffman-1983}.

Knowing $\beta$, the orbital parameters of the dust grains can be inferred using the equations from \cite{wyatt-et-al-1999}.
Assuming that a dust particle is released from a planetesimal which possesses the orbital parameters semi-major axis, $a_\text{p}$, eccentricity, $e_\text{p}$, and true anomaly $f_\text{p}$, the orbit parameters of the dust grain (semi-major axis $a_\text{d}$ and eccentricity $e_\text{d}$)  can be calculated by
\begin{align}
	a_\text{d} &= \frac{a_\text{p}(1-\beta)\,(1-e_\text{p}^2)}{1-e_\text{p}^2-2\beta(1+e_\text{p}\cos(f_\text{p}))}
    \label{eq:semimajor_axis}\\
	e_\text{d}^2 &= \frac{\beta^2 + e_\text{p}^2+ 2\beta e_\text{p}\cos(f_\text{p})}{(1-\beta)^2}
    \label{eq:eccentricity}.
\end{align}
From eq.~(\ref{eq:eccentricity}) we see that the particle's eccentricity reaches a value of larger than one when
\begin{equation*}
    \beta \geq \frac{1 + e_\text{p}}{2\,[1+e_\text{p}\cos(f_\text{p})]}.
\end{equation*}
Assuming that $e_\text{p}$ equals zero, i.e. the planetesimals possess circular orbits, this means that particles with $\beta \geq 1/2$ are expelled from the stellar system on either parabolic ($\beta = 1/2$) or hyperbolic orbits. The particle size where $\beta = 1/2$ is then called \textit{blowout limit}.
In the special case of $\beta \geq 1$, the trajectories of the particles become anomalous hyperbolas for which $e_\text{d}\leq -1$ \citep{wyatt-et-al-1999, krivov-et-al-2006}. 

We emphasise that the particles' orbit parameters (semi-major axis and eccentricity) are determined not only by the bulk density of the material, but also by the optical properties of the grains (e.g., absorption efficiency). There are different methods which can be used to infer those optical properties. The three of them used in this study will be discussed in the following section.

\subsection{Scattered light models}
\label{sec:scattering_models}
The most common approach to calculate isotropic thermal emission is Mie theory \citep[][]{mie-1908, bohren-huffman-1983} where the particles are assumed to be compact spheres. Due to the isotropy of thermal emission, the particles' shape is of no significant importance and disc models are usually in good agreement with observational data \citep[e.g.,][]{matra-et-al-2018, moor-et-al-2020,  pawellek-et-al-2021}.
This looks different for scattered light data where we have to take into account the shape of the dust grains. 
Here the approach with Mie theory often leads to poor modelling results for debris discs, likely because the grains do not possess spherical shape \citep[][see \S~\ref{sec:mie_theory} for details]{pawellek-et-al-2019b}. 

Thus, alternative models are applied.
A common approach is the Henyey-Greenstein (HG) approximation \citep{henyey-greenstein-1941} which does not include any information on the shape of the grains, but there are other methods as well such as the Discrete Dipole Approximation \citep[DDA,][]{purcell-et-al-1973}. We will now introduce HG (\S~\ref{sec:HG_approximation}), Mie (\S~\ref{sec:mie_theory}), and DDA (\S~\ref{sec:dda_theory}) as approaches to model scattered light observations.

\subsubsection{Henyey-Greenstein approximation}
\label{sec:HG_approximation}

The HG approach is used to calculate the scattering phase function, $p$, of the dust material assuming a simple analytical equation:
\begin{equation}
    p(\vartheta) = \frac{1}{4\pi} \frac{1-\langle \cos(\vartheta)\rangle^2}{\left[1+\langle \cos(\vartheta)\rangle^2-2\,\langle \cos(\vartheta)\rangle\,\cos(\vartheta)\right]^{3/2}},
    \label{eq:HG}
\end{equation}
where the asymmetry parameter, $\langle \cos(\vartheta)\rangle$, is fixed to a certain value between $-1$ (back scattering) and $1$ (forward scattering). Isotropic scattering implies $\langle \cos(\vartheta)\rangle=0$. 
Applying this model to scattered light observations of debris discs, some studies infer the best-fitting asymmetry parameter to derive the general scattering properties of the dust material \citep[e.g.,][]{schneider-et-al-2006,millar-blanchaer-et-al-2015,olofsson-et-al-2016, engler-et-al-2017, olofsson-et-al-2020} which helps identify possible dust compositions. Other studies assume isotropic scattering to fit larger samples of discs \citep[e.g.,][]{esposito-et-al-2020}, in order to infer general scattering properties.

The HG approximation usually considers the bulk scattering properties of the dust, rather than considering the behaviour of different grain sizes in the disc, and is usually connected to simple geometric brightness profiles. 
To improve the HG method, some studies combined grain size distributions with HG properties to model debris discs more realistically \citep[e.g.,][]{esposito-et-al-2016, lee-chiang-2016, olofsson-et-al-2016}. 
The disadvantage of this approach is its inconsistency. By fixing the asymmetry parameter, $\langle\cos(\vartheta)\rangle$, the $\beta$ parameter (eq.~\ref{eq:radiationpressureefficiency}) is altered. This is because $\beta$ depends on the radiation pressure efficiency, $Q_\text{pr}$, which depends on $\langle\cos(\vartheta)\rangle$.
Thus, a fixed HG parameter leads to a change in the spatial distribution of the dust \citep[see appendix in][]{pawellek-et-al-2019b}. However, while this mixed approach does not provide reliable grain size information, it does allow the spatial dust distribution to be readily extracted from scattered light images. 

Attempts have been made to scale $\beta$ correctly without taking into account any size information or optical properties of the particles  \citep[e.g.,][]{adam-et-al-2021, olofsson-et-al-2022b}. The advantage is that there is only a small number of free parameters to model, but any information on possible dust compositions remain unused.
So far, a self-consistent calculation of optical (scattered light) and dynamical parameters (particle distribution,  eqs.~\ref{eq:semimajor_axis} and \ref{eq:eccentricity}) is not possible with the HG approach. 
We note that this method is a good approach to analyse the material phase function when not focusing on individual particles though.

\subsubsection{Mie theory}
\label{sec:mie_theory}

A solution to overcome the difficulties of the simple HG approximation is to apply a scattering model which includes different particle shapes e.g., Mie theory assuming compact spherical grains, or the hollow spheres model \citep[][]{min-et-al-2005}.

While Mie theory is easy to implement into a code, it has the disadvantage of overestimating the forward scattering observed especially for large (tens of micron-sized) grains \citep[e.g.,][]{schuerman-et-al-1981, bohren-huffman-1983, weiss-wrana-1983, mugnai-et-al-1986, mcguire-hapke-1995}. 
To circumvent this, it is possible to exclude large grains from the models as done in \cite{pawellek-et-al-2019b} where the maximum size included in the scattered light model was fixed to 10~$\mu$m. However, it is possible that those grains still contribute to the overall flux density of the debris disc (see \S~\ref{sec:size_distribution} for details) and thus, alter the results of the modelling.
This leads to an optimisation problem where we need to find the best maximum size so that all contributing particle sizes are taken into account, and at the same time the overestimation of the forward scattering is minimised.

A possibility to lower the forward scattering is to change the dust material. 
Here we can apply the Effective Medium Theory (EMT) using Bruggeman's mixing rule \citep{bruggeman-1935, bruggeman-1936} to generate mixtures of different sorts of dust.
While Mie theory assumes compact spheres as particles, we can simulate porous material with EMT by generating inclusions of vacuum within the matrix of dust grains. Then the usual Mie calculations can be applied.

\subsubsection{Discrete Dipole approximation}
\label{sec:dda_theory}

Another way is to use a more complex model e.g.,the Discrete Dipole Approximation (DDA) where the optical properties of the grains are calculated by assuming that a particle can be described by a spatial distribution of $N$ discrete polarisable dipoles \citep{purcell-et-al-1973, draine-1988}. With this method the particle shape is not limited to that of a simple sphere but can represent nearly any arbitrary structure including porous agglomerates or fluffy particles.
 
The DDA method is highly flexible and can accommodate a huge variety of particle shapes. Thus, a number of free parameters needs to be introduced, e.g., the dust composition, the grade of porosity, particle shape, etc. In general, these free parameters are barely constrained, however, in combination with information from comets in our own Solar system it is possible to make reasonable assumptions on those parameters (\S~\ref{sec:composition}).  
While the advantage of using DDA to model debris discs in scattered light is evident -- creating a self-consistent model of dynamical and optical properties -- the main disadvantage is its limited applicability to particles of large grain size to wavelength ratio \citep{draine-flatau-2010}. This limit defines a maximum grain size of ${\lesssim}10\,\mu$m for a wavelength around ${\sim}1\,\mu$m \citep{kirchschlager-wolf-2013}.
Another caveat for DDA are highly conducting materials that we will not take into account in this study \citep{michel-et-al-1996}. 

So far, DDA is rarely used to model scattered light of debris discs. 
\cite{kirchschlager-wolf-2013} investigated the influence of grain porosity on the particles' optical properties using DDA. They found that the blowout size significantly increases for porous particles compared to compact grains. The study presented by \cite{brunngraeber-et-al-2017} showed that the minimum grain size and the slope of the grain size distribution are significantly overestimated when modelling debris discs composed of porous dust with a disc model assuming spherical, compact grains. \cite{arnold-et-al-2019} proved that the blowout sizes of agglomerated particles and spherical grains significantly differ but that the dust composition also plays an important role (see \S~\ref{sec:composition} for details).

A few theoretical studies investigate the influence of porosity and irregularity on particles' optical properties  \citep[e.g.,][]{blum-wurm-2008, kirchschlager-wolf-2013,kirchschlager-et-al-2014, ysard-et-al-2018}, while analysis of protoplanetary discs assume non-spherical particles \citep[e.g.,][]{pinte-et-al-2008, birnstiel-et-al-2010, ricci-et-al-2012, min-et-al-2016}. There are also studies analysing the scattered light coming from comets in our Solar System using DDA \citep[e.g.,][]{zubko-2013}. However,
so far there is no direct application of DDA to actual debris disc observations which we want to address in this study.

\subsection{Dust composition}
\label{sec:composition}

While the HG approach does not take into account individual dust compositions such as astrosilicate, water ice, or carbon, the application of Mie theory or DDA makes it possible to choose appropriate compositions freely. Indeed, there is a whole zoo of possible materials \citep[e.g.,][]{zubko-et-al-1996, henning-mutschke-1997, li-greenberg-1998, draine-2003, jaeger-et-al-2008, mutschke-mohr-2019} which makes it necessary to make assumption for the composition based on Solar system data or spectra of debris discs.

Considering spectra, outer planetesimal belts observed with instruments like \textit{Spitzer}/IRS typically do not reveal any solid state features that would allow the dust composition to be constrained \citep{chen-et-al-2006}. The reason is that the dust grains are usually either too cold or too large to generate visible spectral features. 
Of course, there are exceptions for debris discs with hot dust components like HD~172555 \citep{chen-et-al-2006},  HD~36546 \citep{lisse-et-al-2017}, or HD~145263 \citep{lisse-et-al-2020} where silica and carbon-rich material was found. 
On the other hand, studies of comets and asteroids in the Solar system showed that the solid material in our own debris disc often possesses high porosities with values of $P\sim 50$~per cent \citep[e.g.,][]{fulle-et-al-2015, sakatani-et-al-2021}.
We note that it remains debatable whether the dust composition in the inner region (element abundances and porosity) is similar to that in the outer region due to different material processing. 

In terms of our study we want to combine dynamical and optical properties in one model to infer grain sizes. The dust composition is not the main focus here since no spectra or polarimetric data are available for HD~131488 that would allow conclusions on the solid material. 
Thus, reliable size information is more important.
\cite{arnold-et-al-2019} found that for absorptive particles like pure amorphous carbon grains,
porous spheres produce much larger blowout sizes
than dust agglomerates, while for weakly absorbing, pure silicate grains, porous spheres produce slightly smaller
blowout sizes than agglomerates.

To explore the potential of DDA models in comparison to Mie theory, we will use the porous grain model of \cite{kirchschlager-wolf-2013} and consider particles of a basic spherical shape with inclusions of vacuum to reflect grain porosity.
We assume the diameter of the vacuum inclusions (``voids'') to be as large as 1/100 of the grain diameter, 
apply astronomical silicate \citep{draine-2003} and vary the porosity of the dust grains. 
This approach is comparable to that of \cite{arnold-et-al-2019} which also uses a basic spherical shape for the particles. However, \cite{arnold-et-al-2019} uses different void sizes so that the particle structure becomes more complex. With their irregular shape the orientation of the grains within the debris disc also becomes more important. In Appendix~\ref{sec:void_distribution} we show the influence of different void sizes and spatial distributions on the scattering phase function.

We decided against a variation of the voids for our scattered light models of the debris disc around HD~131488 as this would open several dimensions in parameter space (size and spatial distribution of the voids, three directions for the orientation of each particle) making the modelling of the disc complicated and expensive in computational time. Furthermore, from a statistical point of view we would assume that while the orientation of irregular particles is important, averaging over all of them would lead to optical properties similar to a (nearly) spherical particle rendering the computational effort moot. However, we note that based on Appendix~\ref{sec:void_distribution} a dust composition using a different void size might lead to different results than inferred in this study.

\subsection{Comparing HG, Mie and DDA}
\label{sec:comparison}

\subsubsection{Radiation pressure parameter}
\begin{figure}
\includegraphics[width=\columnwidth]{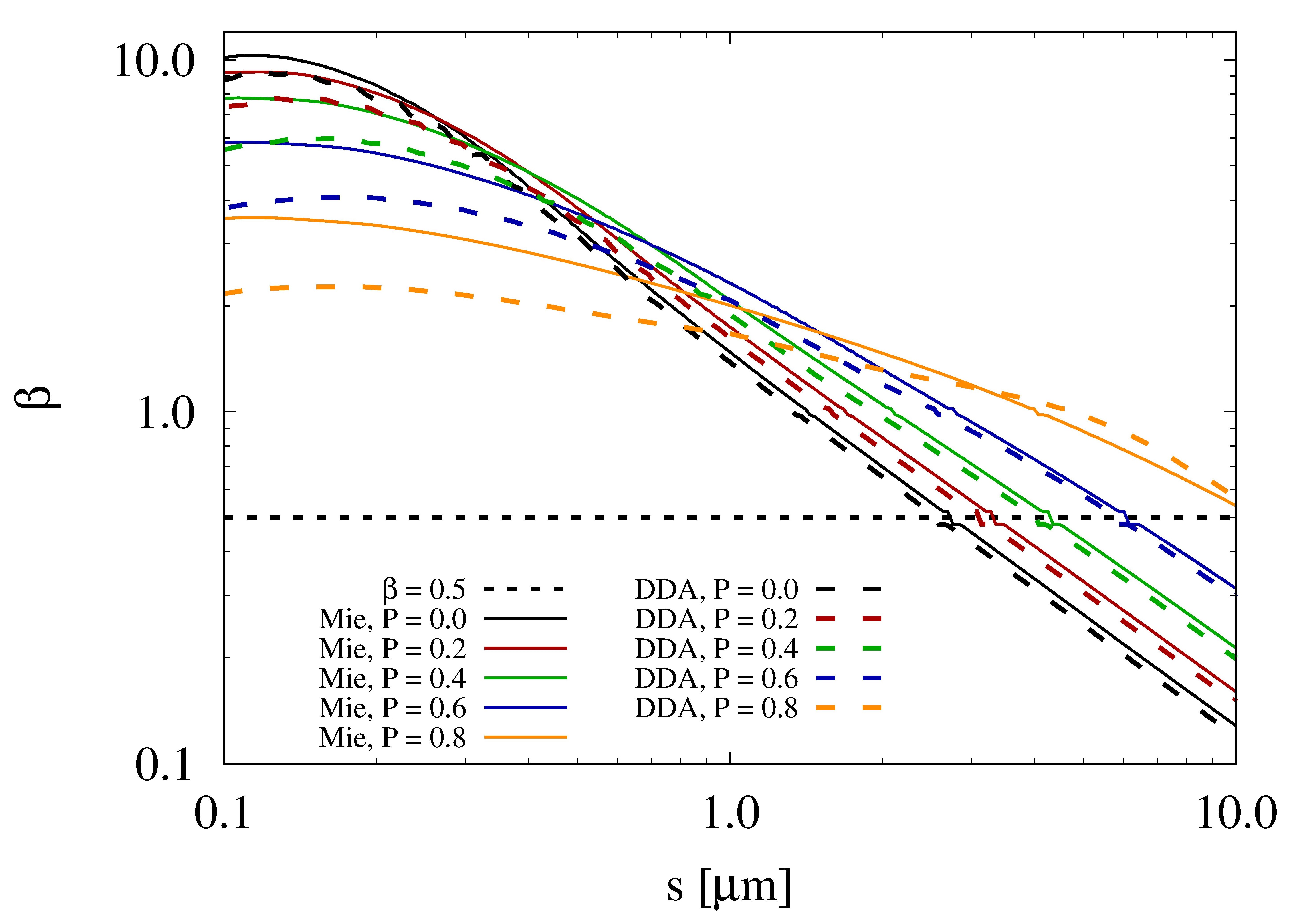}
\caption{Radiation pressure parameter, $\beta$, as a function of grain size, $s$, for different scattering models. The horizontal dotted line shows the blowout limit at $\beta = 0.5$. The solid lines show the results assuming EMT and Mie theory, the dashed lines assuming DDA. }
\label{fig:beta_s}
\end{figure}

In Fig.~\ref{fig:beta_s} we compare $\beta$ (eq.~\ref{eq:beta}) for the case of HD~131488 using Mie theory and DDA assuming a stellar luminosity of 13.9~$L_\odot$ and mass of 1.8~$M_\odot$. We apply the same stellar spectrum 
as used in our modelling of HD~131488 (see \S~\ref{sec:stellar_properties} for details).
We assume particles with a basic spherical shape and varying porosity. In the case of Mie theory, we mix the refractive indices of astrosilicate with those of vacuum to generate porous material (\S~\ref{sec:mie_theory}). 
In the case of DDA, we generate spherical particles of astrosilicate with voids of vacuum and calculate the optical properties directly without using any mixing rules. Hence, applying a porosity of 0 leads to comparable results for DDA and Mie (black lines in Fig.~\ref{fig:beta_s}). 

With increasing porosity the blowout size increases as well for both DDA and Mie theory. Considering HD~131488 and assuming Mie theory to calculate the absorption and scattering efficiencies, the blowout sizes vary between $2.9\mum$ for compact grains and $\sim11\mum$ for particles with a porosity of 0.8. 
The increase of blowout size with increasing porosity is in agreement with results from other studies \citep[e.g.,][]{kirchschlager-wolf-2013,pawellek-krivov-2015, arnold-et-al-2019}. 
The blowout sizes of compact grains inferred from DDA and Mie show small differences which are due to different calculation methods and set-ups as well as averaging $Q_\text{pr}$ over a limited number of wavelengths.

The differences between Mie theory and DDA get more pronounced for sub-blowout grains and for larger porosities. DDA predicts smaller $\beta$-values than Mie theory which will influence the amount and orbits of sub-blowout particles present in the disc. We showed in Appendix~\ref{sec:void_distribution} that different sizes and spatial distributions of voids can influence the scattering phase function of particles. This also leads to changes in the $\beta$ parameter as it depends on the optical properties of the material (see eqs.~\ref{eq:beta} and \ref{eq:radiationpressureefficiency}).

\subsubsection{Scattering properties}
\label{sec:scattering_phasefunction}

\begin{figure*}
    \centering
    \includegraphics[width=0.45\textwidth]{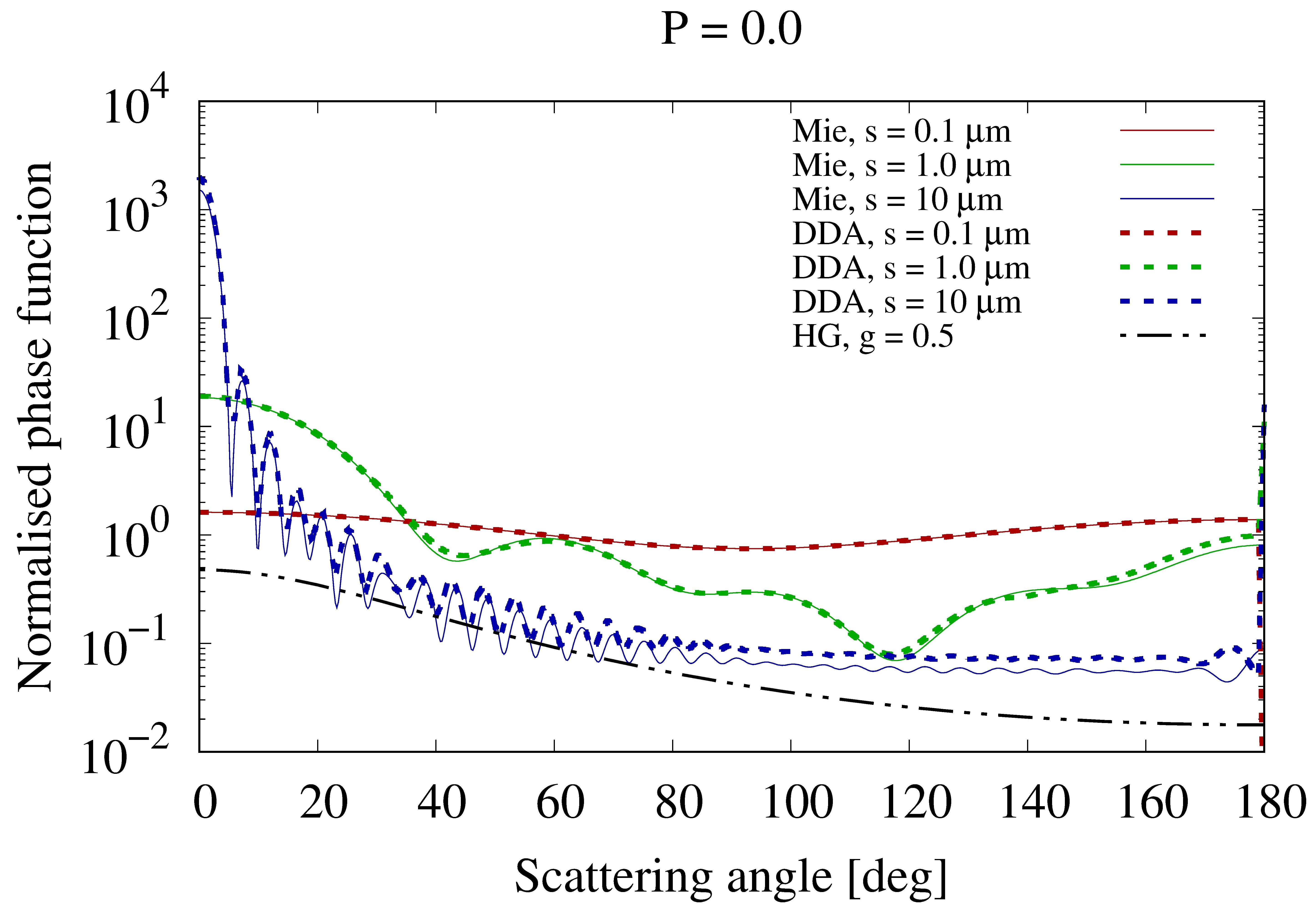}
    \includegraphics[width=0.45\textwidth]{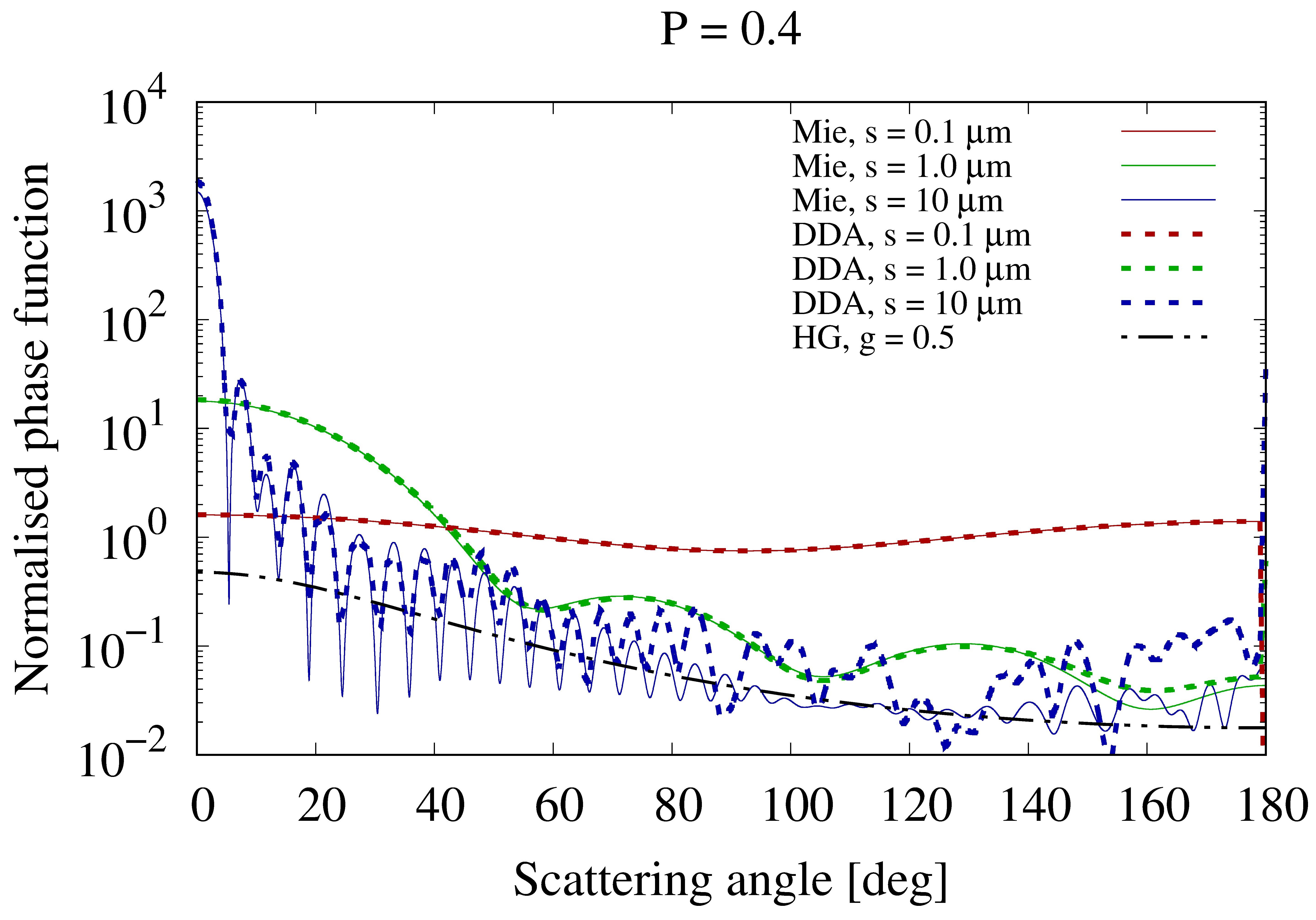}
    \caption{Phase function as a function of scattering angle for different grain sizes and porosities: $P=0.0$ (left), $P=0.4$ (right). Solid lines show results for EMT, dashed lines for DDA. The black dash-dotted line represents HG with $\langle\cos(\vartheta)\rangle=0.5$. Grains with $s=0.1~\mu$m are shown in red, $s=1.0~\mu$m in green, and $s=10~\mu$m in blue.}
    \label{fig:phase_function}
\end{figure*}

Another important aspect are the scattering properties of grains. Fig.~\ref{fig:phase_function} shows the phase function $p(\vartheta)$, also $S_{11}$ from the M\"uller matrix, as function of scattering angle, $\vartheta$, for different particle sizes assuming a porosity of zero (left panel) and 0.4 (right panel), and for Henyey-Greenstein following eq.~(\ref{eq:HG}) with $\langle\cos(\vartheta)\rangle=0.5$. 
We see that in both panels the peak of $S_{11}$ at $\vartheta\sim 0.0$ increases towards larger sizes while the HG approach does not show this behaviour (it is grain size independent). The peak is the aforementioned strong forward scattering for big particles. Comparing EMT (solid lines) and DDA (dashed lines), the phase functions look similar in the case of compact spheres. This is expected since Mie theory can be viewed as a limiting case for both EMT and DDA when assuming compact particles rather than porous grains. 

In the case of $P=0.4$ the phase functions also look similar for (sub-)micron-sized particles indicating that  EMT leads to similar scattering properties as DDA when assuming basic spherical particles with small void sizes. However, for larger grains ($\sim10\mu$m) and scattering angles of $\vartheta \gtrsim 70^\circ$ the deviations of DDA and EMT become more pronounced.

\subsection{Grain Size Distribution}
\label{sec:size_distribution}

For N-body dust models both in scattered light and thermal emission we need to define a size distribution including a minimum and a maximum size of particles that are present in the debris disc. 
A typical size distribution follows a power law
\begin{equation}
    N(s)\,ds = N_0\, \left(\frac{s}{s_0}\right)^{-q}\,ds
    \label{eq:sizedistribution}
\end{equation}
for grains on bound orbits where $N_0$ and $s_0$ are normalisation constants, and $q$ the size distribution index (see Sec.~\ref{sec:SED}) usually set to $3.5$ \citep{dohnanyi-1969}. 
However, when taking into account collisional evolution we find that there is an overabundance of bound grains close to the blowout limit \citep[e.g.,][]{strubbe-chiang-2006, thebault-wu-2008}. This can be explained by the fact that smaller grains become unbound and leave the system. Thus, they cannot act as projectiles to destroy the larger bound grains. We take this into account and apply the correction factor introduced by \cite{strubbe-chiang-2006}, $f(e_\text{d})\propto (1-e_\text{d})^{-3/2}$.

\subsubsection{Minimum size}
\label{sec:minimum_size}
While protoplanetary discs are often modelled applying (sub-)micron-sized dust grains that are coupled to the gas \citep[e.g.,][]{szulagyi-et-al-2019, vorobyov-et-al-2021}, the situation is different in typically gas-depleted debris discs where radiation pressure strongly affects the smaller dust particles that are no longer coupled to the gas.
As shown in Sec.~\ref{sec:orbitparameters}, grains with $\beta \geq 1/2$ are unbound and expelled from the stellar system on very short timescales. However,
models of scattered light data show that sub-blowout grains are often necessary to fit the debris disc data \citep[e.g.,][]{thebault-et-al-2019}. 

Assuming that the debris disc is in a quasi steady state, i.e. the production and destruction rates of grains due to collisions are equal, we can estimate the abundance of sub-blowout grains applying the collisional model from \cite{wyatt-et-al-2007b}. 
The idea is that only unbound (sub-blowout) grains can get lost from the disc so that the total mass loss rate of the dust equals the production rate of the sub-blowout particles. 
The mass loss rate is given by
\begin{equation}
    \dot{M} = \frac{M_\text{dust}}{\tau_\text{max}},
    \label{eq:mass_loss_rate}
\end{equation}
where $\tau_\text{max}$ is the collisional lifetime of the largest grains considered in our model which also contains most of the mass in the distribution $M_\text{dust}$ \citep[for details see e.g.,][]{wyatt-et-al-2007b, loehne-et-al-2007}.
The mass of the unbound grains seen at the moment the disc image is taken has to be equal to the mass loss rate so that we can use a normalisation constant $C$: 
\begin{equation}
    C = \frac{\dot{M}}{M_\text{dust}} = 
    \frac{1}{\tau_\text{max}}.
\end{equation}
So far, we assumed that the number of grains follows eq.~(\ref{eq:sizedistribution}) without taking into account blowout limits or mass loss rates.
Now, we get a corrected number of grains $C\, N_i(s)$ for sub-blowout grains of the ith size based on the production rate not following eq.~(\ref{eq:sizedistribution}) anymore. 
Finally, using the orbital information on each grain (e.g., mean anomaly) we can infer the number of particles produced at each location in the disc.

We investigate the influence of radiation pressure on the minimum grain size by comparing the outcome of the radiation pressure model (RP) to a model that ignores effects of radiation pressure (non-RP). We assume a size distribution of grains between 0.1 and 1000 $\mu$m and a total dust mass of those grains of $0.1M_\oplus$ in both cases. 
The mass loss rate and thus, the production rate of the sub-blowout particles depends on the dynamical excitation of the disc, i.e. the proper eccentricity of the planetesimals. We note that the planetesimal belt as a whole can exhibit an eccentricity of zero while individual planetesimals can deviate from a circular orbit. Only when they do, destructive collisions are possible.
To estimate the mass loss rate in the RP model we assume a proper eccentricity $\langle e\rangle$ of the colliding planetesimals of 0.1 comparable to the classical Edgeworth-Kuiper belt \citep[e.g.,][]{elliot-et-al-2005, vitense-et-al-2010} and a dispersion of inclinations $\langle i \rangle$ of 0.1 for a central radius of 88~au (\S~\ref{sec:radial_profile}) following the approximation $\langle e\rangle \approx \langle i \rangle$ from \cite{wyatt-et-al-2007}. 

\begin{figure*}
    \includegraphics[width=\columnwidth]{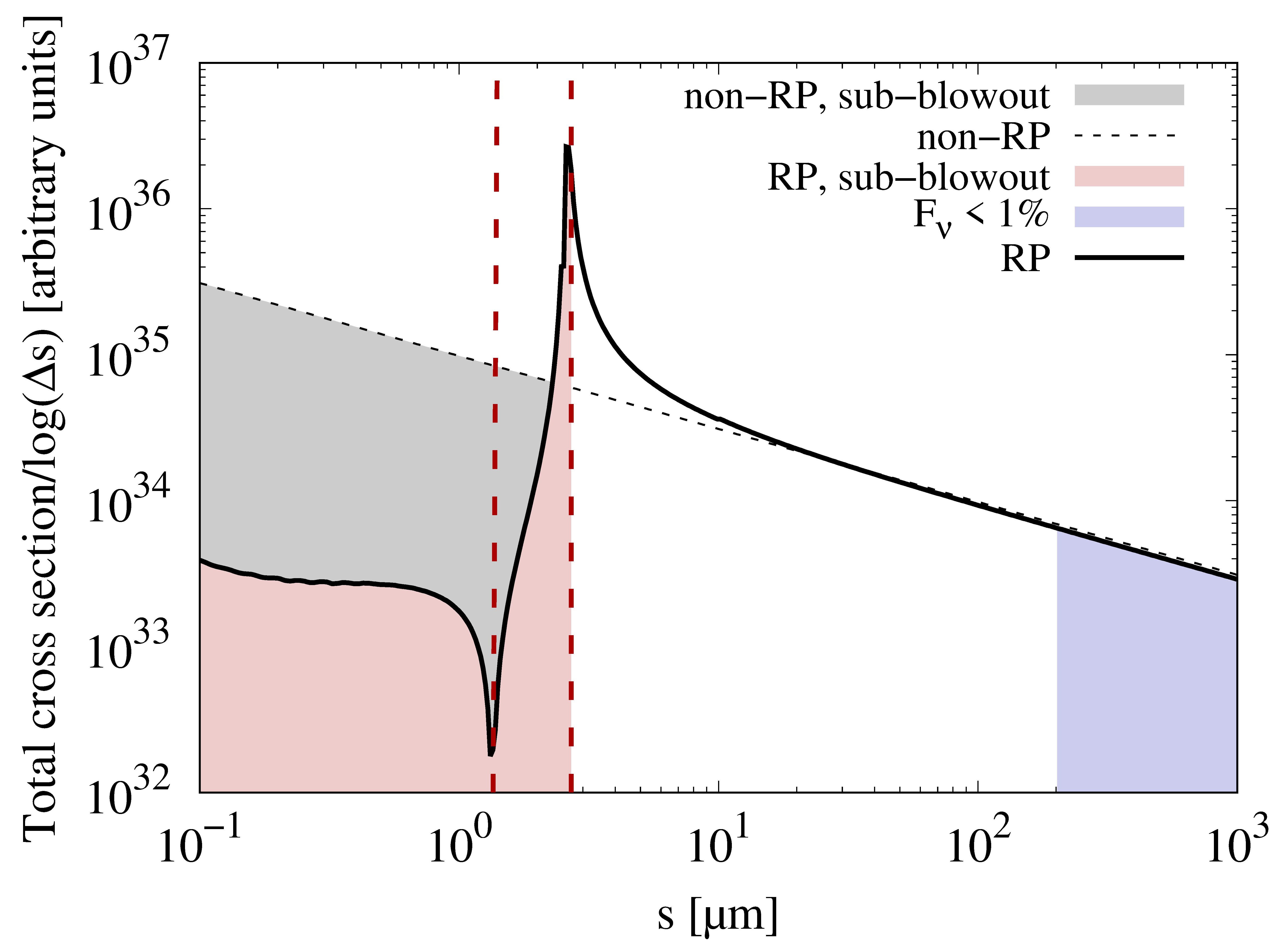}
    \includegraphics[width=\columnwidth]{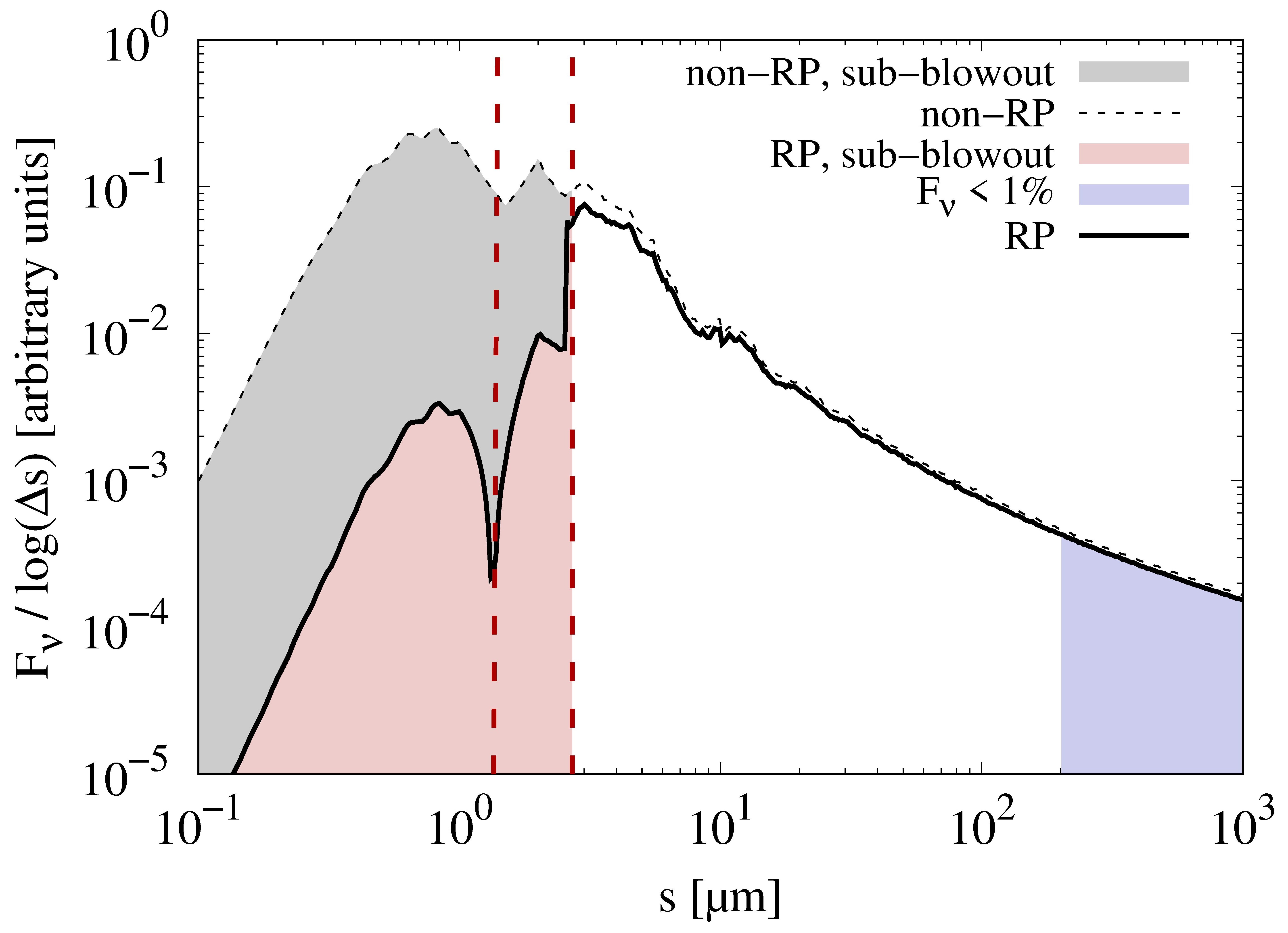}
    \caption{Influence of radiation pressure and collisions on the grain size distribution and the total flux density assuming a size distribution with $q=3.5$, a porosity of $P=0.0$, and a dust mass of $0.5M_\oplus$. 
    Left panel: total particle cross-section as a function of grain size; right panel: total flux density as function of grain size at $\lambda=1.6\,\mu$m.
    Red dashed lines indicate grains with $\beta=1/2$ and $\beta=1$; black dashed line: model without radiation pressure; black solid line: model including radiation pressure. 
    Grey and red shaded areas: area with grains < blowout limit for both models. Blue shaded area: contribution to total flux density < 1 per cent assuming the case of radiation pressure.
    }
    \label{fig:radiationpressure}
\end{figure*}

Fig.~\ref{fig:radiationpressure} shows the influence of radiation pressure on the size distribution and the total flux density for a dust disc made of compact, spherical grains ($P=0.0$) assuming astronomical silicate as dust composition and a dust mass of $0.5M_\oplus$. 
The total flux density seen by the observer is calculated by
\begin{equation}
    F_\nu(s)\,ds = N(s)\, F_{\nu, \text{star}}\, \left(\frac{ R_\text{star}}{d_\text{star}}\right)^2 \, \left(\frac{s}{2\,r}\right)^2 \, p(\vartheta) \, Q_\text{sca}\,ds,
    \label{eq:flux_density}
\end{equation}
where $R_\text{star}$ and $d_\text{star}$ are the stellar radius and the distance to the observer, respectively, $F_{\nu, \text{star}}$ is the flux density of the star at $R_\text{star}$ and at the observational wavelength $\lambda$, $r$ is the distance of the particle from the star, and $\vartheta$ is the scattering angle. 

In the left panel of Fig.~\ref{fig:radiationpressure} the total cross-section of particles is shown as a function of grain size. In the non-RP model (dashed line) the total cross-section increases towards smaller sizes following the power law from  eq.~(\ref{eq:sizedistribution}). Thus, the smallest grains possess the largest cross-section in this model (grey shaded area), and contribute the major part of the total flux density (also grey shaded area in the right panel). More than 95 per cent of the total flux density come from particles of the assumed sub-blowout size in the non-RP model. In the right panel, we also see that the contribution of the smallest grains ($s \lesssim 0.5\mu$m)  to the total flux density decreases again due to the decreasing scattering efficiency of the particles for which $\lambda > s$. The same effect is visible in the RP model. 

Here, the sub-blowout grains leave the system on (anomalous) hyperbolic trajectories, and are re-produced by destructive collisions of larger bodies. As a result, the number of those grains is much smaller than that of bound grains. We see this effect as a steep decrease of the cross-section in the left panel of Fig.~\ref{fig:radiationpressure} for grains between $\sim1$ and $\sim3$~$\mu$m for which $1/2 \leq\beta\leq 1$ (red shaded area). For smaller grains the total cross-section increases again due to the power law distribution given by eq.~(\ref{eq:sizedistribution}). 
Since their total cross-section is much smaller, 
the contribution of the small particles to the total flux density is significantly smaller compared to the non-RP case. Compared to the 95~per cent of the flux density coming from sub-blowout grains in the non-RP case, their contribution is only $\sim10$~per cent in the RP model (red shaded area).
This is in agreement with the results from \cite{thebault-et-al-2019} which investigated the influence of sub-blowout grains on disc modelling results. The fraction of particles with $\beta > 0.5$ is still large enough that we should not exclude them completely, and therefore, we will not fix our size distribution to the blowout limit but to a size of 0.1 $\mu$m to account for the presence of those sub-blowout grains. As can be seen in Fig.~(\ref{fig:radiationpressure}), the contribution of grains smaller than 0.1 $\mu$m is negligible due to the small scattering efficiency of those particles at a wavelength of $\lambda=1.6\,\mu$m.

\subsubsection{Maximum grain size and size distribution index}
\label{sec:maximum_size}

Knowing the maximum size is important since the DDA method limits us to grain sizes smaller than $10\,\mu$m due to the number of dipoles necessary to calculate the optical parameters \citep{kirchschlager-wolf-2013} at an observational wavelength of $\sim1\,\mu$m. Here, we already assume a sphere as simplified basic grain shape and add small vacuum inclusions in order to represent porous grains (\S~\ref{sec:composition}). More complicated shapes would lead to much higher and unfeasible computational times \citep{arnold-et-al-2019} due to a larger number of free parameters.
Thus, we need to consider if we need to fill up our size distribution by adding grains $>10\,\mu$m that were calculated using Mie theory since they might still significantly contribute to the total flux density of the disc. 

Modelling cometary dust using DDA, \cite{zubko-2013} found that grains larger than $15\times \lambda/(2\pi)$ do not significantly contribute to the estimates of back scattering or geometric albedo, but that those particles increase the computational time. 
For HD~131488 this would mean to exclude all grain sizes larger than $\sim 4\,\mu$m (the blowout size for compact spherical particles lies at $\sim3\,\mu$m). 
However, in our study we are more interested in the particles' contribution to the total flux density including physical mechanisms such as radiation pressure, and thus, the estimate from \cite{zubko-2013} might not be valid in our case. Furthermore, assuming that dust grains are produced in destructive collisions of bigger bodies we need to include larger dust sizes.
\cite{pawellek-et-al-2019b} inferred the maximum size considered in the disc models of 49~Cet by estimating the width of the forward scattering peak of compact spherical grains (eq.~7 therein). Following a similar approach for the disc around HD~131488 we would get a maximum size of $\sim9\,\mu$m which is already more than 2 times larger than the estimate from \cite{zubko-2013}. 

However, while the scattering efficiency for grains smaller than the observational wavelength decreases, it stays nearly constant for big particles \citep[see appendix of][]{pawellek-et-al-2019b}. Therefore, their contribution to the total flux density is determined by their size distribution rather than their scattering properties. 
We make a rough estimate and assume that the total flux density coming from a certain size of grains $s$ is given by  $F_\nu\times N_0\, s^{3-q}$ based on  eqs.~(\ref{eq:sizedistribution}) and (\ref{eq:flux_density}). For simplicity we also assume that $Q_\text{sca}$ and $p(\vartheta)$ are constant for large particles. Thus, the ratio of flux densities coming from two different sizes
$s_1$ and $s_2$ is given by 
\begin{equation}
    \frac{F_\nu^1}{F_\nu^2} \approx \left(\frac{s_1}{s_2}\right)^{3-q}.
    \label{eq:flux_ratio}
\end{equation}
For example, grains with a size ratio of 10 and a size distribution index of 3.5 reach a flux density ratio of 3 i.e., the contribution of grains of size $s_1$ is only 3 times larger than that of particles $s_2 = 10\,s_1$. 
This is also visible in Fig.~\ref{fig:radiationpressure} where we assume $q=3.5$. We find that 95~per cent of the total flux density stems from particles smaller than 25~$\mu$m, and 99~per cent from particles smaller than 200~$\mu$m (blue shaded area in Fig.~\ref{fig:radiationpressure}). This means that we need to ``fill up'' our size distribution with spherical grains between 10 and $25\,\mu$m, or $200\,\mu$m, respectively to include the part of the size distribution that still contributes significantly to the total flux density. A similar approach was considered for protoplanetary discs \citep{min-et-al-2016}.
In terms of the size parameter which is defined as $x~=~2\pi\,s/\lambda$, and assuming an observational wavelength of $\lambda=1.6\,\mu$m this means values of either $\sim100$ (for $s=25\,\mu$m) or $\sim800$ (for $s=200\,\mu$m) as upper limits for the size distribution when studying debris discs compared to a value of 15 suggested by \cite{zubko-2013} when studying cometary tails.

As shown by eq.~(\ref{eq:flux_ratio}), the contribution of large particles to the total flux density depends on the size distribution index, and thus, the maximum size also depends on this parameter.
In case of $q=3.5$ (Fig.~\ref{fig:radiationpressure}), we assume an ideal collisional cascade with constant impact velocities and material strength \citep{loehne-2020}.
However, SED modelling of debris discs showed that $q$ often differs from this value. It varies mostly between 3 and 4 \citep[e.g.,][]{loehne-2020, pawellek-et-al-2014, pawellek-et-al-2021}. 
\begin{figure}
    \centering
    \includegraphics[width=\columnwidth]{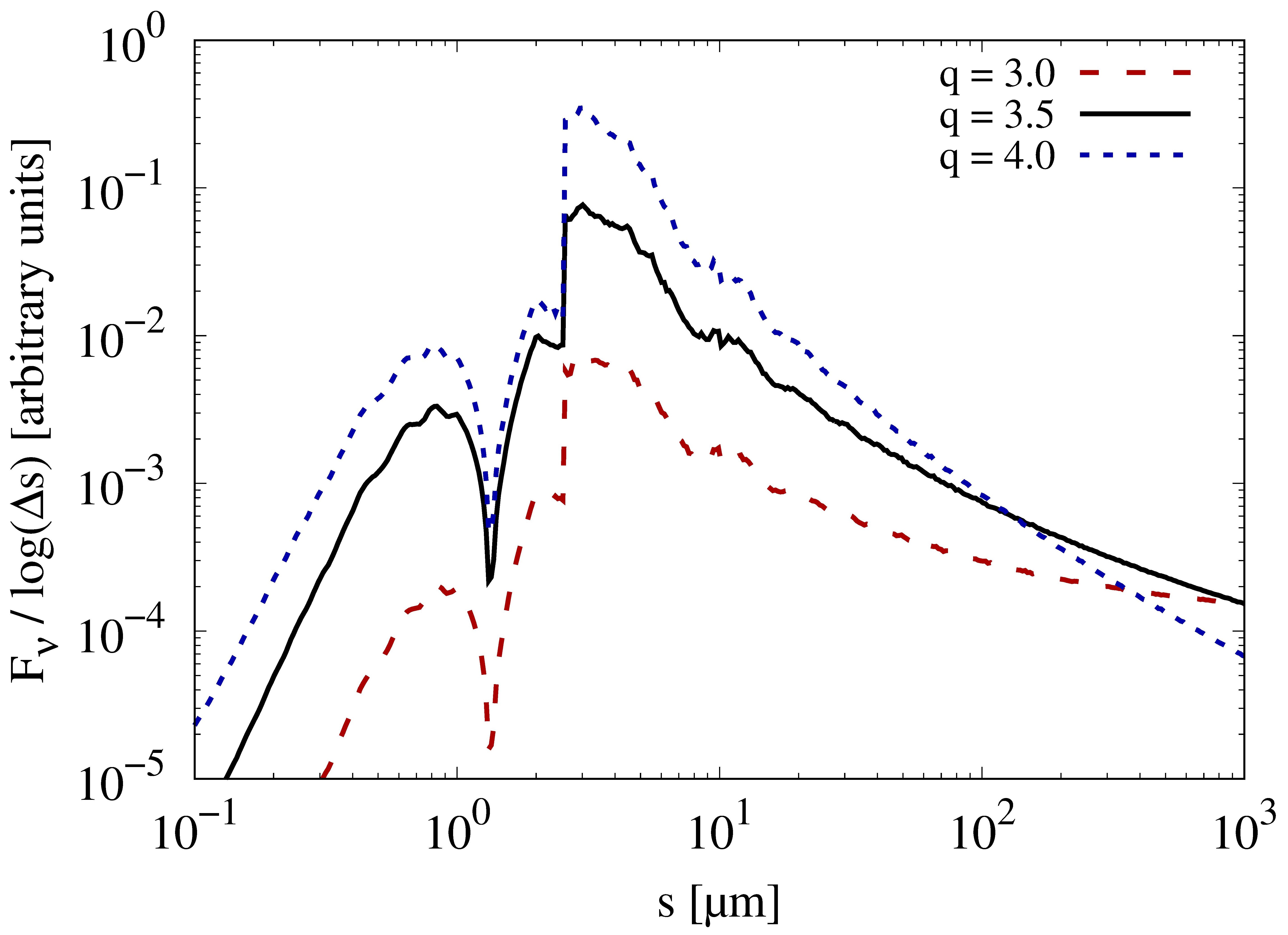}
    \caption{Flux density as a function of grain size for different size distribution indices. The dust mass is fixed to $0.5M_\oplus$.}
    \label{fig:Fnu_s_q}
\end{figure}
If $q=3$, all grain sizes with $\beta<0.5$ contribute the same total cross-section to the distribution (red dashed line in Fig.~\ref{fig:Fnu_s_q}) which results in an equal contribution of flux density based on eqs.~(\ref{eq:flux_density}) and (\ref{eq:flux_ratio}). In this case, the definition of a maximum grain size is rather difficult since larger grains still contribute significantly to the total flux density.
For $q=4$ each size contributes the same mass rather than cross-section, and thus, the total flux density is dominated by particles close to the blowout limit (blue dotted line in Fig.~\ref{fig:Fnu_s_q}). Here, the maximum grain size could be even lower than the $200\,\mu$m suggested for the case $q=3.5$.

To account for most of the cases we will assume a maximum size of $10^4\mu$m which also gives us the opportunity to model the disc's spectral energy distribution with the same size distribution at far-IR wavelengths (see \S~\ref{sec:SED} for details). This means in the case of DDA modelling, we will calculate grains with sizes $s \leq 10\,\mu$m using DDA and grains with $s>10\,\mu$m using Mie theory. We note that a similar approach was considered for protoplanetary discs \citep{min-et-al-2016}. As shown in Fig.~\ref{fig:beta_s} the blowout sizes for Mie theory and DDA are very close when assuming small void sizes as done in this study. Different void sizes and spatial distributions within the particles as used in \cite{zubko-2013} or \cite{arnold-et-al-2019}  would make it difficult to fill up the DDA size distribution with Mie grains and thus lead to inconsistencies in the model.

Considering the sub-blowout grains (\S~\ref{sec:minimum_size}), their contribution to the total flux density is determined by the mass loss rate (eq.~\ref{eq:mass_loss_rate}) which, in return, depends on the total dust mass and the collisional lifetime of the largest particles in the cascade. The collisional lifetime decreases with increasing $q$. Thus, the mass loss rate increases and the contribution of sub-blowout particles to the total flux density increases as seen in Fig.~\ref{fig:Fnu_s_q}.

\subsubsection{Influence of gas on the size distribution}
\label{sec:gas}

The disc around HD~131488 was found to possess a high content of gas \citep[e.g.,][]{moor-et-al-2017, moor-et-al-2019, rebollido-et-al-2022} that could have an impact on the dust distribution in both space and size. 
Depending on the surface density of the gas, particles up to a certain size can be dragged efficiently by the gas while larger grains remain unaffected. 
Thus, strong gas drag might lead to a much higher amount of sub-blowout grains in the disc that are usually expelled from the system.
The (dimensionless) Stokes number gives the timescale necessary to stop a grain from its relative motion towards the gas. It is given by
\begin{equation}
\text{St} = \frac{\pi}{2}\,\frac{s\,\varrho}{\Sigma_\text{gas}},
    \label{eq:stokes}
\end{equation}
where $\Sigma_\text{gas}$ is the surface density of the gas \citep[e.g.,][]{marino-et-al-2020}. If $\text{St}\ll 1$ the dust particle is stopped nearly instantly and follows the motion of the gas. 
\begin{figure}
    \centering
    \includegraphics[width=\columnwidth]{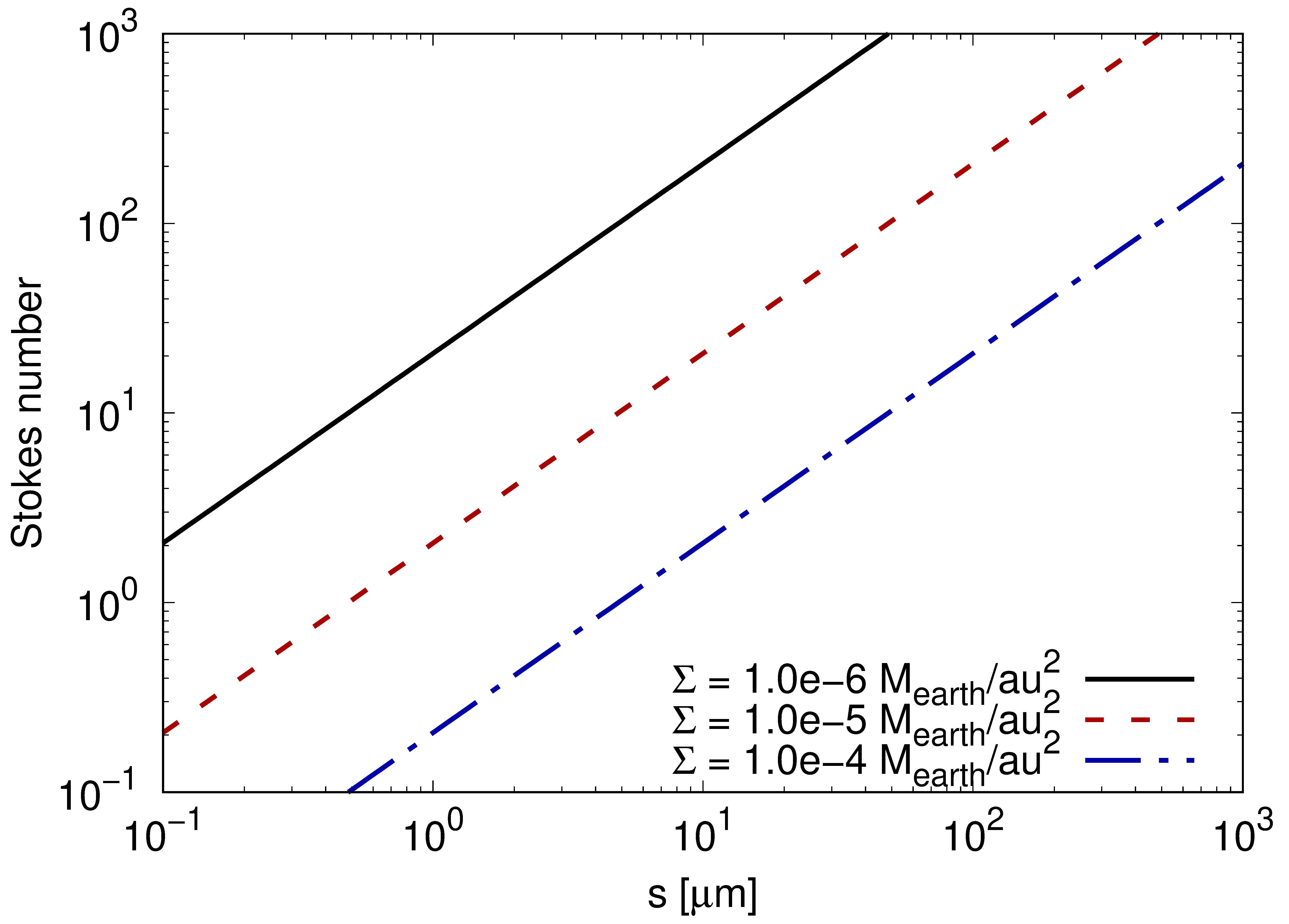}
    \caption{Stokes number as a function of grain size.}
    \label{fig:Stokes}
\end{figure}

Fig.~(\ref{fig:Stokes}) shows the dust particle size as a function of the Stokes number for different gas surface densities. While the surface density of the gas is not well constrained, we can estimate a rough value for the \ce{CO} gas surface density based on the observations presented by \cite{moor-et-al-2017}. The study estimated a \ce{CO} gas mass of $9\times10^{-2}$~$M_\oplus$ and a reanalysis of the data constrains the radial extent of the gas to 30-130~au (see Appendix \ref{sec:pvdiagram}). These translate to an average \ce{CO} gas surface density of $2\times10^{-6}$~$M_\oplus$~au$^{-2}$. 

However, the total gas mass is probably much larger than this. First, the above \ce{CO} mass estimate should be treated as a lower limit, since it was derived assuming an ISM-like abundance of \ce{C^18O}, which is likely to be an underestimate of the true value due to isotope-selective photo-dissociation \citep{moor-et-al-2017}. 
Moreover, this estimate considers only \ce{CO}, which may not be the dominant species. If the gas has a residual primordial nature, then the gas composition is dominated by \ce{H_2} molecules, whose mass exceeds that of CO by orders of magnitude \citep[][and references therein]{miotello-et-al-2023}. 
According to current theories, however, it is more likely that the observed gas is of secondary origin and has been released from icy bodies, e.g. via collisions, sublimation, photo-desorption and/or as an outcome of the thermal evolution of young large icy planetesimals \citep{kral-et-al-2019, marino-et-al-2020, bonsor-et-al-2023}. 
In our Solar system, \ce{H2O}, \ce{CO}, and \ce{CO2} are the most abundant species in the cometary gas \citep{mumma-et-al-2011}. 
While self-shielding and shielding by \ce{C} atoms can substantially increase the photo-dissociation lifetime of \ce{CO} molecules, similar mechanisms are not available for \ce{CO2} and \ce{H2O} molecules, which are therefore rapidly dissociated due to UV photons. 
In order to determine the total gas mass of a \ce{CO}-rich debris disc, we would therefore need to know not only the mass of \ce{CO}, but also the amounts of the various photo-dissociation and photo-ionisation products (\ce{C}, \ce{C+}, \ce{O}, \ce{H}) of the main molecules.
Although, thanks to ALMA, estimates of the \ce{C} content of an increasing number of CO-rich discs are available \citep{cataldi-et-al-2023}, the amounts of \ce{O} and \ce{H} in such discs are not known. 

Based on molecular abundances measured in cometary atmospheres in the Solar system \citep[][]{mumma-et-al-2011}, the mass ratios of above photo-dissociation products to \ce{CO} can vary over a wide range, with an upper bound of $\sim$25 (taking into account that the rapid photo-dissociation of \ce{CO2} results in \ce{CO} gas).
Although this is subject to a number of uncertainties when applied to the disc of HD\,131488 
-- for example, not only is the composition of the ice bodies there unknown, but also the mechanisms that lead to the gas production, 
which can result in different gas mixtures for the same ice composition -- it can be said that the average gas surface density can be as high as several times 10$^{-5}$~$M_\oplus$~au$^{-2}$. Another aspect is that the gas distribution is probably not uniform. It is well possible that the density if gas in the planetesimal belt, which is the main production site for both gas and small dust particles, is higher than elsewhere.

In the case of compact grains, we find that all particles smaller than the blowout size possess a Stokes number significantly smaller than 1 only if $\Sigma_\text{gas}\gtrsim10^{-4}\,M_\oplus/\text{au}^2$, while for coupling grains of size $s\sim0.5\mu$m -- that would be the brightest particles if radiation pressure was inefficient in expelling sub-blowout grains (Fig.~\ref{fig:radiationpressure}) -- needs gas surface densities higher than $\sim10^{-5}\,M_\oplus/\text{au}^2$. These surface densities are much higher than the values estimated for \ce{CO} only, but taking into account the other components, the total gas surface density may be quite similar to them.

\subsection{Images of disc models}
\label{sec:images}

\subsubsection{Disc appearance}
\begin{figure*}
    \centering
    \includegraphics[width=0.8\textwidth]{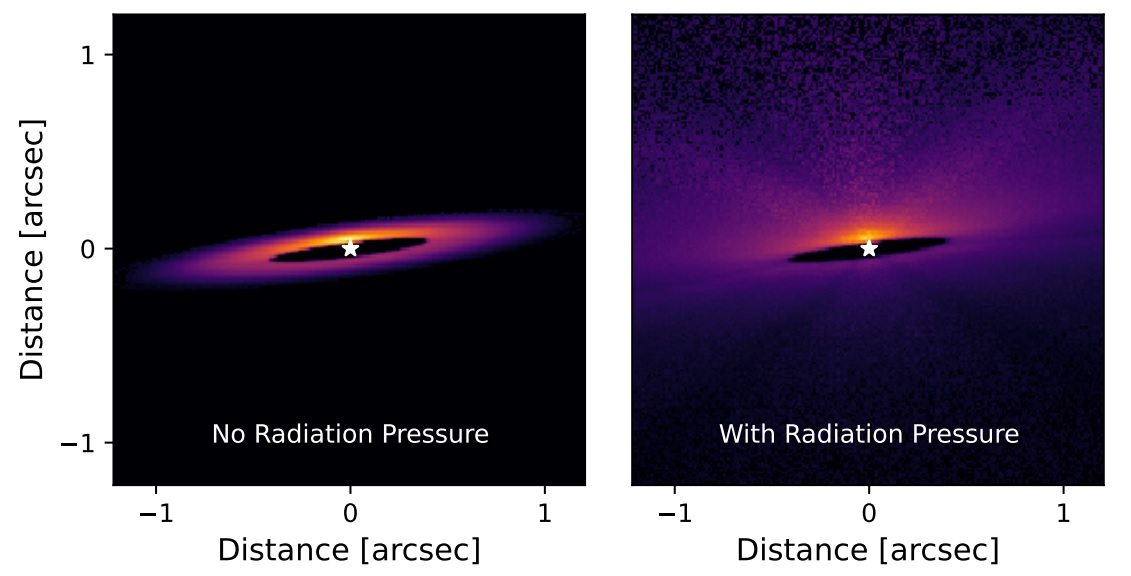}
    \caption{Scattered light models of a debris disc generated with the MODERATO code \citep{pawellek-et-al-2019b}. Left: ignoring radiation pressure. Right: including radiation pressure. The lobes visible in the radiation pressure model can be explained by the scattering phase functions of the dust grains.}
    \label{fig:Model_Discs}
\end{figure*}

We now analyse the influence of the aforementioned aspects (\S~\ref{sec:orbitparameters} - \ref{sec:size_distribution}) on the disc images. 
Fig.~(\ref{fig:Model_Discs}) shows the effect of radiation pressure for a grain size distribution of compact grains between 0.1 and $10^4~\mu$m and with a size distribution index of $q=3.9$ indicating a high fraction of grains at the lower end of the size distribution. 

In the left panel all particles possess $\beta=0$ condition, implying that all grains stay close to their parent bodies on non-eccentric orbits. 
The close proximity of small grains to the planetesimals might be the case for a large surface density of gas (see \S~\ref{sec:gas}). 
While the presence of gas does not change the $\beta$ values, the orbits of the grains will be altered so that the particles stay close to their parent body which can be roughly described with $\beta \approx 0$.
In this case the scattered light is dominated by grains around $\sim0.5\mu$m (Fig.~\ref{fig:radiationpressure}) that are not expelled from the system, i.e. radiation pressure is not efficient. 
These particles possess more or less isotropic scattering properties at $\lambda=1.6\,\mu$m (Fig.~\ref{fig:phase_function}) so that the model disc also shows a more isotropic distribution of scattered light.

In the right panel of Fig.~(\ref{fig:Model_Discs}) radiation pressure is included, i.e. $\beta > 0$. Radiation pressure is effective when the surface density of gas is low.
Grains smaller than the blowout limit are expelled from the system, and only contribute a minor fraction of the dust due to their reproduction by collisions (Fig.~\ref{fig:radiationpressure}). Particles close to the blowout limit are moving on highly eccentric orbits forming a halo of roughly bound grains. Only larger grains for which $\beta \ll 0.5$ stay close to the parent belt and dominate the scattered light i.e., the surface brightness at the centre of the belt. 
The large particles possess strong forward scattering (Fig.~\ref{fig:phase_function}) and thus, the model disc shows a peak close to the star in the centre of the image where the scattering angle is small. Azimuthal changes in brightness in the right panel are caused by large particles. A finer grid of grain sizes will smooth out those variations.

\subsubsection{Total flux density}
Radiation pressure not only influences the appearance of the disc, but also affects the total flux density of the model disc which is connected to the disc mass. As shown in Fig.~(\ref{fig:radiationpressure}), the highest contribution to scattered light comes from small grains with sizes of $\sim0.5~\mu$m when radiation pressure is not taken into account. These grains do not significantly contribute to the total dust mass and thus, only a small amount of material is needed to generate a high total flux density when such small grains are present within the disc.

If radiation pressure is included, these dominating sub-blowout grains are expelled from the system, and the majority of the scattered light comes from bound particles close to the blowout limit (Fig.~\ref{fig:radiationpressure}). 
If we use the same total dust mass in both cases (RP and non-RP), the flux density of the bound grains is much smaller compared to that of the sub-micron-sized grains.
Therefore, we would need a much larger dust mass to generate the same level of flux density when taking radiation pressure into account. 
This effect is even stronger for porous dust grains. The amount of scattered light coming from a compact grain is larger than the light coming from a porous particle of the same size (\S~\ref{sec:flux_densities}).

\section{Modelling results}
\label{sec:results}

\subsection{Stellar photosphere and dust composition}
\label{sec:stellar_properties}
For all our approaches we apply an ATLAS9 model \citep{castelli-kurucz-2004} as stellar photosphere to determine the influence of the host star HD~131488. Here, the stellar temperature, metallicity, and surface gravity provided by \cite{rebollido-et-al-2018} are taken into account to generate the synthetic spectrum. 
We assume a dust composition of pure astronomical silicate with a bulk density of 3.3~g/cm$^{3}$, and use porosities of
$P=0.0$ (for compact grains), 0.2, 0.4, 0.6, and 0.8  by using  Bruggeman's mixing rule of EMT \citep{bruggeman-1935,bruggeman-1936}.

\subsection{Fitting approach}

To find the best fit model, we use a $\chi^2$-minimisation
assuming that an ideal residual image should only contain white noise in each pixel.
The $\chi^2$-parameter is then computed for each pixel by
\begin{equation}
\chi^2 =  \sum\limits_{i=1}^{N_\text{pixel}} { \left(\frac{F_\text{i, residual}}{F_\text{i, noise}}\right)^2}
\label{eq:chi2}
\end{equation}
The noise is estimated by computing a disc-free image with the same noise distribution as in the PCA-reduced image. This is done by derotating the IRDIS images in the opposite direction, compared to the correct reduction. The faint disc signal present in individual images is therefore diluted when the images are stacked, to produce a final disc-free image containing only residual noise.

We use the $\chi^2$ to estimate the uncertainties of our free parameters. Assuming a confidence level of 95~per cent, we infer the critical $\chi^2$-value for which we need to reject the hypothesis that our model represents the observations. In terms of a reduced $\chi^2$ this means a value of $\leq 1.05$ in our case.
Based on the best-fit parameter values we change each free parameter individually until the final $\chi^2$ gets larger than the critical value.

\subsection{Comparing scattered light models}

\subsubsection{Henyey-Greenstein model}
\label{sec:HG_results}

\begin{figure*}
\includegraphics[width=\textwidth]{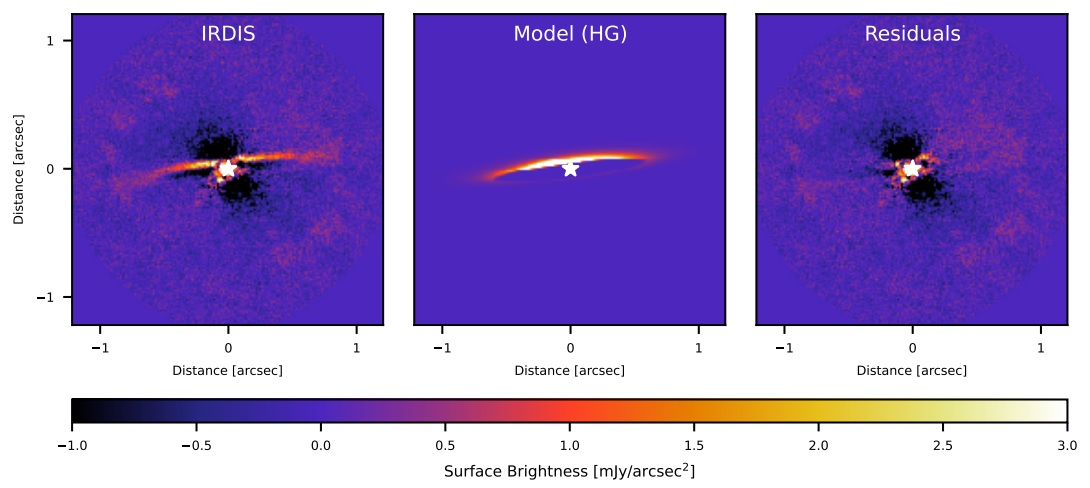}
\caption{Best fit model assuming Henyey-Greenstein approximation. From left to right: PCA-reduced SPHERE/IRDIS observations; HG model; Residual image.  The color scale is given in mJy per arcsec$^2$.
}
\label{fig:HG_Fit}
\end{figure*}

The HG approach is useful to infer general scattering properties of the dust material by modelling the phase function. Fig.~\ref{fig:HG_Fit} shows the results of this procedure.
We use a simple geometric model with the free parameters PA, inclination, $g$, central radius $r_0$, and flux density. We apply a two-part power law as radial profile centered at $r_0$. The slopes of the power law were fixed to $\alpha_1=12$ and $\alpha_2=-12$ so that the disc is narrow and $r_0$ corresponding to the peak density of the disc. 
The best fit value for $r_0$ is then found for ($110\pm25$)~au but remains rather uncertain. 
While \cite{moor-et-al-2017} found a best fit value of ($88\pm3$)~au it seems that in scattered light the disc peaks at a larger distance. We will discuss this issue in \S~\ref{sec:extent}.

We included an adhoc phase function and find a best fitting value for the asymmetry parameter $g = \langle\cos(\vartheta)\rangle$ of ($0.67\pm0.07$), a PA of ($97\pm2$)$^\circ$ and an inclination of $(84^{+1.5}_{-2.0})^\circ$ (see \S~\ref{sec:radial_profile}). We assume a zero disc eccentricity.
The positive $g$-parameter indicates a material of forward scattering particles which is comparable to the results of other debris disc studies \citep[e.g.,][]{millar-blanchaer-et-al-2015, olofsson-et-al-2016, engler-et-al-2017, engler-et-al-2019, olofsson-et-al-2019}.
It also indicates that the particles might not resemble compact spherical bodies since we would expect an even stronger forward scattering around $g=0.9$ when applying Mie theory.

\subsubsection{Mie model}
\label{sec:mie_model}

Now, we generate the semi-dynamical disc models using the MODERATO code. 
The code assumes surface density profiles for the parent belt following a Gaussian distribution. We use the following free parameters: disc width $\Delta r$, dust mass $M_\text{d}$, size distribution index $q$, and dust porosity $P$. We fix PA and $i$ to the values inferred from the HG approach to keep the fitting process fast.
Each planetesimal in the belt releases grains of different sizes following a power law (eq.~\ref{eq:sizedistribution}). Sub-blowout grains are produced following the collisional model of \cite{wyatt-et-al-2007b}.
Then the orbits of the individual dust particles are calculated and from their position the light scattered in the direction of the observer is inferred. 

\begin{figure*}
    \centering
    \includegraphics[width=\textwidth]{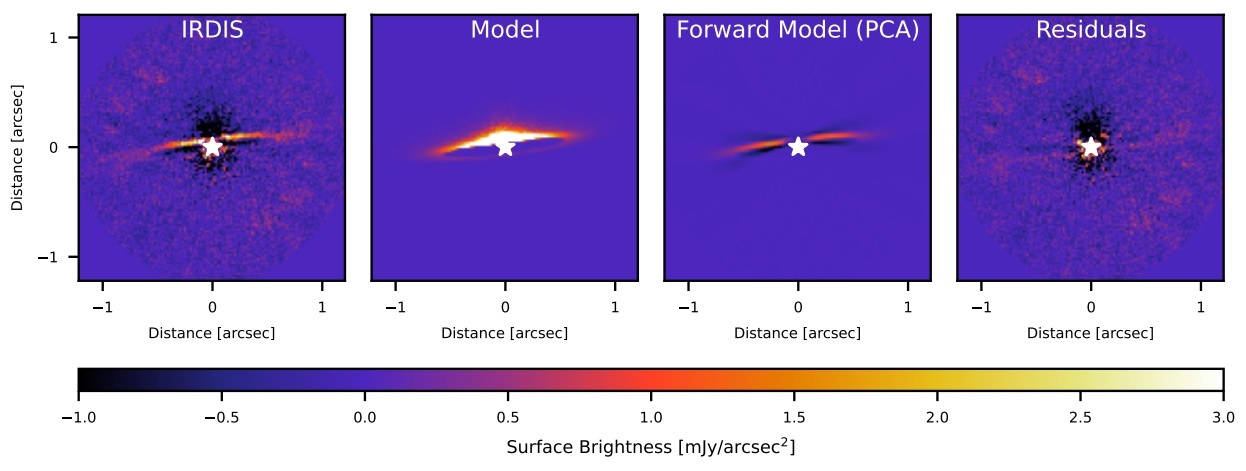}
    \caption{Best fit model assuming Mie theory and radiation pressure. From left to right: PCA-reduced SPHERE/IRDIS observations; Model generated by MODERATO assuming $P=0.6$ and $q=3.0$; Forward model (PCA);  Residual image. 
    }
    \label{fig:best_fit_model_mie}
\end{figure*}

\medskip
For our first model we consider grains calculated by Mie theory that range from compact spheres to high porosity ($0.0 \leq P \leq 0.8$ in steps of 0.2) and assume a size distribution index smaller than 4 (for $q=4$ each size bin would contribute the same mass to the total dust mass). The particles are produced in a single axisymmetric ring, and we do only take into account the effect of radiation pressure on the particles' orbits. 

We find a best fitting porosity of 0.6. However, we cannot exclude porosities between 0.0 and 0.4 as they also lead to results within the confidence interval. Assuming a porosity of 0.8 did not lead to a well-fitting model.
The best-fitting size distribution index is $q=3.0\pm 0.1$. For grains with sizes between $0.1\,\mu\text{m}\leq s\leq 10^4\,\mu\text{m}$ we need a dust mass of $\sim4.5\pm0.8\,M_\oplus$ to fit the scattering flux density observed. 
In Fig.~(\ref{fig:best_fit_model_mie}) we show the results for the best fitting model. 

Based on ALMA observations \cite{moor-et-al-2017} used a Gaussian radial profile and found a best fitting value for the central radius of ($88\pm3$)~au and a total disc width of ($46\pm12$)~au (after updating the distance of HD~131488 by Gaia data, \S~\ref{sec:radial_profile}). 
In contrast to that we find that a symmetric Gaussian radial profile does not lead to a fit within the expected $\chi^2$ confidence level of the scattered light data, i.e. $\chi^2_\text{reduced}>1.05$. We found that in this case the dust at radial distances smaller than 88~au dominate the scattered light and lead to the disc appearing smaller than observed.

\begin{figure}
    \centering
    \includegraphics[width=\columnwidth]{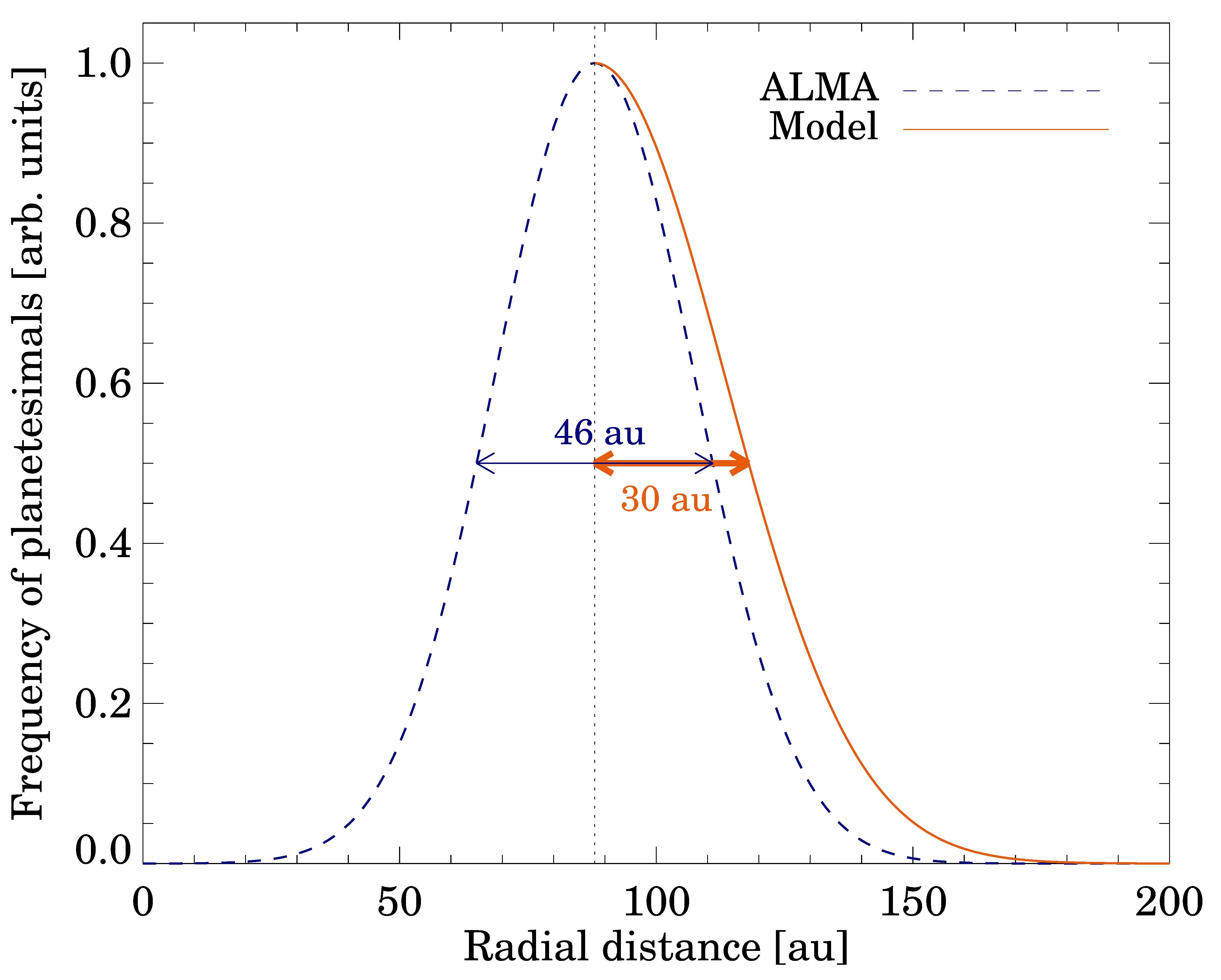}
    \caption{Radial profiles of the planetesimal belt. Dashed line: profile inferred from ALMA data. Solid line: profile used to model scattered light data. Both profiles use a Gaussian.}
    \label{fig:radial_profile_belt}
\end{figure}

We tested different values as inner boundary and found that only when moving the inner boundary to ($88\pm5$)~au i.e., ignoring the inner part of the Gaussian and only taking the outer part into account, our models could reach a $\chi^2$-value within the appropriate confidence level ($\chi^2_\text{reduced}\leq 1.05)$. With this approach we find a best fitting disc width of ($30\pm3$)~au (Fig.~\ref{fig:radial_profile_belt}).
We will discuss this discrepancy in disc width in more detail in \S~\ref{sec:extent}.

\subsubsection{DDA model}
\label{sec:rp_model}

\begin{figure*}
    \centering
    \includegraphics[width=\textwidth]{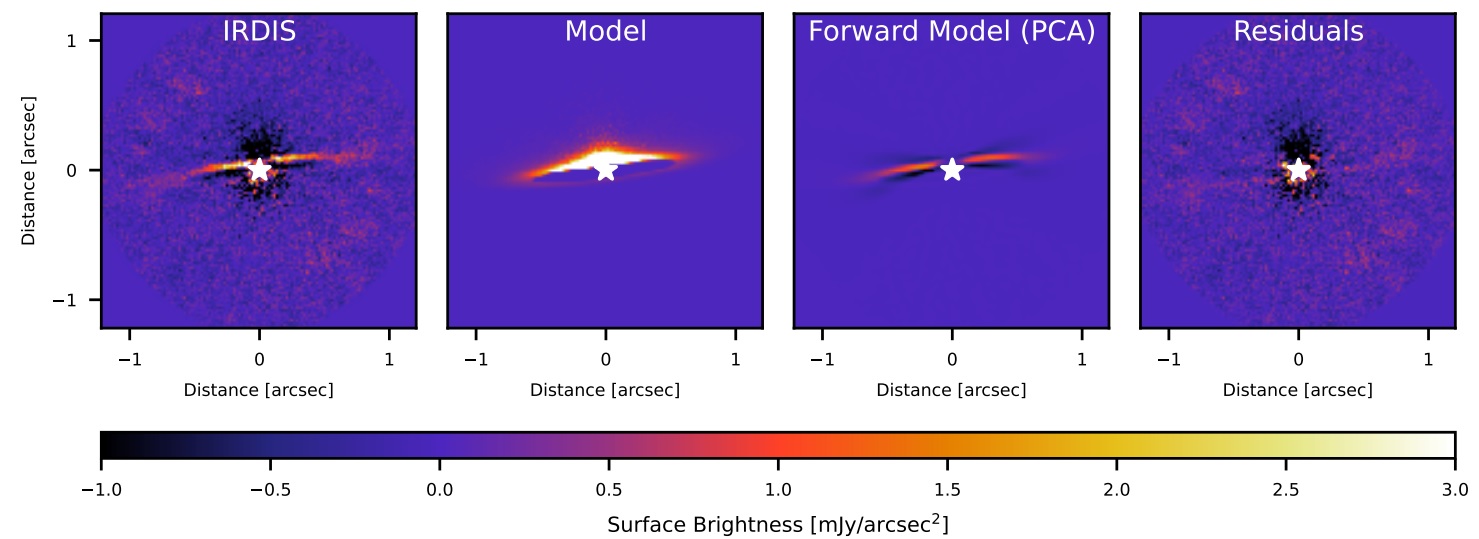}
    \caption{Best fit model assuming radiation pressure. From left to right: PCA-reduced SPHERE/IRDIS observations; Model generated by MODERATO assuming $P=0.6$ and $q=3.0$; Forward-model (PCA);  Residual image. The differences between the residuals of the DDA and the Mie model (Fig.~\ref{fig:best_fit_model_mie}) are subtle. The Mie model leaves slightly more dark areas in the eastern part of the disc.
    }
    \label{fig:best_fit_model}
\end{figure*}

In a similar approach as described in \S~\ref{sec:mie_model} we now apply optical dust properties inferred from DDA. Again we use the free parameters: disc width $\Delta r$, dust mass $M_\text{d}$, size distribution index $q$, and dust porosity $P$.
We consider grains ranging from compact spheres to high porosity ($0.0 \leq P \leq 0.8$ in steps of 0.2) and assume a size distribution index smaller than 4. The results for the best-fit model are shown in Fig.~(\ref{fig:best_fit_model}).
Similar to the Mie model we find a best fit for $P=0.6$. Again, we cannot exclude lower porosities ($P=$ 0.0, 0.2, 0.4) as they also lead to models within the confidence interval, but with larger $\chi^2$. The higher porosity of 0.8 can be excluded as it leads to best fits outside the confidence interval. The size distribution index is found as $q=3.0\pm 0.2$, and the dust mass as $M_\text{dust} \approx 4.4\pm0.7~M_\oplus$. Also the asymmetric radial profile with a disc width of $30\pm3$~au is similar to that found in \S~\ref{sec:mie_model} (see \S~\ref{sec:extent} for details). 

Based on the scattering phase function and blow-out sizes (Figs.~\ref{fig:beta_s} and \ref{fig:phase_function}) the differences between the Mie and DDA approach for the pure radiation pressure model were found to be minor when assuming grains of basic spherical shape and small sizes of the vacuum inclusions (\S~\ref{sec:dda_theory}). This is now confirmed by a similar quality of our Mie and DDA models ($\chi^2$ values are similar). This leaves us with the question whether the time consuming DDA approach is useful to model debris discs. We will discuss this question in \S~\ref{sec:Mie_vs_DDA}. 

\subsection{Spectral Energy Distribution}
\label{sec:SED}

\subsubsection{Model set-up}
We now compare the scattering flux density inferred from the radiation pressure model
with the results from modelling the thermal emission of the disc around HD~131488 at longer wavelengths.
With the MODERATO code we generate thermal emission images at wavelengths smaller than $10^4\mu$m, and then calculate the spectral energy distribution (SED).
The SED is calculated by the same approach as the SONATA code \citep[][]{pawellek-et-al-2014, pawellek-krivov-2015, pawellek-et-al-2021}, but now includes the effect of radiation pressure on particles of different sizes. This approach guarantees that the models for thermal emission and scattered light are self-consistent.
The photometric data of the dust continuum used to constrain the thermal emission model are given in Tab.~\ref{tab:photometry}.

\begin{table}
\caption{Continuum flux density. 
\label{tab:photometry}}
\tabcolsep 2pt
\centering
\begin{tabular}{rrclcc}

Wavelength      & \multicolumn{3}{c}{Flux density}       & Instrument & Reference				\\
$[\mum]$        & \multicolumn{3}{c}{[mJy]}              &            &        			\\
\toprule
0.42            & 2229.66 & $\pm$ & 34.95                & TYCHO B       & 1			\\
0.43            & 2558.05 & $\pm$ & 23.47                & APASS B        & 2			\\
0.47            & 2671.94 & $\pm$ & 24.51                & APASS G        & 2			\\
0.51            & 2045.98 & $\pm$ & 19.39                & Gaia BP       & 3			\\
0.53            & 2335.20 & $\pm$ & 30.67                & TYCHO V            & 1			\\
0.54            & 2631.62 & $\pm$ & 24.14                & APASS V        & 2			\\
0.62            & 2165.84 & $\pm$ & 19.87                & APASS R        & 2			\\
0.64            & 2052.12 & $\pm$ & 18.87                & Gaia G        & 3			\\
0.78            & 1718.90 & $\pm$ & 16.76                & Gaia RP       & 3			\\
0.79            & 1527.60 & $\pm$ & 31.34                & DENIS I        & 4			\\
1.24            & 1123.65 & $\pm$ & 36.53                & 2MASS J       & 5			\\
1.65            & 763.25 & $\pm$ & 33.65                 & 2MASS H       & 5			\\
2.16            & 500.87 & $\pm$ & 15.84                 & 2MASS Ks      & 5			\\
3.38            & 240.65 & $\pm$ & 8.31                  & WISE          & 6			\\
4.63            & 163.45 & $\pm$ & 5.16                  & WISE          & 6			\\
8.98            & 164.20 & $\pm$ & 7.16                  & AKARI         & 7			\\
12.33           & 111.12 & $\pm$ & 4.80                  & WISE          & 6			\\
22.25           & 153.15 & $\pm$ & 8.80                  & WISE          & 6			\\
101.40          & 331.20 & $\pm$ & 19.84                 & Herschel/PACS & 8			\\
163.60          & 184.80 & $\pm$ & 25.54                 & Herschel/PACS & 8			\\
1322.42         &   2.91 & $\pm$ &  0.31                 & ALMA          & 9 			\\
1652.22         &   1.64 & $\pm$ &  0.17                 & ALMA          & 10			\\
8750            &   0.0595 & $\pm$& 0.0124               & ATCA          & 11\\
\bottomrule

\end{tabular}

\noindent
{\em References:}
[1] - \cite{hog-et-al-2000};
[2] - \cite{henden-et-al-2016};
[3] - \cite{gaia-et-al-2018};
[4] - \cite{DENIS-et-al-2005};
[5] - \cite{cutri-et-al-2003};
[6] - \cite{wright-et-al-2010}; 
[7] - \cite{ishihara-et-al-2010};
[8] - \cite{marton-et-al-2017};
[9] - \cite{moor-et-al-2017};
[10] - This work;
[11] - \cite{norfolk-et-al-2021}
\end{table}

\subsubsection{Results}

\begin{figure}
\includegraphics[width=\columnwidth]{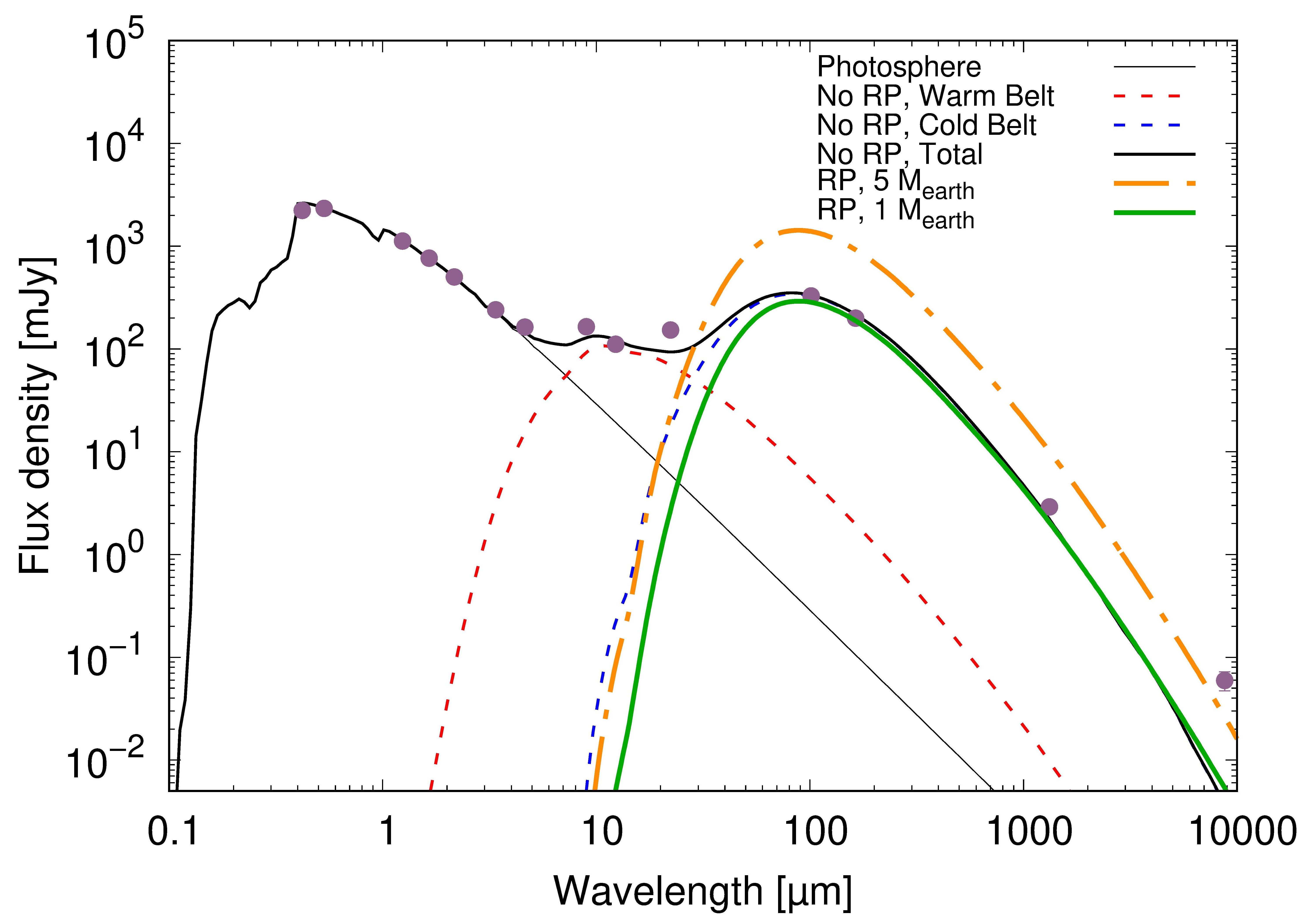}
\caption{SED of HD~131488. The purple circles show the measured flux density at different wavelengths. The lines show models using astronomical silicate with a bulk density of 1.3~g/cm$^3$ for $P=0.6$. The blue and red dashed lines show the two component model done by the SONATA code ignoring radiation pressure. The green solid and the orange dash-dotted lines show the results from the MODERATO code where no second component was taken into account.}
\label{fig:SED}
\end{figure}

In Fig.~(\ref{fig:SED}) we show the SED of HD~131488. 
In a first approach we use the SONATA code to fit a two component model to the observational data (red and blue dashed lines in Fig.~\ref{fig:SED}), but we note that this code does not take into account radiation pressure. We find a best-fitting model for a dust mass of $1.0\pm0.2~M_\oplus$ assuming the same porosity (0.6) and size distribution index (3.0) found by our scattered light models.

In a second approach we use the MODERATO code which now includes the effects of radiation pressure. In the scattered light images we do not detect warm dust close to the host star since the coronagraph of the SPHERE instrument is blocking out the inner region. Hence, we do not model a possible Asteroid belt analogue with MODERATO but only the outer Kuiper belt analogue (green solid line and orange dash-dotted line in Fig.~\ref{fig:SED}).
When applying the best-fitting parameters inferred from the scattered light model ($P=0.6$, $q=3.0$, $M_\text{d}=5~M_\oplus$, $r_0=88$~au, $\Delta r = 30$~au) we get the orange dash-dotted line seen in Fig.~(\ref{fig:SED}). This line is not consistent with the observational data at far-IR wavelengths.
However, we find that the observational data can be fitted when using a dust mass of $1\,M_\oplus$ instead of $5\,M_\oplus$ (green solid line in Fig.~(\ref{fig:SED}) which is a comparable dust mass as found by the SONATA model.

The ATCA point at 8.7\,mm shows a higher flux density than predicted by our models. One reason is that at a wavelength of $\sim1$\,cm particles with sizes larger than our applied maximum size of $10^4\,\mu$m contribute to the emission. Furthermore a simple power law size distribution might not be valid at long wavelengths. A higher amount of large particles might be present in the disc. This seems to be a common occurrence based on observational data from other debris discs \citep{lestrade-et-al-2020} and data from our own Solar system \citep[e.g.,][]{morbidelli-et-al-2021}. 

As mentioned before, the dust mass necessary to model the thermal emission data is a factor 5 lower than the prediction made from scattered light models only. The dust mass is well constrained by the SED so that we need to find a scattered light model that can reproduce the total flux density with this mass value. This suggests that a higher amount of sub-blowout grains is retained than our pure radiation pressure model predicts. A possible explanation might be the presence of gas that we will investigate in \S~\ref{sec:gas_model}.

\subsection{Combining SED and scattered light results}
\label{sec:gas_model}

As found for the pure radiation pressure model, a dust mass of $4.4\pm0.7\,M_\oplus$ would be necessary to generate the amount of scattered light observed for HD~131488. However, SED models predict a dust mass of only $1.0\pm0.2\,M_\oplus$. 
A possible explanation for the much higher flux density at short wavelengths might be the presence of gas within the disc that could increase the amount of sub-blowout grains dominating the scattered light. 
We showed in \S~\ref{sec:gas} that a gas surface density of $10^{-4}M_\oplus/$au$^2$ would be enough to strongly couple all the sub-blowout grains to the gas, and that a surface density of at least $\sim10^{-5}M_\oplus/\text{au}^2$ seems realistic for HD~131488. We note that this is a rough estimate without uncertainties as we do not have enough data on gas species present in the disc.
We now assume that sub-blowout grains up to a certain size are efficiently coupled to the gas and not expelled by radiation pressure. In a simple approach we vary the gas surface density and assume that particles for which the Stokes number is St$\leq1$ (eq.~\ref{eq:stokes}) are retained within the disc by setting the $\beta$ parameter to zero.  
We note that in reality $\beta$ is not changed by the presence of gas, but the grains' orbits are. 
\begin{figure*}
    \centering
    \includegraphics[width=\textwidth]{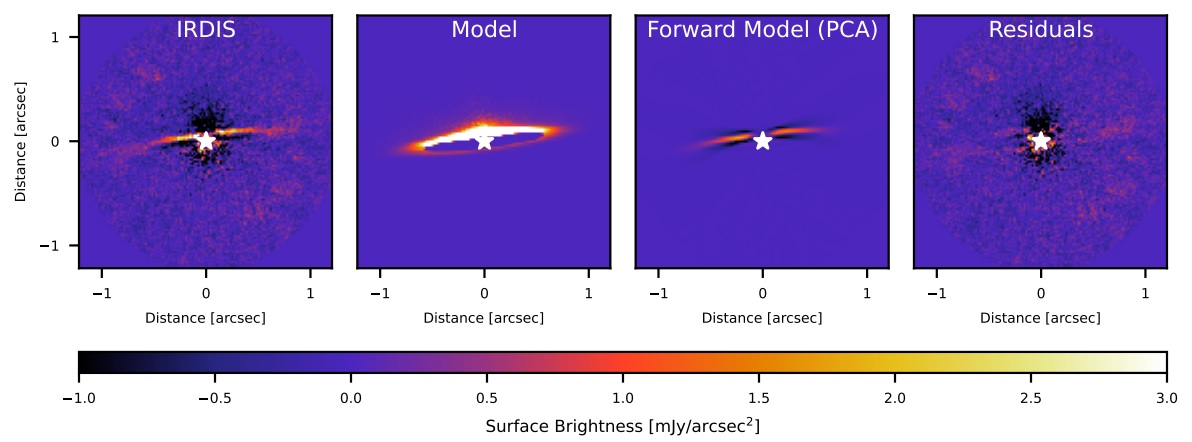}
    \caption{Best fit model assuming radiation pressure and gas drag with a surface density of $\Sigma=2\times10^{-5}M_\oplus/$au$^2$. From left to right: PCA-reduced SPHERE/IRDIS observations; Model generated by MODERATO assuming $P=0.6$ and $q=3.0$; Forward-model (PCA);  Residual image.}
    \label{fig:best_fit_model_gas}
\end{figure*}

In Fig.~(\ref{fig:best_fit_model_gas}) we show the resulting best-fit model for which the dust masses found in scattered light and thermal emission are equal ($1~M_\oplus$).
We find that for a size distribution index of $q=3.0$ we need a surface density of $\Sigma = (2.0\pm0.1)\times 10^{-5}\,M_\oplus/$au$^2$ to fit both the SED and the scattered light data. As explained in \S~\ref{sec:gas} this value is in agreement with expectations/estimates from ALMA observations.

\subsection{Deriving dust properties}

\subsubsection{Average phase function}

In \S~\ref{sec:scattering_phasefunction} and Appendix~\ref{sec:phase_function} we analyse the scattering phase function for different porosities and grain sizes independent of actual debris disc models. 
We find that for increasing porosity the phase function of specific sizes decreases for larger scattering angles, $\vartheta$ (Fig.~\ref{fig:S11_theta_grains}). At small angles ($\vartheta<10^\circ$) we see an increase of the phase function for grains with sizes of $s\geq 1\mu$m which we mentioned before as \textit{forward scattering}. 
In Fig.~(\ref{fig:Model_Discs}) the forward scattering is visible in the right panel where radiation pressure expels grains smaller than the blowout size leading to a dominance of grains with $s\sim10\mu$m. In the left panel it is not visible as the size distribution is dominated by 0.1$\mum$-sized grains that do not show the forward scattering peak in the phase function (Fig.~\ref{fig:S11_theta_grains}).

Now we combine those results on the scattering behaviour of the individual dust particles with the semi-dynamical disc model and infer an \textit{average phase function} for the debris disc model. 
To do so we infer the scattering angle and flux density (eq.~\ref{eq:flux_density}) of each dust grain of size $s$ at position $r$ in the disc.  
Then we multiply the value by the number of grains within the same size bin and at the same location (eq.~\ref{eq:sizedistribution}) and sum over all particle sizes and distances to get a flux density that only depends on the scattering angle, $\vartheta$.
Finally, we normalise this flux density by the total flux density of the model to get the average phase function.

\begin{figure}
    \centering
    \includegraphics[width=\columnwidth]{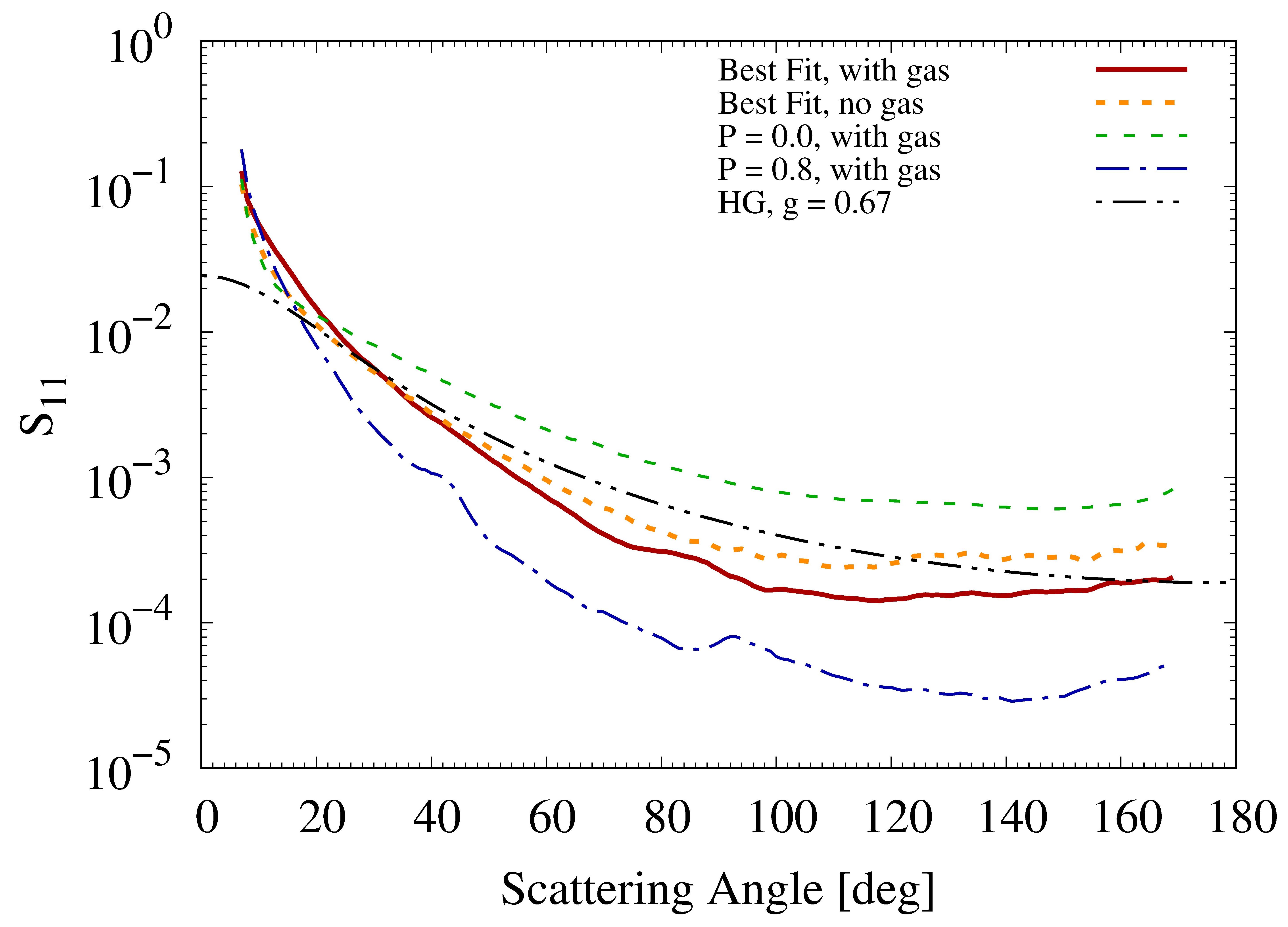}
    \caption{Average phase function as function of scattering angle. The parameter $P$ gives the porosity of the dust composition, HG the phase function assuming the Henyey-Greenstein approximation. Best Fit models use $P=0.6$. All models use $q=3.0$. 
    }
    \label{fig:average_phasefunction}
\end{figure}

In Fig.~(\ref{fig:average_phasefunction}) we show the resulting average phase function for our best fitting models (pure radiation pressure, gas drag) and for different porosities. 
Similar to the results for individual grains the average phase function shows smaller values at larger scattering angles when the porosity increases. Also, at small angles the phase function increases.
Comparing the grain model to the Henyey-Greenstein model (black dash-double dotted line in Fig.~\ref{fig:average_phasefunction}) we find that a porosity of $P=0.6$ leads to the closest match between Mie, DDA and HG approach. For $P=0.0$ the phase function shows a larger contribution at larger scattering angles while for $P=0.8$ the contribution is smaller. Thus, the $g$ parameter gives a hint on the general porosity of the material being larger than 0.0. The very porous material of 0.8 could be excluded by our scattered light model already.

\subsubsection{Reflectance}
\label{sec:reflectance}
We do not only possess the SPHERE/IRDIS data for the disc around HD~131488 at $\lambda=1.6\,\mu$m that were used to generate the scattered light models in this paper, but we also have the data available from IFS at $\lambda = 1.04, 1.18$, and $1.29\,\mu$m (see \S~\ref{sec:observations}). 
This gives us the opportunity to analyse the reflectance, i.e. the fraction of stellar light that is scattered by the dust of the disc at all these wavelengths. 
The reflectance gives us information on the colour of the debris disc in scattered light. We assume that the colour is determined by the grain size distribution, i.e. the amount of small grains compared to large grains. 

\begin{figure}
    \centering
    \includegraphics[width=\columnwidth]{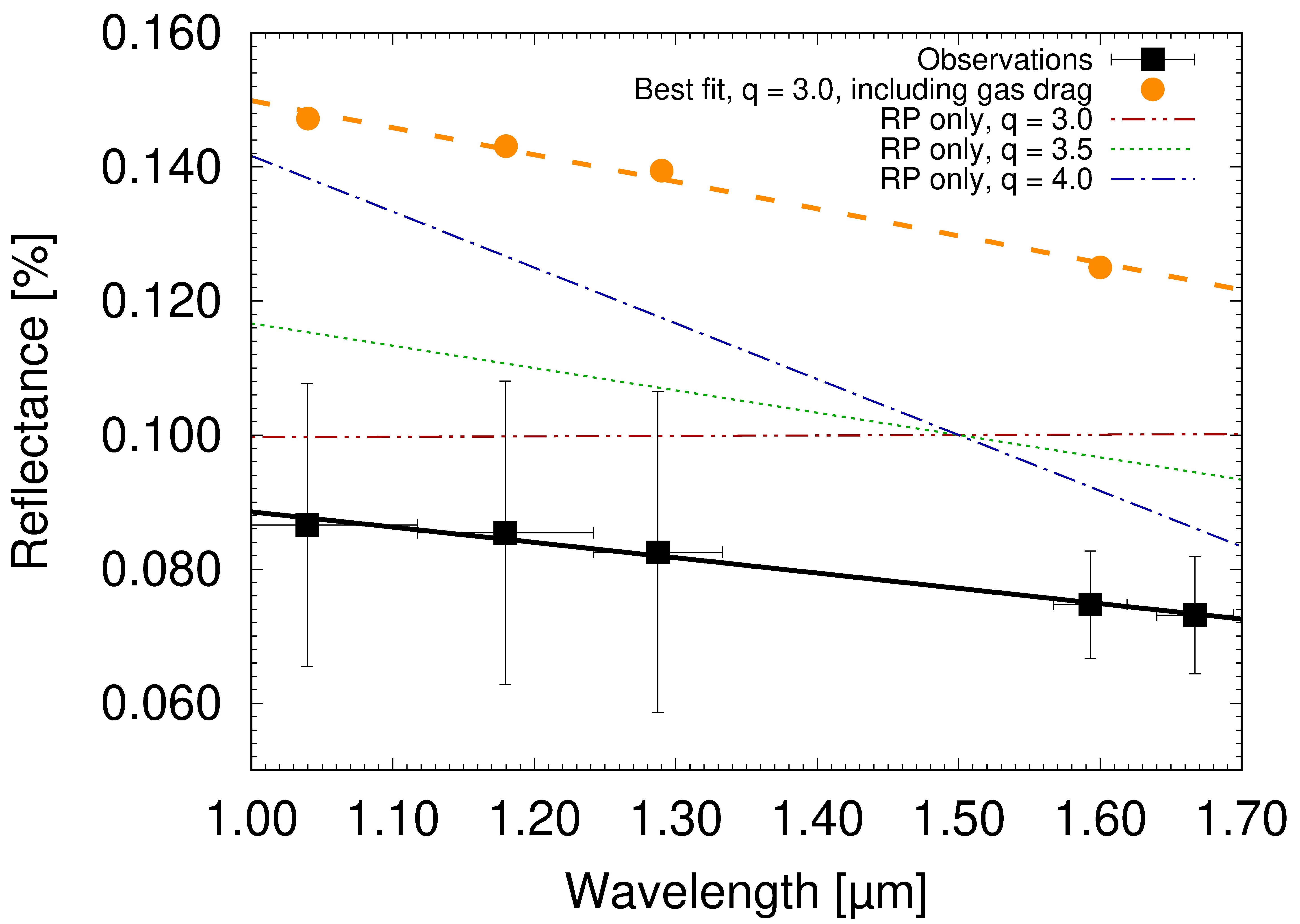}
    \caption{Reflectance for the different wavelength bands used by VLT/SPHERE. Black squares show the values inferred from observations including $1\sigma$ errorbars, orange circles those predicted from the best fit model including gas drag. Additional lines show the predictions for models using radiation pressure only and different size distribution indices. These lines were scaled to a value of 0.1~per cent at 1.5\,$\mu$m.}
    \label{fig:reflectance}
\end{figure}

To derive the reflectance we inferred the total flux density of the disc from the observational images at five different wavelengths (IFS: 1.04, 1.18, 1.29\,$\mu$m; IRIDS: 1.593, 1.667\,$\mu$m). For this we used the HG approach (\S~\ref{sec:HG_results}) with $g=0.67$ and optimised a scaling factor to minimise the residuals in the five spectral channels. 
For our DDA model predictions we only used four wavelengths (1.04, 1.18, 1.29, 1.6\,$\mu$m) as the IRDIS wavelengths are very close to each other. 
We applied the best-fitting values ($q$ and $P$) and calculated the expected flux density. Then we divided the total flux density by the respective stellar flux density at the respective wavelength. 
The stellar flux density of the observations was inferred directly by the PSF while for the models we interpolated the stellar spectrum used in the modelling process (\S~\ref{sec:stellar_properties}).
The result is shown in Fig.~(\ref{fig:reflectance}). 

The observations (black squares) show a weak decrease of reflectance with increasing wavelength, but the uncertainties are large so that the colour of the disc -- while suggested to be blue -- remains uncertain. Recent studies on debris disc colours \citep[e.g.,][]{thebault-et-al-2019, ren-et-al-2023} show that most debris discs seem to be blue in scattered light at wavelengths longer than 1~$\mu$m.  
\cite{ren-et-al-2023} studied discs in the optical and near-IR and also found some discs to possess a red colour at wavelengths shorter than 1\,$\mu$m. Based on the right panel of Fig.~\ref{fig:radiationpressure} we would expect this as sub-micron sized grains sensitively traced at visual wavelengths become less abundant than their still bound micron-sized counterparts. 

To get an idea on how the reflectance would change for different size distribution indices we generated models that show the difference for flat ($q=3.0$), intermediate ($q=3.5$) and steep ($q=4.0$) size distributions. We only assumed radiation pressure to influence the grains. Gas drag was ignored. The results are shown as red, green and blue lines in Fig.~(\ref{fig:reflectance}).
The scaling for those models was done so that the lines cross 0.1 per cent at a wavelength of $1.5\mu$m to keep the plot readable. It is not connected to the actual dust mass. 

We find that for a flat size distribution (red line) there is no change in reflectance with wavelength. This is understandable as for $q=3.0$ all particles independent of their size contribute the same total cross-section and thus, the same amount of light should be scattered at all wavelengths.
For $q>3.0$ we find the reflectance decreasing with increasing wavelength. The larger $q$, the steeper the decrease becomes as the ratio between small and large particles increases as well. 
For values of $q<3.0$ the reflectance is expected to increase since here the total cross-section of big grains is larger than for small particles. 
Due to the large uncertainties of the slope we find that a constraint on the size distribution based on reflectance is not possible.

The orange line in Fig.~(\ref{fig:reflectance}) shows the result for our best fit model including our simple assumptions on gas drag. This model and the observations are scaled correctly. 
Firstly, we see that our best fit model in general predicts a reflectance that is significantly higher than the observational reflectance even taking into account the large uncertainties. 
This can be explained by the differences in scattering phase function (Fig.~\ref{fig:phase_function}). For angles smaller than 20$^\circ$ the phase function for the DDA model is larger than the HG phase function by more than a factor of two. The total flux density is dominated by grains with small scattering angles. However, for small angles observations are rather uncertain as for example the PCA reduction leads to strong over-subtraction. Therefore, we do not put too much emphasise on the different levels of reflectance, but are more interested in its slope. 

We get a similar slope of radiation pressure model and gas drag model when assuming $q\sim3.4$ for the former and $q=3.0$ for the latter.
This shows that even for a size distribution index of $q = 3.0$ we can get a decreasing slope and thus, a blue colour of a disc when including the effects of gas drag. However, due to the large uncertainties a comparison with the observations is not reliable.
Based on the results of our SED and scattered light models we assume that the reflectance indicates a slightly blue colour for the disc around HD~131488.

\section{Discussion}
\label{sec:discussion}

\subsection{Mie vs DDA}
\label{sec:Mie_vs_DDA}

Modelling scattered light observations of debris discs is often difficult as optical dust models such as Mie theory were found to give poor modelling results \citep[e.g.,][]{krist-et-al-2010, milli-et-al-2017, pawellek-et-al-2019b}.
In this paper we analysed the benefit of using DDA to model scattered light observations of debris discs. This approach is not new as other studies investigated circumstellar discs applying DDA \citep[e.g.,][]{min-et-al-2016, arnold-et-al-2019, arnold-et-al-2022, audu-et-al-2023} already. 
However, with this study we present the first analysis using a semi-dynamical disc model and thus, combining optical properties with dust dynamics to generate scattered light models. 

Instead of changing the dust material we varied the level of porosity to investigate the resulting disc models in both scattered light and thermal emission. Other studies showed that porosity influences the modelling outcome e.g., the ratio between dominant grain size and blowout size which hints at the disc's dynamical excitation \citep{pawellek-krivov-2015, brunngraeber-et-al-2017}.
In this study we stayed comparable to Mie theory when applying DDA. We assumed particles of basic spherical shape. For Mie grains we applied Effective Medium Theory (EMT) to generate porous material. For DDA we generated small inclusions of vacuum with a size of 1/100(\S~\ref{sec:dda_theory}). Since we cannot use DDA for grains with $s\geq 10\mu$m, this approach allowed us to fill up the size distribution with Mie particles. To use more complex particle structures in DDA more work is needed to infer possible ways of filling up the size distribution.

We find that when assuming grains of spherical shape and small void sizes, DDA and Mie lead to similar results for pure radiation pressure models. Deviations between the models can be explained by different set-ups (e.g., different blowout sizes, phase functions, $\beta$, etc). 
This outcome is somewhat expected as Mie theory is a limiting case for DDA when assuming spherical shapes. 
For the special case of HD~131488 we find that particles of basic spherical shape and small vacuum inclusions can reproduce the observations very well. Thus, Mie theory seems a valid approach to model the scattered light data for this disc. As mentioned before, this seems not the case for many debris discs, although there are studies reaching a similar result \citep[e.g.,][]{ertel-et-al-2011}.

While the benefit of DDA is not particularly obvious for this study, we emphasise that we are now able of introducing arbitrarily shaped grains or dust aggregates similar to \citet{zubko-2013} or \cite{min-et-al-2016} into the MODERATO code to model discs where Mie theory is not a good approximation for the dust particles. However, we note that a transition between DDA and Mie will be necessary to cover the whole grain size range.

\subsection{Radial extent}
\label{sec:extent}

\subsubsection{ALMA vs SPHERE}
The MODERATO code uses the location of the planetesimal belt as input to calculate images at different wavelengths. The largest dust grains traced by ALMA are barely affected by radiation pressure or other transport mechanisms \citep[e.g.,][]{pawellek-et-al-2019a} so that we can assume the radial extent inferred from ALMA data to reflect the actual planetesimal belt location. 
Also assuming that the dust grains are produced in mutual collisions within the planetesimal belt, the radial extent of the disc at shorter wavelengths should be in agreement with those dust grains put on eccentric orbits due to radiation pressure.

While the radial profile inferred from ALMA favours a Gaussian with a central radius of $(88~\pm3)$~au and a total disc width of $(46\pm 12)$~au \citep{moor-et-al-2017}, all three modelling approaches for scattered light  prefer a radial profile that deviates from the ALMA data. 
The HG~approach (\S~\ref{sec:HG_results}) used a narrow ring with a peak in surface brightness at $110\pm25$~au. Both DDA and Mie used an asymmetric Gaussian starting at the peak fixed to 88~au (similar to ALMA) and a disc width of 30~au (Fig.~\ref{fig:radial_profile_belt}).
Despite the differences in profile set-ups we found that HG, DDA, and Mie predict that within 88~au the amount of dust material has to be small in order to fit the scattered light data (see \S\ref{sec:PR-drag} and \S\ref{sec:transport_gas} for details). 

Due to the low spatial resolution of the ALMA data and the nearly edge-on orientation of the disc, the region within 88~au is not well constrained at long wavelengths. The uncertainty in disc width ($46\pm12$~au) inferred from ALMA is one indicator for this. 
Ignoring the inner region, we find that the radial profile for the outer region ($r>88$~au) seems to be consistent in ALMA and scattered light models (width for ALMA: $(23\pm 6)$~au; width for SPHERE: $(30\pm 5)$~au).

\subsubsection{Inward transport - PR-drag}
\label{sec:PR-drag}

We found that all scattered light models predict that the amount of dust within 88~au is low. In this section we investigate how large the amount is that we would expect at these regions due to transport processes, and whether the dust should be visible in scattered light.
Similar to \cite{pawellek-et-al-2019a} we analyse the effect of Poynting-Robertson drag on the radial distribution to estimate the amount of dust drifting inwards from the planetesimal belt. 
To do so we use the collisional code ACE \citep{loehne-et-al-2017} for a planetesimal belt between 88 and 118~au and bodies up to 40\,km in radius. We collisionally evolve the belt for several Myr and infer the surface density of the dust as function of radius.
From this we then calculate the flux density applying eq.~(\ref{eq:flux_density}). 
\begin{figure}
    \centering
    \includegraphics[width=\columnwidth]{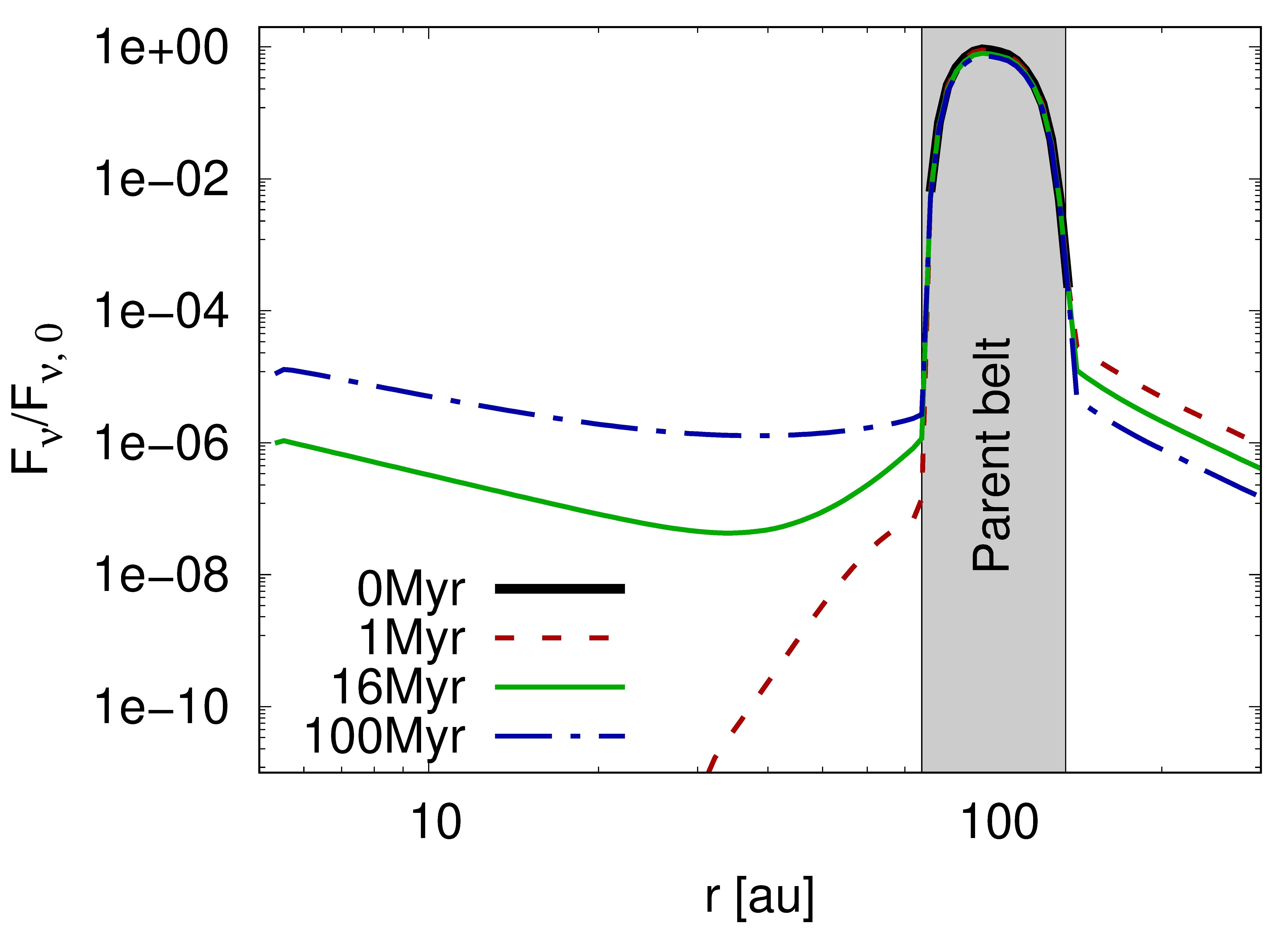}
    \caption{Normalised flux density as function of radius inferred by the collisional code ACE for a planetesimal belt of different ages.}
    \label{fig:PR-drag}
\end{figure}

In Fig.~(\ref{fig:PR-drag}) we show the flux density as function of radius normalised to the maximum found at the location of the planetesimal belt.
At the age of 16\,Myr -- the proposed age of HD~131488 -- the amount of dust drifting inwards due to PR-drag leads to a flux density between 6 and 7 orders of magnitudes lower than that of the planetesimal belt which is well below the detection limit of VLT/SPHERE. Even for older systems the flux density within 88~au would be 5 orders of magnitude lower. 
The low amount of dust expected for distances close to the planetesimal belt is in agreement with the scattered light models suggesting that particles moving inwards due to PR-drag do not significantly contribute to the scattered light.

\subsubsection{Inward transport - gas drag}
\label{sec:transport_gas}

In addition to PR-drag the gas present in the disc might cause the dust grains to migrate. For HD~131488 we found gas between 30 and 130~au (\S\ref{sec:gas}) so that it seems possible to have dust as close as 30~au. We use the classical approach of \cite{takeuchi-artymowicz-2001} to estimate the amount of dust within 88~au.

The ratio of the force supporting the gas against stellar gravity to the gravity force is given by the parameter $\eta$. The gas can add (remove) angular momentum to (from) the dust grains. As a result the particles migrate outwards (inwards) until they reach a certain stability distance, $r_\text{s}$. At this distance the gas pressure gradient and the stellar radiation pressure balance each other \citep{takeuchi-artymowicz-2001} so that 
\begin{equation}
    \beta(s) = \eta(s, r_\text{s}).
\end{equation}
We are interested in the particles that migrate inwards. For those $\beta<\eta$. In a similar approach to \cite{krivov-et-al-2009} we inferred the $\beta$-values for which this relation is fulfilled \citep[see Fig.~2 in][]{krivov-et-al-2009}. Note that $\eta$ is only a function of the gas surface density profile, and the gas temperature, $T$, \citep[eq.~10 in ][]{krivov-et-al-2009}.
For both we assume power laws with typical exponents: $\Sigma\propto r^-{3/2}$ and $T\propto r^{-1/2}$. We assume that $\eta \propto L_\text{star}^{-0.25}/M_\text{star}$.
At a distance of $88$~au it is expected that only grains with $\beta<0.05$ are dragged inwards. All particles with $\beta>0.05$ are expected to drift outwards. This means that there seems to be no inward-drift of (sub-)blowout particles for which $\beta\gtrsim0.5$ (Fig~\ref{fig:beta_s}).

Additionally, we need to take into account the drag force of the gas which determines the migration timescale of the large dust grains and thus, how many of them we would expect to drift inwards. Following \cite{marino-et-al-2020} we found that for particles with $\beta\leq0.05$ the migration timescale would be longer than the collision timescale. Thus, for these grains we do not expect inward-migration due to gas-drag.

We conclude that the effect of PR-drag is more dominant than gas drag when analysing the inner region of HD~131488 in scattered light. 
As shown in \S\ref{sec:PR-drag}, the flux density coming from grains migrating inwards due to PR-drag is several orders of magnitude lower compared to that of the planetesimal belt. 
Based on our findings we assume the effect of gas drag to be lower than that of PR-drag, and thus, we assume that the total amount of dust migrating inwards from the planetesimal belt is small and remains unseen in scattered light.

\subsubsection{Projection effects}

\begin{figure*}
    \includegraphics[width=\textwidth]{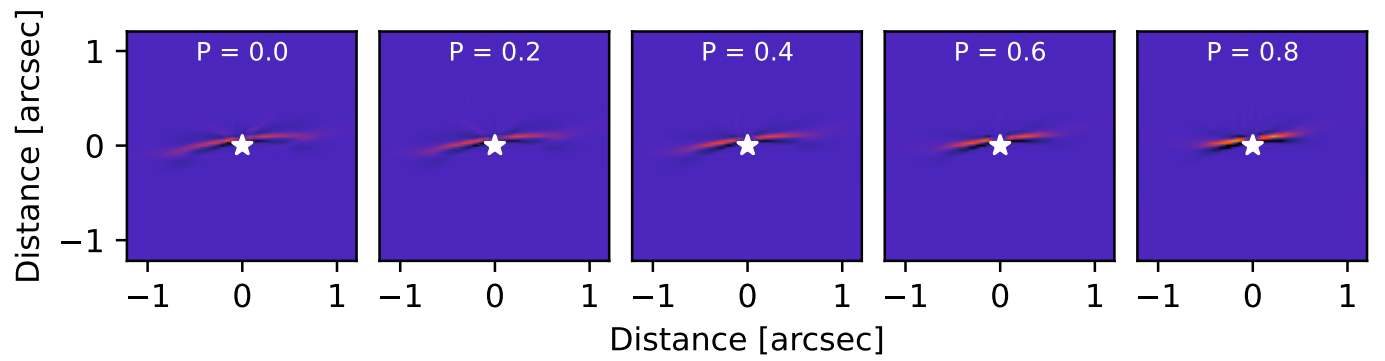}
    \caption{Disc models for a disc radius of 100~au and a disc width of 20~au applying different porosities $P$.}
    \label{fig:Plot_Model_P}
\end{figure*}

In Fig.~(\ref{fig:Plot_Model_P}) we show a disc model where $r_0 = 100$~au,  $\Delta r = 20$~au, $i = 84^\circ$, and $M_d=1~M_\oplus$ for different levels of porosity.
We find that for an edge-on disc the radial extent of surface brightness seems to decrease with increasing porosity. 

An explanation might be given by the scattering phase function. The lower the disc inclination the lower the range of scattering angles we can observe. For a face-on disc we only get particles with a scattering angle of $\vartheta = 90^\circ$. Thus, if the material becomes more porous, the total flux density of the disc decreases (Fig.~\ref{fig:phase_function}, Appendix~\ref{sec:flux_densities}, \citealt{samra-et-al-2022}). As a result, the sensitivity limit of our instrument is reached at smaller distances from the star already so that the radial extent might seem smaller.

If the disc is now close to edge-on, we nearly cover all scattering angles between 0 and 180$^\circ$. For more porous material the phase function shows a steeper decrease at larger scattering angles compared to compact materials. Thus, with increasing porosity the particles at larger scattering angles contribute less to the total flux density which results in an apparently decreasing disc extent shown in Fig~\ref{fig:Plot_Model_P}.

\subsection{Dust properties}
\label{sec:dust_properties}

\subsubsection{Scattering phase function}

Our first scattered light image was generated by using the HG approximation where we found $g = 0.67$ to give the best-fitting model (\S~\ref{sec:HG_results}). This indicates a relatively high level of forward scattering compared to other debris discs that were modelled with HG in scattered light \citep[e.g.,][]{engler-et-al-2020, stark-et-al-2023}. 
A strong forward scattering is expected for spherical particles which is the reason that our Mie models were well suited to fit the scattered light data.

Assuming that the results from the HG model give the best approximation of the ``real'' scattering phase function of HD~131488, we find that a porosity of $P=0.6$ is the closest fit to the HG function for both DDA and Mie models (Fig.~\ref{fig:average_phasefunction}, \S~\ref{sec:phase_function}), but that we cannot rule out porosities of 0.2 and 0.4. A porosity of 0.0 would lead to a higher fraction of backward scattering, a porosity of 0.8 would not exhibit enough backward scattering compared to the HG function.

\subsubsection{SED}
We modelled the SED of HD~131488 for the different porosities and found that for $P=0.8$ the size distribution parameter is not well constrained as $q$ reaches a value of $-8.2\pm2.7$ (note the negative sign). 
The size distribution index is determined by the long-wavelength data (far-IR to mm). For higher porosities the decrease becomes steeper even if $q$ stays constant. A negative $q$-value shows the dominance of the largest particles in the size distribution ($10^4~\mu$m) in order to fit the long-wavelength data. This is not consistent with collisional evolution but rather a pure outcome of the fitting procedure.

Since lower porosities result in reasonable fits of the photometric data, and are in agreement with collisional evolution, we conclude that a very high porosity of $P=0.8$ seems unrealistic for the material in the disc of HD~131488 based on the SED. This is confirmed by the scattered light models.

\subsubsection{Conclusion}

Combining the results from the scattering phase function, and SED modelling, we conclude that the material in the disc around HD~131488 probably possesses a porosity between 0.2 and 0.6. While we cannot rule out smaller porosities, we find that porosities larger than 0.6 seem unlikely.
Porosities of 0.6 are consistent with results from our Solar system where porosities of $\sim$50\% were found for ``rubble-pile'' asteroids \citep[e.g.,][]{weidling-et-al-2009, walsh-2018, omura-et-al-2021, sakatani-et-al-2021}.

In comparison a study of the debris disc around AU~Mic found a significantly higher porosity of 76\% \citep{arnold-et-al-2022}. 
The difference in porosity between the late and early-type stars might indicate a direct influence of dynamical excitation on the dust material. 
\cite{pawellek-krivov-2015} found that debris discs around earlier-type stars such as HD~131488 possess a higher excitation i.e., a higher collision velocity than discs around late-type stars such as AU~Mic. 
Collisions between planetesimals lead to compaction of the material and thus might decrease the overall porosity down to 40\% \citep[e.g.,][]{housen-et-al-2018, walsh-2018}. This might be the case for HD~131488. 
In contrast to that, AU~Mic might be less dynamically excited so that the material might not be compacted to the same degree as HD~131488. 
So far, we are lacking a study investigating the porosity of debris discs as function of stellar luminosity to make any conclusive remarks on relations between disc excitation and porosity.

\subsection{Size distribution index}

Our scattered light and SED model predict a size distribution index of $q = 3.0$ which suggests that all particle size bins contribute the same cross-section so that even large particles contribute to the near-IR scattered light image (Fig.~\ref{fig:Fnu_s_q}).
Assuming a collisional cascade we would expect $q$ to lie between 3 and 4 (\S~\ref{sec:size_distribution}) with an ideal collisional cascade at $q=3.5$ \citep{dohnanyi-1969}. 
This puts the size distribution index of our debris disc to the lower boundary for such a cascade. 
While not very common, such low $q$-values can also be found for other debris discs such as HD~32297 and HD~131835 \citep{norfolk-et-al-2021, loehne-2020}.
Both of these discs contain a significant amount of gas \citep{moor-et-al-2019} so that we cannot rule out a link between gas content and size distribution index. 
On the other hand, other CO-rich debris discs did not show such a low $q$-value \citep[e.g., HD 9672, HD~21997,][]{pawellek-et-al-2014} so that a more thorough study is needed to draw any conclusions on a possible link between those parameters. 
Studying a sample of 22 debris discs including both gaseous and gas-poor discs, \cite{norfolk-et-al-2021} found that for gaseous discs the $q$-value tends to be lower compared to that of gas-poor discs. However, this is based on a small number of gaseous discs so that an actual trend is still debatable. 

Considering the results on the reflectance (\S\ref{sec:reflectance}), we were not able to constrain the size distribution index due to the large uncertainties.

SED models of debris discs that included mm-data suggest that a single power law as size distribution might not be realistic \citep[e.g., this work,][]{lestrade-et-al-2020}. Furthermore, from our own Solar system we know that the size distributions of the Asteroid and Edgeworth-Kuiper belt change with size and also take values of  $q < 3$ \citep[e.g.,][]{yoshida-et-al-2007, morbidelli-et-al-2021}. Thus, it is likely that HD~131488 also possesses a more complex size distribution.

\subsection{Combining scattered light and SED models}
\label{sec:SED_scattered_light}

There are several studies that analysed debris discs at several wavelengths \citep[e.g.,][]{ertel-et-al-2011, macgregor-et-al-2015, ballering-et-al-2016, pawellek-et-al-2019b, thebault-et-al-2019, thebault-et-al-2023, esposito-et-al-2020}.
The study of \cite{schneider-et-al-2006} investigated the debris disc around HD~181327 and tried to combine results from thermal emission and scattered light. In their Fig.~(14) the study shows that there is nearly no overlap between SED and scattered light models.
With our study we were able to generate a self-consistent model fitting both thermal emission and scattered light for the first time with a semi-dynamical disc model.

While we were able to fit the scattered light data with a pure radiation pressure model, we found that we would need an amount of dust five times higher than required by thermal emission data when assuming a proper eccentricity of the planetesimals of $\langle e\rangle = 0.1$ (see \S~\ref{sec:minimum_size}).

\begin{figure}
    \centering
    \includegraphics[width=\columnwidth]{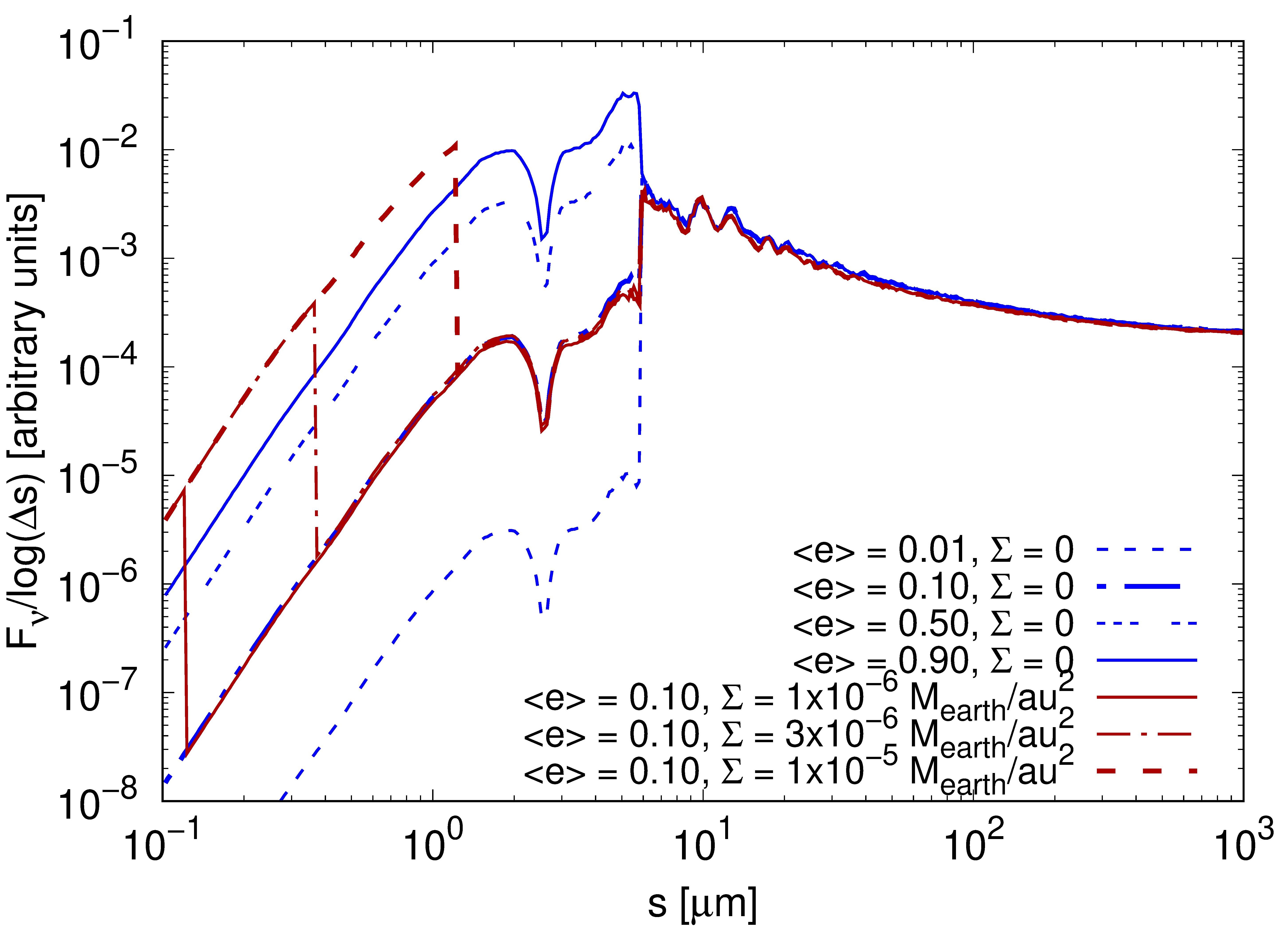}
    \caption{Flux density as function of grain size. Blue lines indicate different proper eccentricities $\langle e\rangle$, red lines different levels of gas surface density $\Sigma$.}
    \label{fig:excitation}
\end{figure}

We investigated the influence of dynamical excitation on dust mass. In Fig.~(\ref{fig:excitation}) we show the total flux density per size bin for different levels of proper eccentricity $\langle e\rangle$ in blue. All models assume a porosity of 0.6.
We see that a higher dynamical excitation increases the flux density per size bin as a higher amount of sub-blowout particles is produced by collisions.
We also find that for a pure radiation pressure model with dynamical excitation of $\langle e\rangle = 0.9$ we only need a total dust mass of $1.3\,M_\oplus$ to reproduce the scattered light observations. 
This is close to the mass inferred by the SED ($1.0\pm0.2\,M_\oplus$), but firstly, the $\chi^2$-value for this model is larger than the critical value and thus, does not lead to a good fit. And secondly, a proper eccentricity of 0.9 leads to a very short collisional timescale for the planetesimals.
For example for a km-sized body at a radius of 88~au the lifetime is $\sim30$\,Myr assuming $\langle e\rangle = 0.1$. Assuming $\langle e\rangle = 0.9$ instead this shortens to 0.5\,Myr for the same body \citep{wyatt-et-al-2007b, loehne-et-al-2007}.
Thus, for 16~Myr-old HD~131488, it would be likely that the debris disc was collisionally depleted if $\langle e\rangle = 0.9$. In that case, the disc would not be detectable for our instruments anymore. As a result, we need a model with a lower dynamical excitation but with a mechanism retaining the sub-blowout grains (in our case gas drag) to explain the observational data of HD~131488.

In Fig.~(\ref{fig:excitation}) we see that a higher gas surface density increases the size of the grains coupled to the gas i.e., the number of size bins, but not the flux density per size bin. Thus, gas drag and dynamical excitation are degenerate.
Compared to our best-fit model with $\langle e\rangle=0.1$ and $\Sigma=2\times10^{-5}\,M_\oplus/$au$^2$ we find that for a very high dynamical excitation of $\langle e \rangle = 0.5$ we still need a gas surface density of $\Sigma=1\times10^{-5}\,M_\oplus/$au$^2$ to reproduce the observations. The model also gives a $\chi^2$-value below the critical value and thus, a good fit.

While this model can reproduce the data, there are several arguments against it.
Firstly, the dynamical excitation is related to the size distribution index $q$ which, in an ideal case, is given as 3.5. The larger $q$, the higher the amount of micron-sized particles compared to mm-sized ones (Fig~\ref{fig:Fnu_s_q}). This implies that for larger $q$ the disc might possess a higher dynamical excitation as more small grains are produced during collisions of the large counterparts. However, the SED model of HD~131488 and its scattered light data both led to a best-fit of $q=3.0\pm0.1$. This is a low value indicating a \textit{small} dynamical excitation.

Secondly, the high amount of gas in the disc leads to damping of the particles' eccentricities which we use in our gas drag model. And thirdly, other dynamically excited (``self-stirred'') discs were found to possess proper eccentricities of $\langle e\rangle\lesssim0.2$ rather than 0.5 \citep[e.g.,][]{krivov-et-al-2006, thebault-augereau-2007, loehne-et-al-2012b, pawellek-krivov-2015, schueppler-et-al-2015, geiler-et-al-2019, daley-et-al-2019, matra-et-al-2019b}. 
We therefore prefer the model with $\langle e\rangle = 0.1$ and $\Sigma=2\times10^{-5}\,M_\oplus/\text{au}^2$. Anyhow, we cannot rule out smaller proper eccentricities.

\subsection{Influence of gas}

As mentioned in \S~\ref{sec:SED_scattered_light}, a pure radiation pressure model is not able to reproduce the observational data of HD~131488. Only a retaining mechanism such as gas drag led to consistent scattered light and thermal emission models.

We found a gas surface density for \ce{CO} of $2\times10^{-6}~M_\oplus\,$au$^2$ but assume that other gas species might add to this. 
\cite{smirnov-et-al-2022} tried to find additional molecules in \ce{CO}-rich debris discs but did not detect any. It is likely that the species targeted by this study are not shielded from the stellar UV radiation and thus, dissociate very quickly \citep[e.g.,][]{matra-et-al-2018}. Additionally, the selected molecules are not thought to be dominant components in the gas mixture so that they would hardly contribute to the total gas mass. We would need detections of \ce{C}, \ce{O}, and \ce{H} to constrain the total gas mass reliably assuming a secondary origin for the gas. In the case of a primordial origin, a detection of \ce{H2} would help constraining the gas mass.

A total surface density of $10^{-5}~M_\oplus\,$au$^{-2}$ for our debris disc seems possible(\S~\ref{sec:gas}), but we cannot offer a strong constraint on this number. However, 
with our simple assumptions on gas drag we found that a gas surface density of $2\times 10^{-5}~M_\oplus\,$au$^{-2}$ is sufficient to couple a high amount of sub-blowout grains so that we can fit the SED and scattered light data with the same amount of dust. While this is an first order agreement with the rough estimate from ALMA, a more detailed work is needed to get reliable constraints on the total gas mass.

What we can say is that, when ignoring a mechanism that can retain small dust particles (in our case gas) it seems not possible to fit both thermal emission and scattered light of the disc around HD~131488 at the same time. 
Matching the amount of dust needed to fit the SED and the scattered light data opened a way to roughly estimate the surface density of the gas.

\section{Summary}
\label{sec:summary}

In this study we analysed the scattered light and thermal emission data of the debris disc around HD~131488 applying a semi-dynamical disc model in combination with HG, DDA, and Mie theory.
The SPHERE/IRDIS and IFS data of HD~131488 were presented for the first time. The modelling results are summarised in Tab.~(\ref{tab:fitting_results}).

\begin{table}
    \centering
    \begin{tabular}{c|cc}
    Parameter & Best Fit & Reference section\\
    \midrule
        $R_\text{central}$ [au]      & $88\pm5$    & \S~\ref{sec:mie_model}\\
        $\Delta R$ [au]              & $30\pm3$    & \S~\ref{sec:mie_model}\\
        $s_\text{min}$ [$\mu$m]      & $10^{-1}$   & \S~\ref{sec:minimum_size}   \\ 
        $s_\text{max}$ [$\mu$m]      & $10^4$      & \S~\ref{sec:maximum_size}   \\ 
        $q$                          & $3.0\pm0.1$ & \S~\ref{sec:mie_model}, \ref{sec:rp_model}, \ref{sec:SED}  \\
        $M_\text{dust}$ [$M_\oplus$] & $1.0\pm0.2$ & \S~\ref{sec:SED}  \\
        $P$                          & 0.2\ldots0.6& \S~\ref{sec:dust_properties}  \\
        $\Sigma_\text{gas}$ [$M_\oplus/$au$^2$]& $(2.0\pm0.1) \times 10^{-5}$ & \S~\ref{sec:gas_model} 
    \end{tabular}
    \caption{Summary of best fitting  results for HD~131488 and the sections of the paper where they were discussed.}
    \label{tab:fitting_results}
\end{table}

The high amount of \ce{CO}-gas found in the disc \citep{moor-et-al-2017} is capable of retaining a large fraction of sub-blowout grains. 
Only if we take into account these particles, we are able to generate a model that can fit all data available for this debris disc (thermal emission and scattered light). 
This opens a way of roughly estimating the amount of gas necessary to fit all data. 

The radial profile of the planetesimal belt preferred by the scattered light models seems to deviate from the one inferred by ALMA observations in the way that in scattered light we do not expect a significant amount of dust within the central radius of 88~au. This is in agreement with expectations from PR and gas drag models.
However, the deviation of the profiles might be attributed to the low spatial resolution of the ALMA data.

The disc possesses a flat size distribution ($q=3$) and moderate level of porosity of $\sim20\ldots60\%$ which is in agreement with a collisional cascade and results from Asteroid observations suggesting a pebble pile scenario for planetesimal growth within this system.
Compared to the disc around AU~Mic the material might have been more compacted by collisions. 
While the reflectance might indicate a slightly blue colour of the debris disc, the uncertainties of the observations are too large to draw any conclusions. However, the modelling results including gas drag indicate a blue colour as well and thus, might be in agreement with the observational results.

The modelling approach of DDA and Mie led to similar results when assuming particles of basic spherical shape and small sizes for vacuum inclusions. For HD~131488 Mie theory leads to well-fitting models indicating that the dust particles possess a scattering behaviour similar to spheres.
To study the influence of more complex particle shapes more work is needed as DDA is limited to small sizes making Mie grains necessary to fill-up the size distribution.

\section*{Acknowledgements}

We thank the anonymous referee for their constructive criticism and help improving the presentation of this study.

NP is grateful to Torsten L\"ohne, Philippe Th\'ebault, J\"urgen Blum and Peter Woitke for many useful discussions. NP also thanks Kevin Wagner and Benoit Pairet for comments on data analysis.  
AMH is supported by a Cottrell Scholar Award from the Research Corporation for Science Advancement. SM is funded by the Royal Society through a University Research Fellowship.

This work has made use of data from the European Space Agency (ESA) mission
{\it Gaia} (\url{https://www.cosmos.esa.int/gaia}), processed by the {\it Gaia}
Data Processing and Analysis Consortium (DPAC,
\url{https://www.cosmos.esa.int/web/gaia/dpac/consortium}). Funding for the DPAC
has been provided by national institutions, in particular the institutions
participating in the {\it Gaia} Multilateral Agreement.

SPHERE is an instrument designed and built by a consortium
consisting of IPAG (Grenoble, France), MPIA (Heidelberg, Germany), LAM
(Marseille, France), LESIA (Paris, France), Laboratoire Lagrange
(Nice, France), INAF–Osservatorio di Padova (Italy), Observatoire de
Genève (Switzerland), ETH Zurich (Switzerland), NOVA (Netherlands),
ONERA (France) and ASTRON (Netherlands) in collaboration with
ESO. SPHERE was funded by ESO, with additional contributions from CNRS
(France), MPIA (Germany), INAF (Italy), FINES (Switzerland) and NOVA
(Netherlands).  SPHERE also received funding from the European
Commission Sixth and Seventh Framework Programmes as part of the
Optical Infrared Coordination Network for Astronomy (OPTICON) under
grant number RII3-Ct-2004-001566 for FP6 (2004–2008), grant number
226604 for FP7 (2009–2012) and grant number 312430 for FP7
(2013–2016). We also acknowledge financial support from the Programme National de
Planétologie (PNP) and the Programme National de Physique Stellaire
(PNPS) of CNRS-INSU in France. This work has also been supported by a grant from
the French Labex OSUG@2020 (Investissements d’avenir – ANR10 LABX56).
The project is supported by CNRS, by the Agence Nationale de la
Recherche (ANR-14-CE33-0018). It has also been carried out within the frame of the National Centre for Competence in 
Research PlanetS supported by the Swiss National Science Foundation (SNSF). MRM, HMS, and SD are pleased 
to acknowledge this financial support of the SNSF.
Finally, this work has made use of the the SPHERE Data Centre, jointly operated by OSUG/IPAG (Grenoble), PYTHEAS/LAM/CESAM (Marseille), OCA/Lagrange (Nice) and Observatoire de Paris/LESIA (Paris) and is supported by a grant from 
Labex OSUG@2020 (Investissements d'avenir - ANR10 LABX56). 
This work has made use of the High Contrast Data Centre, jointly operated by OSUG/IPAG (Grenoble), PYTHEAS/LAM/CeSAM (Marseille), OCA/Lagrange (Nice), Observatoire de Paris/LESIA (Paris), and Observatoire de Lyon/CRAL, and supported by a grant from Labex OSUG@2020 (Investissements d’avenir – ANR10 LABX56). We thank P. Delorme and E. Lagadec (High Contrast Data Centre) for their efficient help during the data reduction
process. 

This paper makes use of the following ALMA data: ADS/JAO.ALMA\#2015.1.01243.S. ALMA is a partnership of ESO (representing its member states), NSF (USA) and NINS (Japan), together with NRC (Canada), MOST and ASIAA (Taiwan), and KASI (Republic of Korea), in cooperation with the Republic of Chile. The Joint ALMA Observatory is operated by ESO, AUI/NRAO and NAOJ

\section*{Data Availability}

The data underlying this article will be shared on request to the corresponding author. The ALMA and VLT/SPHERE data are publicly available and can be queried and downloaded directly from the ALMA archive: https://almascience.nrao.edu/asax/ and the SPHERE archive: https://archive.eso.org/wdb/wdb/eso/sphere/.





\newcommand{\AAp}      {A\& A}
\newcommand{\AApR}     {Astron. Astrophys. Rev.}
\newcommand{\AApS}     {AApS}
\newcommand{\AApSS}    {AApSS}
\newcommand{\AApT}     {Astron. Astrophys. Trans.}
\newcommand{\AdvSR}    {Adv. Space Res.}
\newcommand{\AJ}       {AJ}
\newcommand{\AN}       {AN}
\newcommand{\AO}       {App. Optics}
\newcommand{\ApJ}      {ApJ}
\newcommand{\ApJL}     {ApJL}
\newcommand{\ApJS}     {ApJS}
\newcommand{\ApSS}     {Astrophys. Space Sci.}
\newcommand{\ARAA}     {ARA\& A}
\newcommand{\ARevEPS}  {Ann. Rev. Earth Planet. Sci.}
\newcommand{\BAAS}     {BAAS}
\newcommand{\CelMech}  {Celest. Mech. Dynam. Astron.}
\newcommand{\EMP}      {Earth, Moon and Planets}
\newcommand{\EPS}      {Earth, Planets and Space}
\newcommand{\GRL}      {Geophys. Res. Lett.}
\newcommand{\JGR}      {J. Geophys. Res.}
\newcommand{\JOSAA}    {J. Opt. Soc. Am. A}
\newcommand{\MemSAI}   {Mem. Societa Astronomica Italiana}
\newcommand{\MNRAS}    {MNRAS}
\newcommand{\NAT}      {Nature Astronomy}
\newcommand{\PASJ}     {PASJ}
\newcommand{\PASP}     {PASP}
\newcommand{\psj}      {PSJ}
\newcommand{\PSS}      {Planet. Space Sci.}
\newcommand{\RAA}      {Research in Astron. Astrophys.}
\newcommand{\SolPhys}  {Sol. Phys.}
\newcommand{\SolSysRes}{Sol. Sys. Res.}
\newcommand{\SSR}      {Space Sci. Rev.}

\bibliographystyle{mnras}
\bibliography{Englisch}



\appendix

\section{Scattering phase functions}

\subsection{Size and spatial distribution of voids}
\label{sec:void_distribution}

\begin{figure}
    \includegraphics[width=\columnwidth]{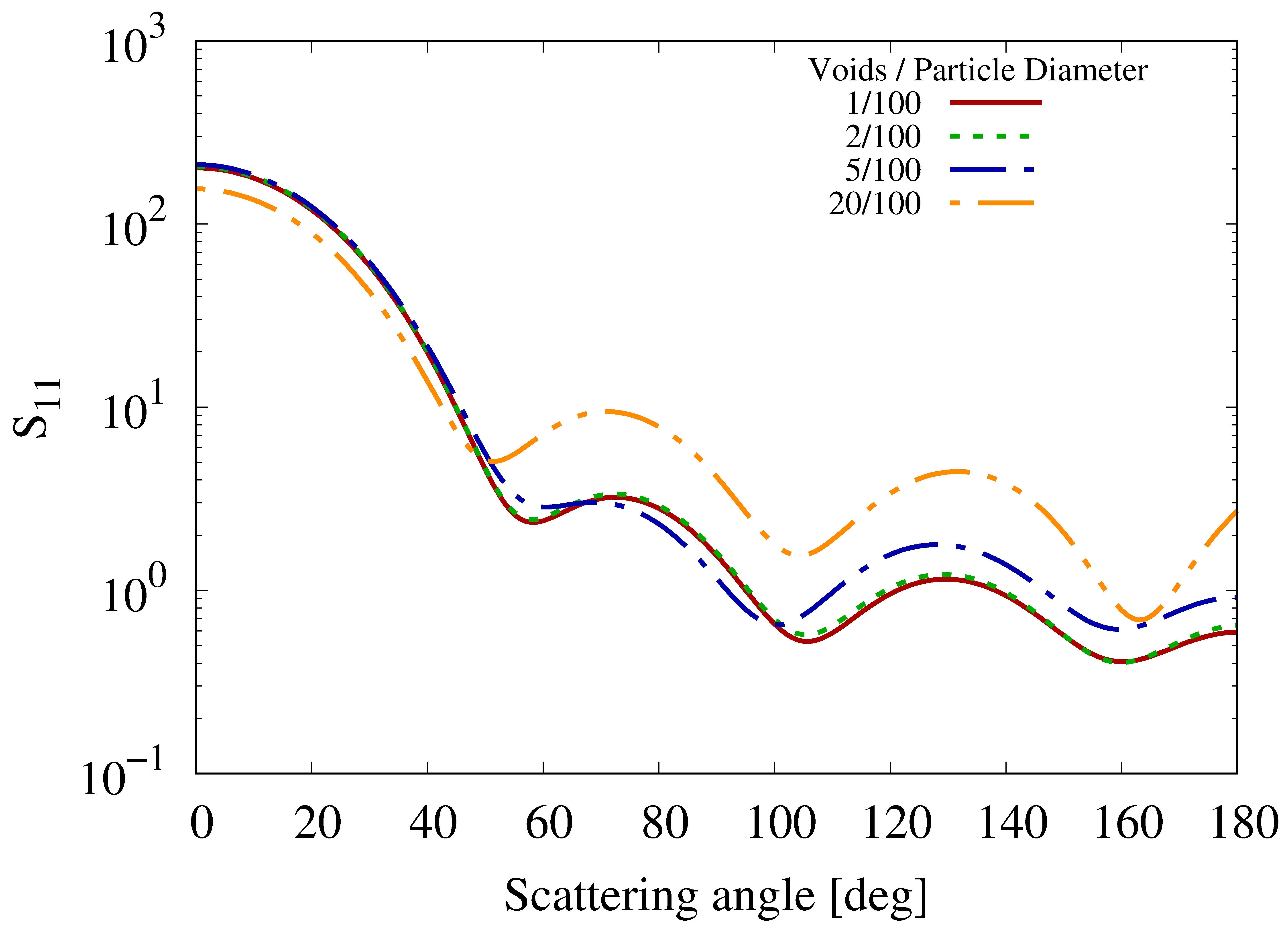}
    \includegraphics[width=\columnwidth]{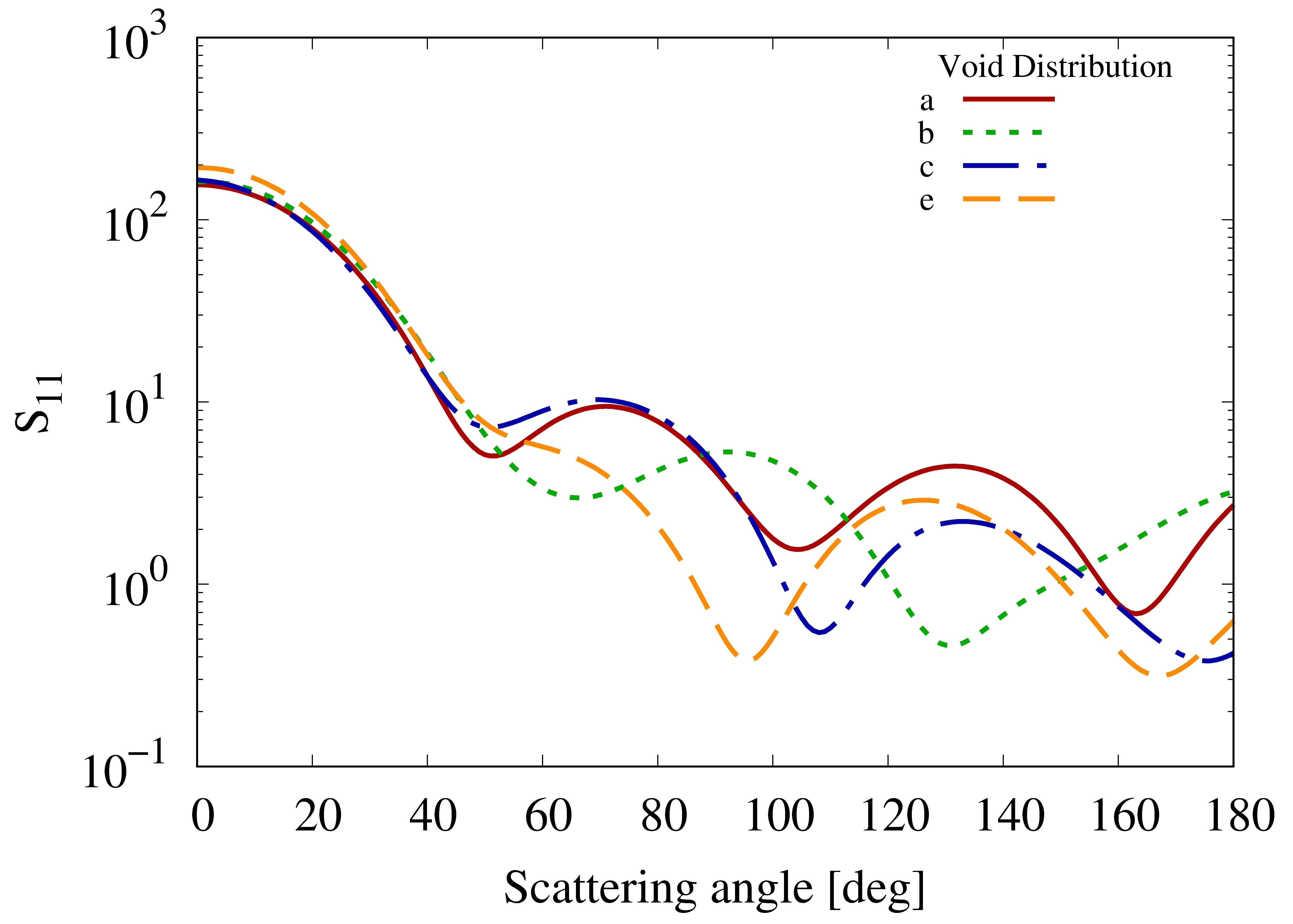}
    \caption{Scattering phase function as function of scattering angle for a particle of radius $1\,\mu$m, assuming a porosity of $P=0.4$ at a wavelength of $1.6\,\mu$m. Top panel: Change of the void size but keeping the spatial distribution constant. Bottom panel: Change of the spatial distribution of voids with constant sizes of 20/100 voids per particle diameter. The different random distributions are called a, b, c, and d.}
    \label{fig:S11_theta_voids}
\end{figure}

In Fig.~(\ref{fig:S11_theta_voids}) we show the scattering phase function as function of scattering angle analysing the influence of different void sizes (inclusions of vacuum) and spatial distributions of those voids.
The void size of 1/100 was applied in all DDA scattered light models of this study (red solid line, top panel). 
We see that for small sizes (1/100 and 2/100) the phase function does not change significantly, but that the changes become more pronounced with larger sizes (5/100 and 20/100) which is in agreement with results from studies using more complex particle structures \citep[e.g.,][]{arnold-et-al-2019}.

In a similar fashion we kept the void size constant (20/100, bottom panel of Fig.~\ref{fig:S11_theta_voids}) and analysed the influence of the spatial distribution of the vacuum inclusions. We see that for a large size even their spatial distribution can change the phase function significantly.

\subsection{Different Porosities}
\label{sec:phase_function}

\begin{figure*}
    \includegraphics[width=0.45\textwidth]{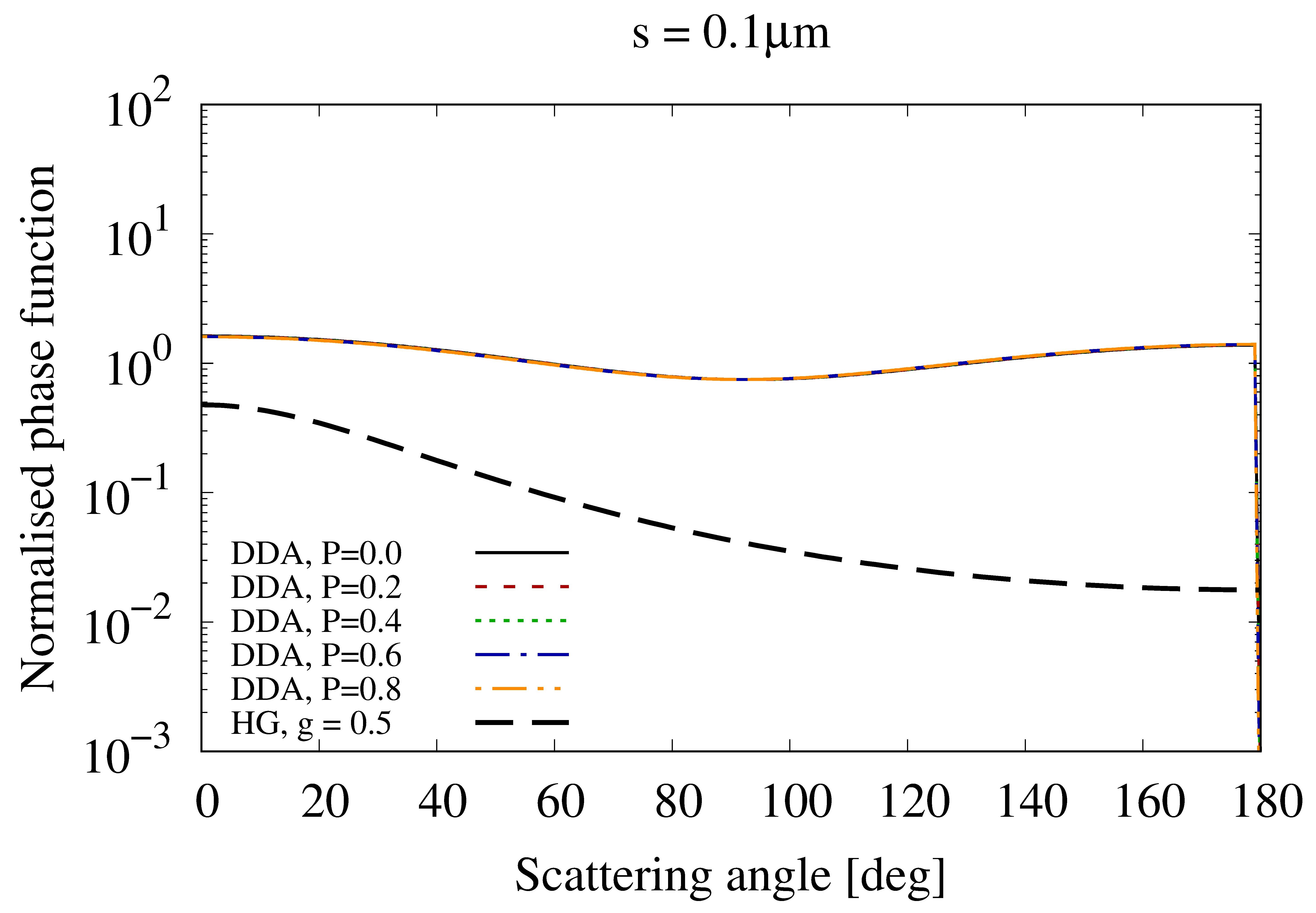}
    \includegraphics[width=0.45\textwidth]{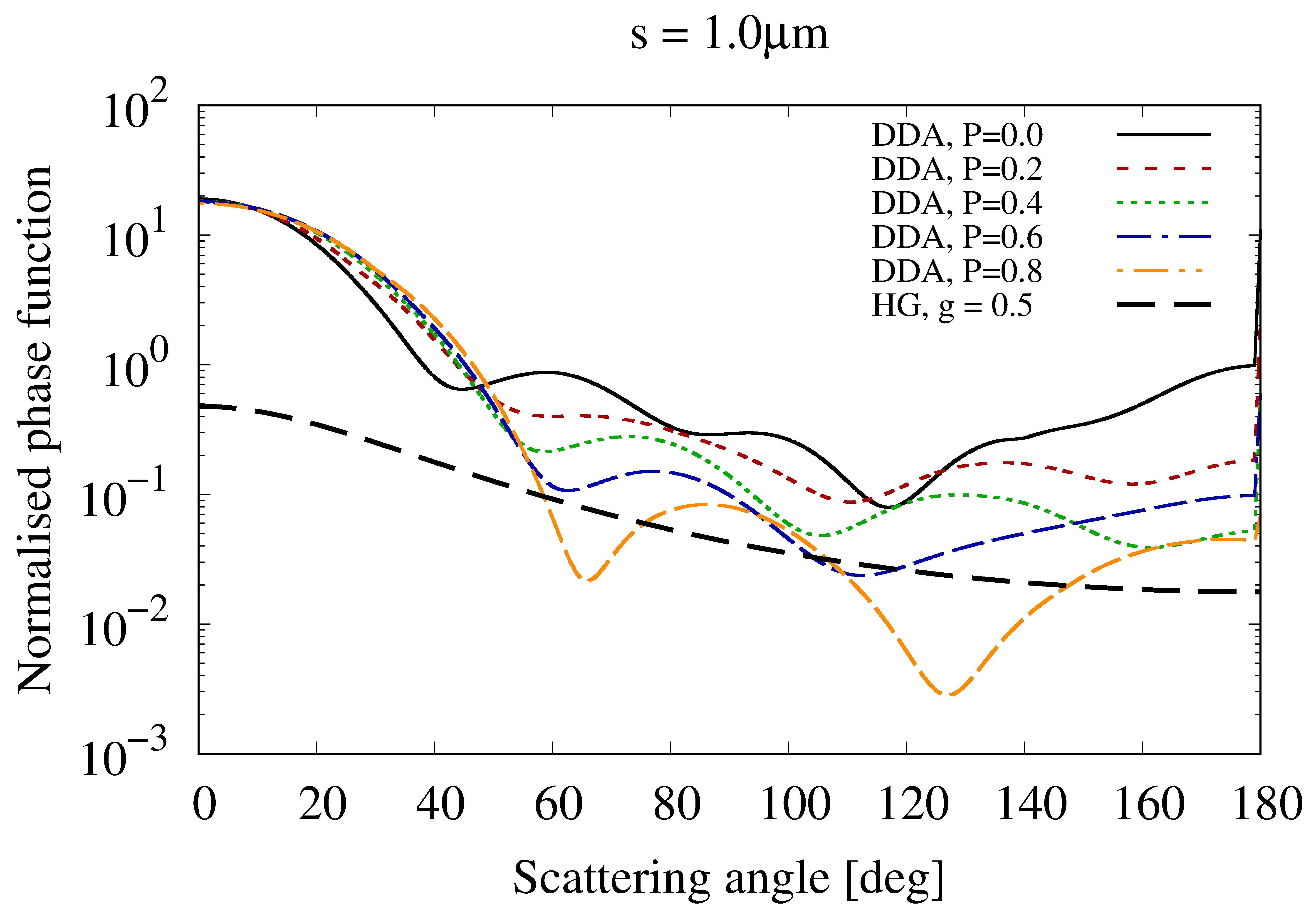}
    \includegraphics[width=0.45\textwidth]{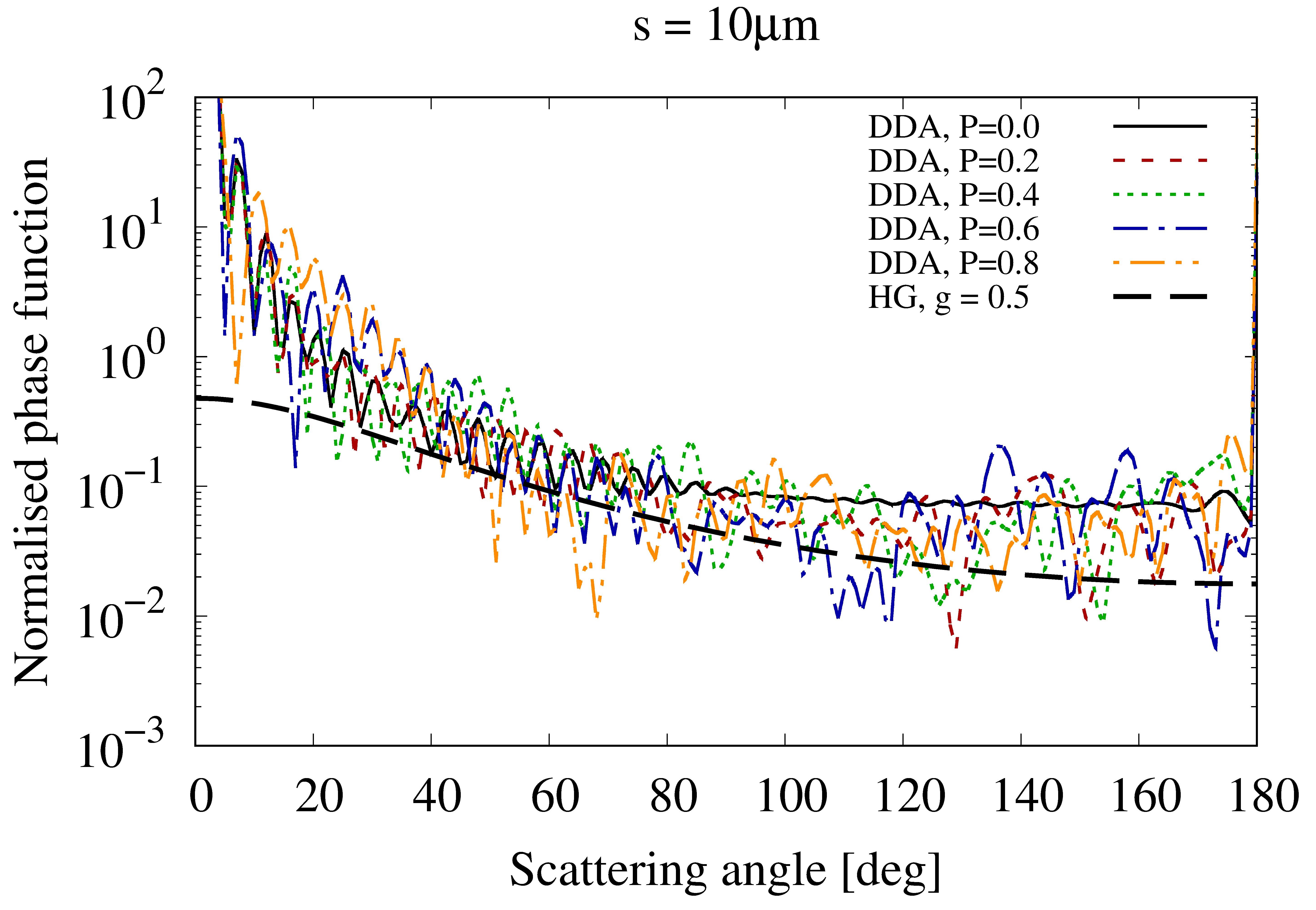}
    \includegraphics[width=0.45\textwidth]{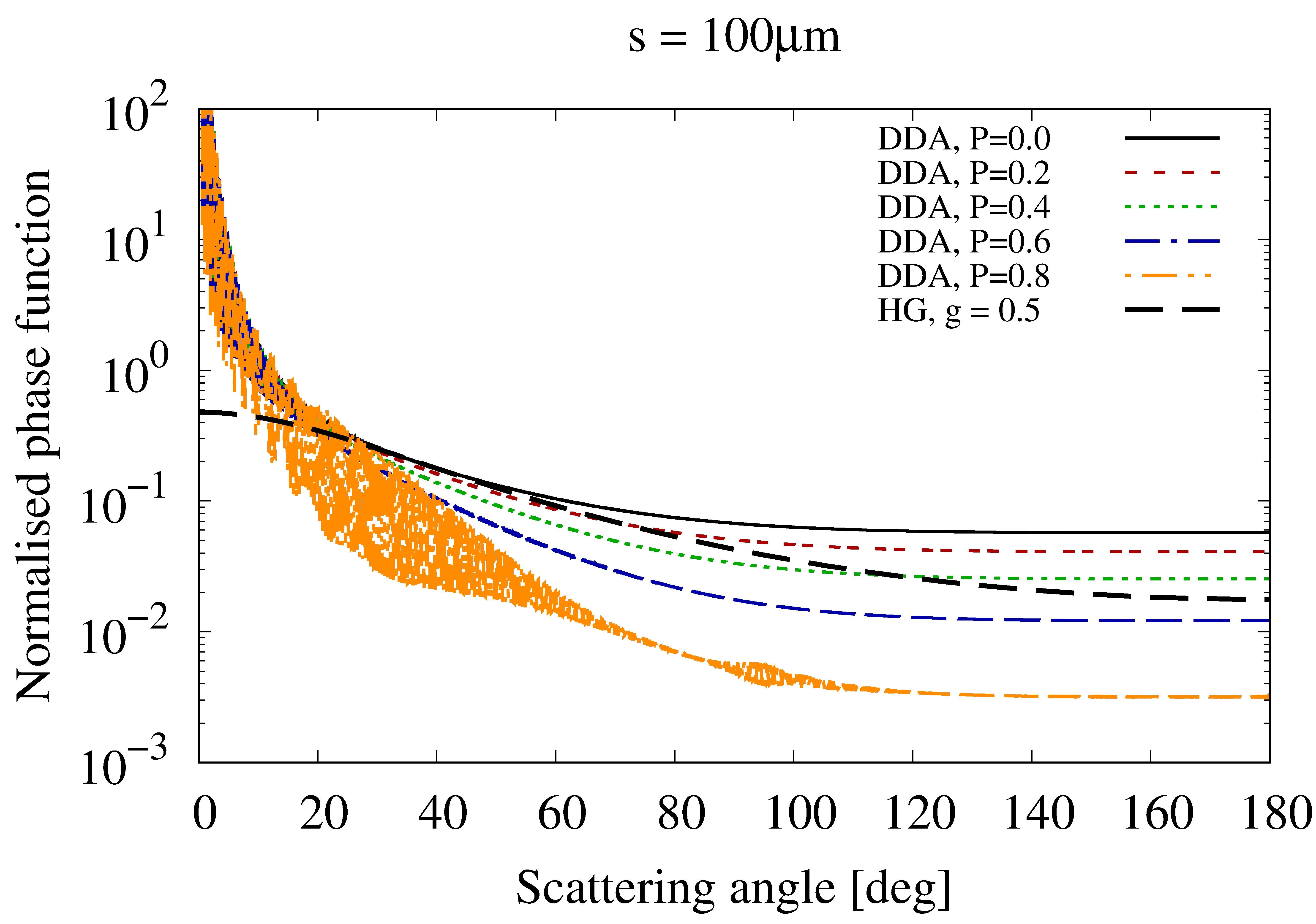}
    \includegraphics[width=0.45\textwidth]{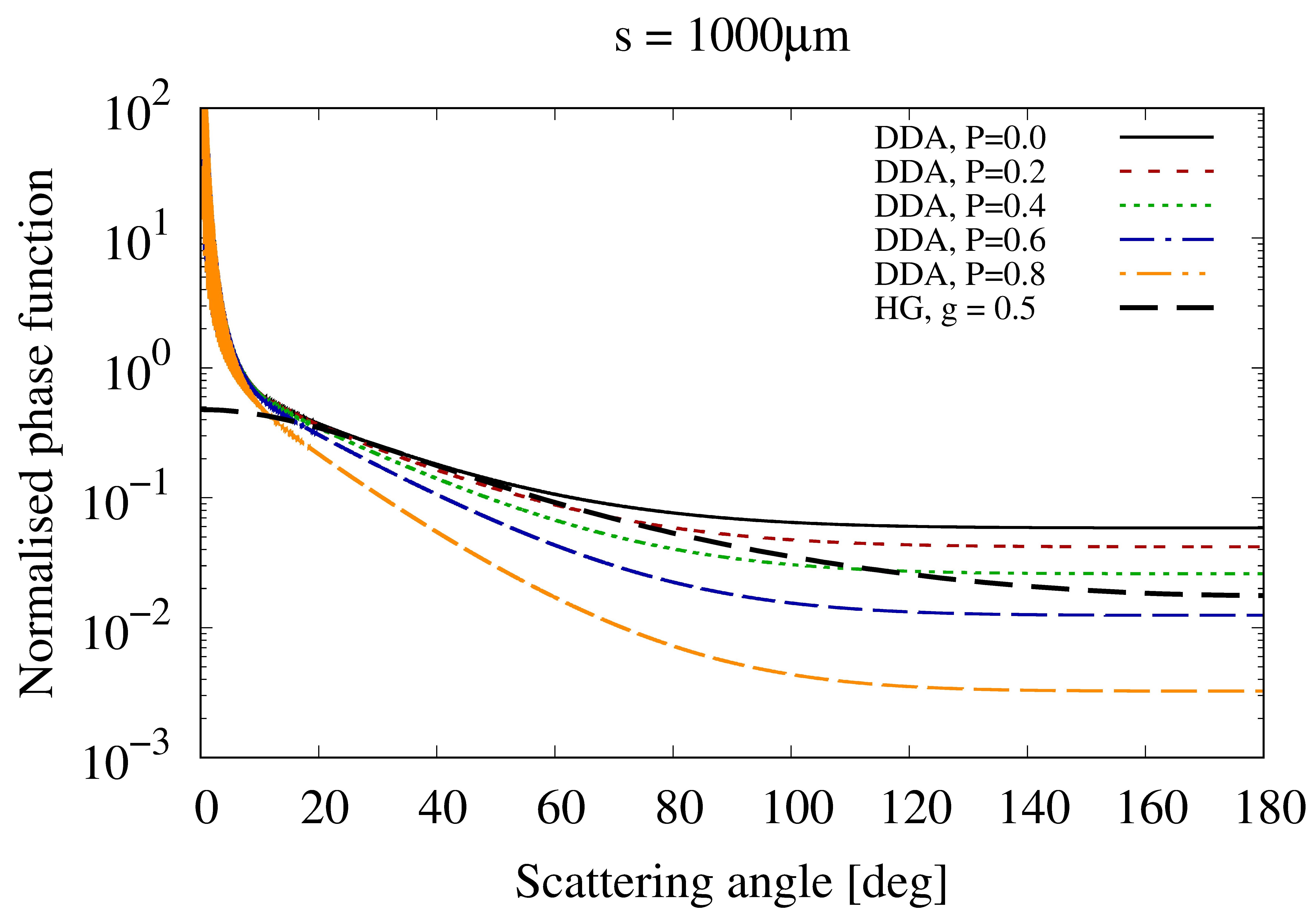}
    \caption{Scattering phase function as function of scattering angle for different porosities at a wavelength of 1.6$\mu$m. The different panels show the results for different grain sizes.}
    \label{fig:S11_theta_grains}
\end{figure*}

In Fig.~(\ref{fig:S11_theta_grains}) we show the scattering phase function for different porosities as inferred from our models: for $s\leq 10\mu$m we use DDA, for $s>10\mu$m we use Mie theory.

We see that for sub-micron-sized dust the scattering is more or less isotropic and the behaviour independent of the level of porosity.
This changes for micron-sized and larger grains. 
Up to 10$\mu$m-sized grains the phase functions become more complex in structure. 

For increasing porosity the values for back-scattering ($\vartheta\sim180^\circ$) decreases for all grains. However, we note that for $s=10\mu$m we cannot differentiate between the different phase functions. Interestingly, the forward-scattering ($\vartheta < 5^\circ$) does not change with porosity.

\section{Flux densities}
\label{sec:flux_densities}
\begin{figure}
    \includegraphics[width=\columnwidth]{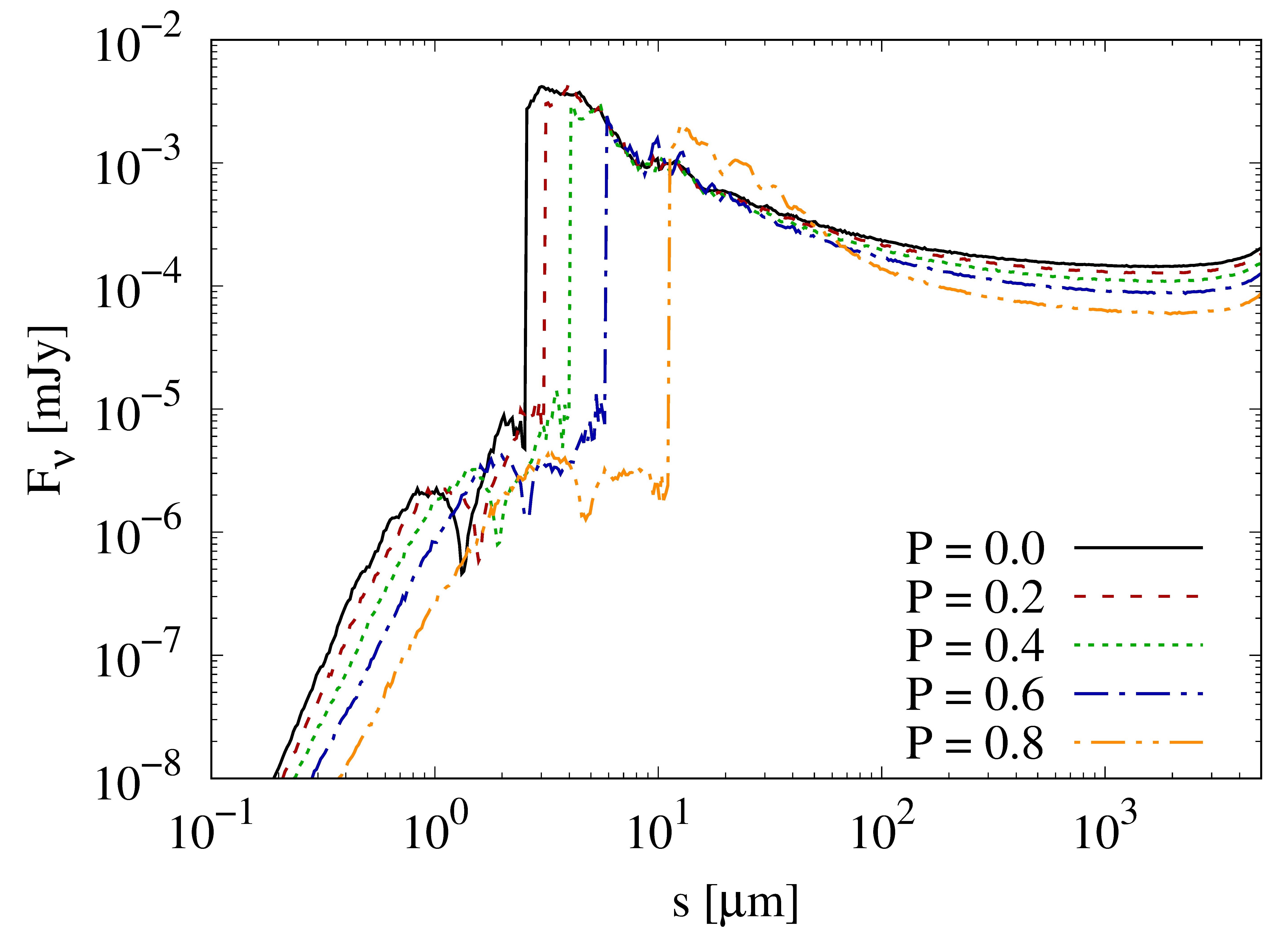}
    \caption{Flux density as function of grain size for different porosities.}
    \label{fig:Fnu_s_porosity}
\end{figure}

In Fig.~(\ref{fig:Fnu_s_porosity}) we show the contribution to the total flux density per size bin for different porosities. As expected from Fig.~(\ref{fig:beta_s}) we see an increase of the blowout size with increasing porosity. 
For bound grains the level of flux density is comparable. 
The total flux density is decreasing  with increasing porosity.

\section{$^{12}$CO position-velocity diagram} \label{sec:pvdiagram}

\begin{figure}
    \centering
    \includegraphics[width=1.0\columnwidth]{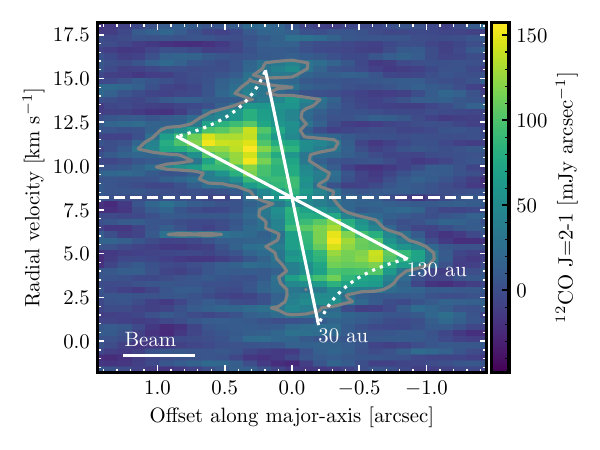}
    \caption{Position-velocity diagram of $^{12}$CO $J=2-1$ emission. The grey contours represent emission at $3\sigma$. The diagonal white solid lines represent the line-of-sight velocity of gas at a fixed orbital radius and in Keplerian rotation as a function of projected separation. The white dotted lines show the maximum velocity along the line of sight for a Keplerian rotational profile and as a function of separation. The horizontal white line at the bottom left represents the beam FWHM of 0.51 arcsec in the direction of the disc PA.}
    \label{fig:PVCO}
\end{figure}

Here we reanalyze the $^{12}$CO $J=2-1$ emission reported by \cite{moor-et-al-2017} to constrain the extent of the gas. Figure \ref{fig:PVCO} shows a positional-velocity diagram of $^{12}$CO obtained assuming the inclination and position angle derived from the scattered light images, and a stellar mass of 1.8~$M_{\odot}$ \citep{matra-et-al-2018}. We can constrain the radial distribution of CO by overlaying two diagonal lines representing the line-of-sight velocities as a function of separation along the major axis, at two fixed orbital radii, and assuming Keplerian rotation. The curves in white dotted lines connecting the two diagonal lines show the maximum line-of-sight velocities as a function of projected separation. By varying the two orbital radii such that the white wedges enclose most of the CO emission, we find that the CO gas is mostly contained between 30 and 130~au. The significant emission just outside the white wedges is due to the large beam size that smooths the radial extent of CO.    

\bsp	
\label{lastpage}
\end{document}